%% file: Planck_2015_XXVIII_astroph.tex
\documentclass[twocolumn,traditabstract,longauth]{aa}
\usepackage{graphicx}
\usepackage[breaklinks, colorlinks, citecolor=blue]{hyperref}
\usepackage{natbib}
\usepackage{subfigure}
\usepackage{epsfig}
\usepackage{multirow}
\usepackage{array}
\usepackage{psfrag}
\usepackage{color}
\usepackage{txfonts}

\usepackage{url}
\usepackage{ascii}
\usepackage[T1]{fontenc}
\usepackage{rotating}

\newcolumntype{L}[1]{>{\raggedright\let\newline\\\arraybackslash\hspace{0pt}}m{#1}}
\newcolumntype{C}[1]{>{\centering\let\newline\\\arraybackslash\hspace{0pt}}m{#1}}
\newcolumntype{R}[1]{>{\raggedleft\let\newline\\\arraybackslash\hspace{0pt}}m{#1}}

\input{Planck.tex}
\input{A37_Catalogue_cold_clumps_authors_and_institutes.tex}

\definecolor{Blue}{rgb}{0.,0.,1.}
\definecolor{LightSkyBlue}{rgb}{0.691,0.827,1.}
\definecolor{Red}{rgb}{1.,0.,0.}
\definecolor{Green}{rgb}{0.,1.,0.}
\definecolor{Purple}{rgb}{0.5, 0., 0.5}
\definecolor{Try}{rgb}{0.15,0.,1}
\definecolor{Black}{rgb}{0., 0., 0.}

\def\endPlancktable{\tablewidth=\columnwidth $$\hss\copy\tablebox\hss$$ \vskip-\lastskip\vskip -2pt}

\def\tablenote#1 #2\par{\begingroup \parindent=0.8em \abovedisplayshortskip=0pt\belowdisplayshortskip=0pt \noindent $$\hss\vbox{\hsize\tablewidth \hangindent=\parindent \hangafter=1 \noindent \hbox to \parindent{$^#1$\hss}\strut#2\strut\par}\hss$$
\endgroup}

\title{ {\Planck} 2015 results. XXVIII. \\ The Planck Catalogue of Galactic Cold Clumps}

\include{A37_Catalogue_cold_clumps_authors_and_institutes}
\abstract{We present the  Planck Catalogue of  Galactic Cold Clumps (PGCC), an all-sky catalogue of Galactic cold clump candidates detected by {\Planck}. 
This catalogue is the full version of the Early Cold Core (ECC) catalogue, 
which was made available in 2011 with the Early Release Compact Source Catalogue (ERCSC) and contained 915 high signal-to-noise sources.
It is based on the {\Planck} 48\,months mission data that are currently being released to the astronomical community. The
PGCC catalogue is an observational catalogue consisting exclusively of Galactic cold sources. The three highest {\Planck} bands (857, 545, 353\,GHz)
have been combined with IRAS data at 3\,THz to perform a multi-frequency detection 
of sources colder than their local environment. After rejection of possible extragalactic contaminants,  
the PGCC catalogue contains 13188 Galactic sources 
spread across the whole sky, i.e.,  from the Galactic plane to high latitudes, following the spatial distribution of the main molecular cloud complexes.
The median temperature of PGCC sources lies between 13 and 14.5\,K, depending on the quality of the flux density measurements, 
with a temperature ranging from 5.8 to 20\,K after removing sources with the 1\% largest temperature estimates. 
Using seven independent methods, reliable distance estimates have been obtained 
for 5574 sources, which allows us to
derive their physical properties such as their mass, physical size, mean density and luminosity. The PGCC sources are 
located mainly in the solar neighbourhood, up to a distance of 10.5\,kpc towards the Galactic centre, and range from 
low-mass cores to large molecular clouds. Because of this diversity and because the PGCC catalogue contains 
sources in very different environments, the catalogue is useful to investigate the evolution from molecular clouds to cores. 
Finally, the catalogue also includes 54 additional sources located in the Small 
and Large Magellanic Clouds.
}


\begin{document}
\authorrunning{{\Planck} Collaboration}
\titlerunning{PGCC catalogue}
\maketitle

\begin{figure*}[th!]
\center
\includegraphics[width=8cm, angle=90, viewport=35 20 420 730]{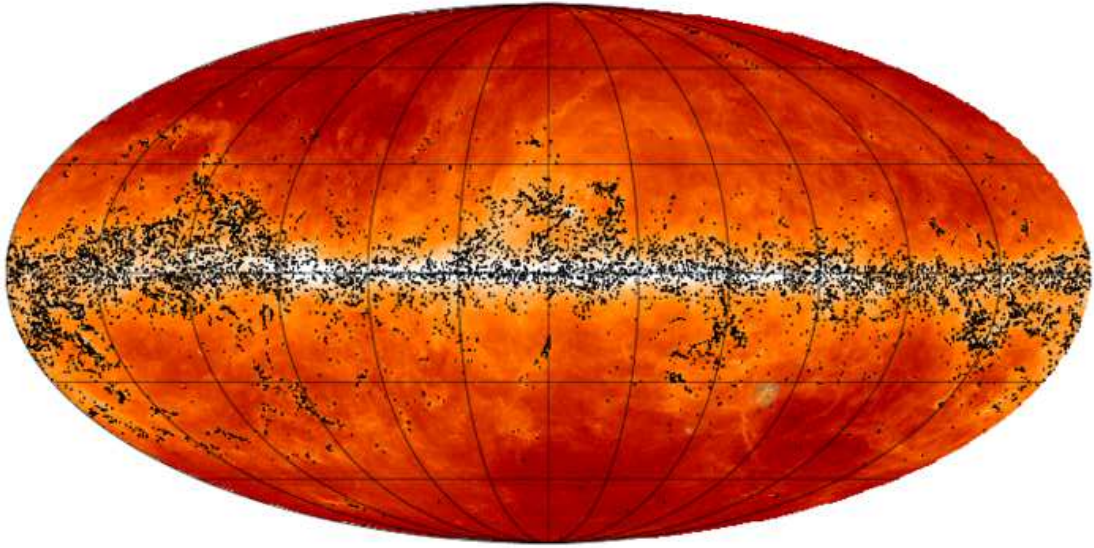} 
\caption{All-sky distribution of the  PGCC sources: 13188 Galactic clumps ({\it black dots}), plus 54 
 Large Magellanic Cloud (LMC) and Small Magellanic Cloud (SMC) clumps ({\it grey dots}) . 
  The source distribution is overlaid on the 857\,GHz {\Planck} map, shown in logarithmic scale between $10^{-2}$ to $10^2$\,$\rm{MJy sr^{-1}}$.}
\label{fig:allsky}
\end{figure*}

\alltwentyfifteenresultspapers

\section{Introduction}

The all-sky {\Planck}\footnote{\Planck\ (\url{http://www.esa.int/Planck}) is a project of the European Space Agency  (ESA) with instruments provided by two scientific consortia funded by ESA member states and led by Principal Investigators from France and Italy, telescope reflectors provided through a collaboration between ESA and a scientific consortium led and funded by Denmark, and additional contributions from NASA (USA).} mission 
has opened up the possibility of carrying out comprehensive investigations of the Galactic emission components. 
With its high sensitivity and wide wavelength coverage, {\Planck} is providing all-sky maps 
of the thermal dust emission and, in particular, of the emission arising from cold dust. Because cold dust is
mainly associated with dense regions within molecular clouds, these observations are  relevant for studies of the
early phases of star formation, in particular to explore how star formation depends on the physical
conditions provided by the parent cloud.  To this end, it is necessary to investigate the spatial distribution and physical properties of dense clumps in different
Galactic environments, and this objective can be attained only by extended surveys, which can cover the full range of scales encompassed 
by the star formation process, i.e.,  from subparsec to several kpc. 

During the past decade, new insights on the study of cold sources have been provided by sophisticated numerical 
modeling and by the development of sensitive millimetre and submillimetre detectors, operating both from space and from the ground, and with 
both imaging or spectroscopic capabilities \citep[e.g.,  Sect.~1 in][]{planck2011-7.7b}. 
By combining the highest frequency channels of the {\Planck} survey (353--857\,GHz, 350--850\,$\mu$m) 
with the far-infrared IRAS \citep{Neugebauer1984} data, and by applying a dedicated source detection method, which leverages on the cold sources spectral signature, we can 
obtain an all-sky census of the Galactic coldest objects. In particular, the method of \citet{Montier2010} 
makes it possible to separate cold and warm dust emission components, and to derive the physical properties (flux density, size of the emitting region,
temperature) of the cold component. Furthermore, {\Planck} has provided the first
uniform submillimetre surveys that cover both the Galactic plane and regions at intermediate and high latitudes, which allows us to expand the physical parameter space 
probed by the previously known cold sources. The {\Planck} detected sources span a broad range of temperature, mean density, mass, and size. 
The most compact and nearby sources have a linear diameter of $\sim0.1$\,pc. At large distances, though, and because of the limited instrument resolution,  
many sources have an intrinsic size of tens of parsec. More importantly, the average {\Planck} cold clump, with a linear diameter of 1\,pc, is typically characterized 
by the presence of sub-structures, each corresponding to individual cores, as revealed by the {\it Herschel} follow-up    
\citep{Juvela2010, Juvela2011, Juvela2012a, planck2011-7.7a, Montillaud2015}. The {\it Herschel} observations also highlighted that the {\Planck}  
sources likely correspond to different evolutionary stages, with  half of the targeted fields showing signs of 
active star formation, as indicated by the presence of mid-infrared point sources. In addition, the {\it Herschel} high-angular resolution has allowed us to shed light 
on the filamentary nature of a substantial fraction of {\Planck} clumps and has evidenced that, in one case out of ten, the clumps have a cometary shape 
or a sharp boundary indicative of compression by an external force \citep{Juvela2012a}. 

As part of the first {\Planck} data release, the sample of the most robust {\Planck} detections has been already delivered to the
astronomical community. This Early Cold Clump sample (ECC) included 915 {\Planck} cold clumps (at $T<14$\,K) that are 
distributed over the whole sky \citep{planck2011-1.10, planck2011-1.10sup}. We are now providing the entire catalogue of cold sources, i.e.,  the
Planck Catalogue of Galactic Cold Clumps  (PGCC), based on the full  {\Planck} 2014 data release over the all-sky and shown in Fig.~\ref{fig:allsky}. 

In this paper, we describe the generation and content of the PGCC catalogue. A detailed analysis of the cold source 
population contained in the catalogue will be presented in forthcoming papers. 
In Sect.~\ref{sec:method} we describe the data as well as the source detection and extraction method.  
In Sect.~\ref{sec:catalogue_building} we discuss the generation of the catalogue, including the applied quality flags for the source selection, and the 
photometric measurements. In Sect.~\ref{sec:statistical_validation} we discuss the source validation process of the detection algorithm, based on a statistical analysis.
In Sect.~\ref{sec:distance} we present the different methods used to derive distance estimates for the clumps. 
In Sect.~\ref{sec:physical_properties} we describe the derivation of other physical properties of the sources such as mass and luminosity. 
Finally, in Sect.~\ref{sec:ancillary_validation}, we provide details  on the cross-matching of the final catalogue with ancillary catalogues and complementary 
data sets.

\section{Source detection and photometry}
\label{sec:method}

\subsection{Data set}
\label{sec:input_data}

This paper is based on the whole observing time of the {\Planck} mission, corresponding to the 5 all-sky surveys. 
Here we approximate the {\Planck} beams by using effective circular Gaussians
\citep{planck2013-p02d, planck2013-p03}. The Full-Width-Half-Maximum (FWHM) at each frequency channel is given in Table~\ref{tab:fwhm}. 
In addition, in this work, we focus on the three highest {\Planck} frequency channels, i.e.,   
857, 545, and 353\,GHz, which are designed to cover the Galactic cold dust emission peak.  
The 217\,GHz band has not been included in our analysis, despite being characterized by an angular resolution comparable to the 
other bands, and this is for two reasons: i) this band is strongly contaminated by the CO $J$=2$\rightarrow$1 emission line, as described in \citet{planck2013-p03d}, and this 
contribution is expected to be significant towards dense regions, given their associations with molecular material; ii) the contamination by the 
cosmic microwave background may become problematic in this band at high latitude, implying a complex component separation issue.
The noise in the channel maps is assumed to be Gaussian with a standard 
deviation of 1.55$\times$$10^{-2}$, 1.49$\times$$10^{-2}$ and 1.4$\times$$10^{-3}$\,MJy sr$^{-1}$ at 857, 545, and 353\,GHz, respectively 
\citep{planck2013-p01}. The absolute gain calibration of High Frequency Instrument (HFI) maps is about
 1.2\%, 6.08\%, and 6.33\% at 353, 545, and 857\,GHz, respectively \citep[see Table 6 in ][]{planck2013-p01}. 
Further details on the data reduction, {\Planck} frequency maps and the calibration scheme can be found in \citet{planck2013-p03}.

The {\Planck} data are combined with the IRIS all-sky data \citep{Miville2005}, i.e.,  a reprocessed version of the IRAS data. 
As described in \citet{planck2011-1.10} and \citet{planck2011-7.7b}, 
the IRIS 3\,THz (100 $\mu$m) data have been chosen because they allow us to complement the {\Planck} data. In fact: 
i) 3\,THz is a very good tracer of Galactic warm ($\sim20$\,K) dust;  ii) the emission from small grains does not contribute substantially at this 
frequency; iii) the IRIS (4{\parcm}3) and {\Planck} data angular resolutions are very similar (see Table~\ref{tab:fwhm}). 
Note that the IRAS survey coverage presents two gaps, which in total account for 2\% of the whole sky. In IRIS data, these gaps were  
filled in by using lower angular resolution {\it DIRBE} data ($\sim40${\arcmin}). Due to the discrepancy in resolution between IRIS and {\it DIRBE}, these regions have been excluded from 
our analysis. Furthermore, sources detected by our algorithm close to the location of the gaps were carefully examined, since they might be contaminated by 
noisy features in the IRIS 3\,THz map. 

All {\Planck} and IRIS maps have been convolved to the same resolution, 5{\arcmin} FWHM, before performing source detection and extraction.

\begin{table}
\caption{FWHM of the effective beam of the IRAS and {\Planck} high frequency channel maps.}
\label{tab:fwhm}
\nointerlineskip
\setbox\tablebox=\vbox{
\newdimen\digitwidth 
\setbox0=\hbox{\rm 0} 
\digitwidth=\wd0 
\catcode`*=\active 
\def*{\kern\digitwidth} 
\newdimen\signwidth 
\setbox0=\hbox{+} 
\signwidth=\wd0 
\catcode`!=\active 
\def!{\kern\signwidth} 
\newdimen\pointwidth 
\setbox0=\hbox{.} 
\pointwidth=\wd0 
\catcode`?=\active 
\def?{\kern\pointwidth} 
\halign{\hbox to 0.6 in{\hfil#\hfil}\tabskip=1.0em&
\hfil#\hfil&
\hfil#\hfil\tabskip=0pt\cr
\noalign{\doubleline}
Frequency  &  Wavelength &  FWHM  \cr
$[$GHz$]$ & $[\mu$m$]$ & $[$'$]$ \cr
\noalign{\vskip 3pt\hrule\vskip 4pt}
 *353 & 850 & 4.818*$\pm$*0.024 \cr
*545 & 550 & 4.682*$\pm$*0.044 \cr
*857 & 350 & 4.325*$\pm$*0.055 \cr
3000 & 100 & 4.300*$\pm$*0.200 \cr
\noalign{\vskip 3pt\hrule\vskip 4pt}
}}
\endPlancktable
\end{table}

\subsection{Detection method}
\label{sec:detection_method}

To detect cold sources in the combined {\Planck} and IRIS 3\,THz maps, we 
applied the {\tt CoCoCoDeT} detection algorithm presented in  \citet{planck2011-7.7b}, and described in \citet{Montier2010}. 
This algorithm is based on a multi-frequency approach which exploits the specific colour properties of this type of sources. 
The detection is performed independently at 857, 545, and 353\,GHz using the {\it cold residual} maps, which are 
built by subtracting a {\it warm} component from each frequency map. 
This {\it warm} component is estimated separately in each pixel by extrapolating a {\it warm} template, i.e.,  the IRIS 3\,THz map, 
to a given {\Planck} frequency $\nu$, using the local average background colour estimated at 3\,THz and $\nu$, and 
computed in an annulus from 5{\arcmin} to 15{\arcmin} centred on the pixel. The catalogues obtained in each of the three {\Planck} bands, 857, 545, and 353\,GHz, 
are then merged by requiring: i) a detection in each band on the {\it cold residual} maps; ii) a signal-to-noise ratio (S/N) greater than 4 in all bands; 
iii) a maximum distance between the centres of the three detections of 5{\arcmin}. These criteria assure cross-band detection consistency as well as source compactness. 
{ More details can be found in  \citet{planck2011-7.7b}. }

We emphasize that our method differs from classical detection algorithms that typically perform the detection directly on frequency maps, as for instance 
is the case for the Planck Catalogue of Compact Sources \citep[PCCS][]{planck2013-p05}. 
Our method allows a detection in {\it temperature}: cold sources show a positive signal in the {\it cold residual} maps, while warm sources
show a negative signal. More precisely, this technique allows us to enhance sources having a temperature lower than the local background.
This does not automatically imply that the detected sources are intrinsically cold: for example, 
a source could be detected as {\it cold} simply because it is seen against a very warm background (or foreground). This is the typical case 
of objects located along the line-of-sight of active star forming regions. In building the catalogue, this effect has been taken into account, 
as discussed in Sect.~\ref{sec:nearby_hot_sources}.

\subsection{Photometry}
\label{sec:photometry}

The flux density of the cold clumps has been estimated from the IRIS and {\Planck} bands using the algorithm described in \citet{planck2011-7.7b}. 
Here we recall the main steps of the method: i) determination of the clump size and position by means of an elliptical Gaussian fit of the 
{\it cold residual} map at 857\,GHz at the location where the S/N is the highest; 
ii) polynomial fit of the background surface at 3\,THz and removal of the cold component from the 3\,THz warm template 
before extrapolation to the {\Planck} bands; iii) aperture photometry in all bands using the elliptical profile. 
We provide two estimates of the flux density in each band, one for the cold clump and one for its associated {\it warm background}. 
The estimate for the clump is based on the {\it cold residual} maps, after subtraction of the warm component, while 
the {\it warm background} flux density is computed from the warm component, extrapolated from 3\,THz and integrated over a solid angle with the same size as the cold clump. 
The sum of these two values gives the total flux density of the cold clump in the original IRIS and {\Planck} maps. 

The uncertainty on the flux density estimates have been obtained by performing Monte-Carlo simulations. In particular, 
for each source we inject one at a time, in an annulus extending from 10{\arcmin} to 30{\arcmin} centred on the source itself and in both the IRIS and {\Planck} {\it cold residual} maps,  
artificial sources with the same flux density and elliptical shape. This operation allows us to preserve the instrumental noise and confusion 
level of the original maps. The uncertainty on the true source photometric measurement is then given by the standard deviation of the flux densities of the artificial sources 
when the same photometric steps are applied to both the true and artificial sources. We emphasize that our flux density uncertainties are likely slightly conservative, as 
the confusion arising from injecting artificial sources in the proximity of the true source generates additional noise and this term is included in our calculations. 
 More details can be found in  \citet{planck2011-7.7b}. 
We have performed a Monte Carlo Quality Assessment (MCQA) to evaluate the overall quality of our photometric measurements and this is 
presented in Sect.~\ref{sec:statistical_validation}. As in \citet{planck2011-7.7b}, we assign to the catalogued sources quality flags to indicate the accuracy on the estimated 
flux densities and sizes, and these flags are used to divide the sources into three categories of increasing flux quality, as described in Sect.~\ref{sec:photometry_quality}.

Notice that a minimum distance of 5{\arcmin} between two sources was required for detections at 857, 545, and 353\,GHz.
However, band-merging and elliptical Gaussian fitting both modify the final centroid coordinates of the clumps, which therefore may not longer satisfy the
5{\arcmin} criterion. Furthermore, elongated clumps may partially overlap even at a distance greater than 5{\arcmin} between them. 
Because Galactic very sources are preferentially  found highly structured regions of the Galaxy, where confusion is significant, 
we have to face severe blending issues.
After obtaining the elliptical Gaussian profile of all the sources, we compute the overlap between a given source and all its neighbours
located within 15{\arcmin}. All sources with a non-zero overlap are flagged ({\asciifamily FLUX\_BLENDING}), 
and further information is provided, as detailed in Sect.~\ref{sec:deblending}.

\section{Catalogue generation}
\label{sec:catalogue_building}

\begin{figure*}[hp!]
\center
\begin{tabular}{c}
\includegraphics[width=8cm, angle=90, viewport=35 20 420 730]{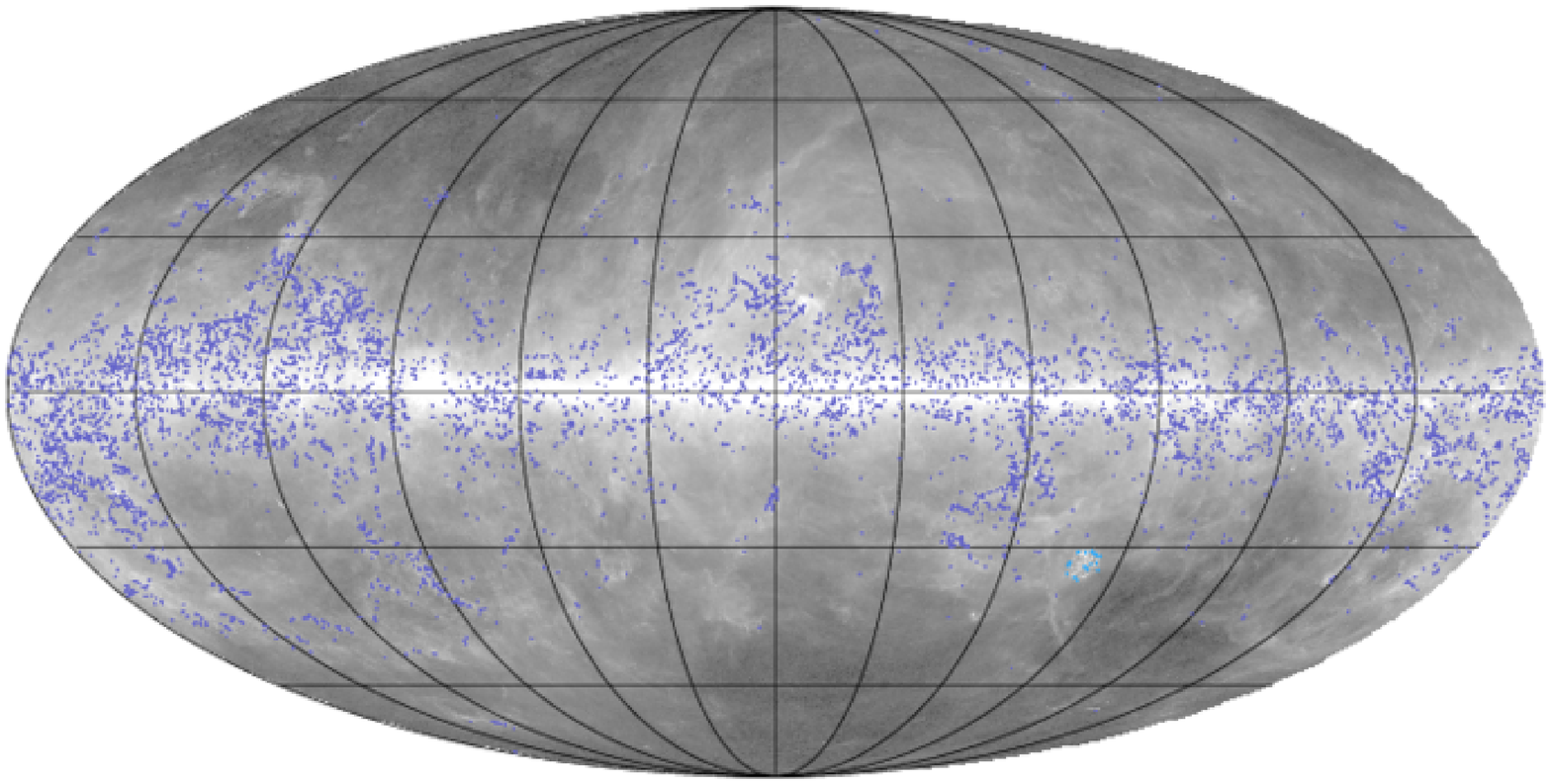} \\
\includegraphics[width=8cm, angle=90, viewport=35 20 420 730]{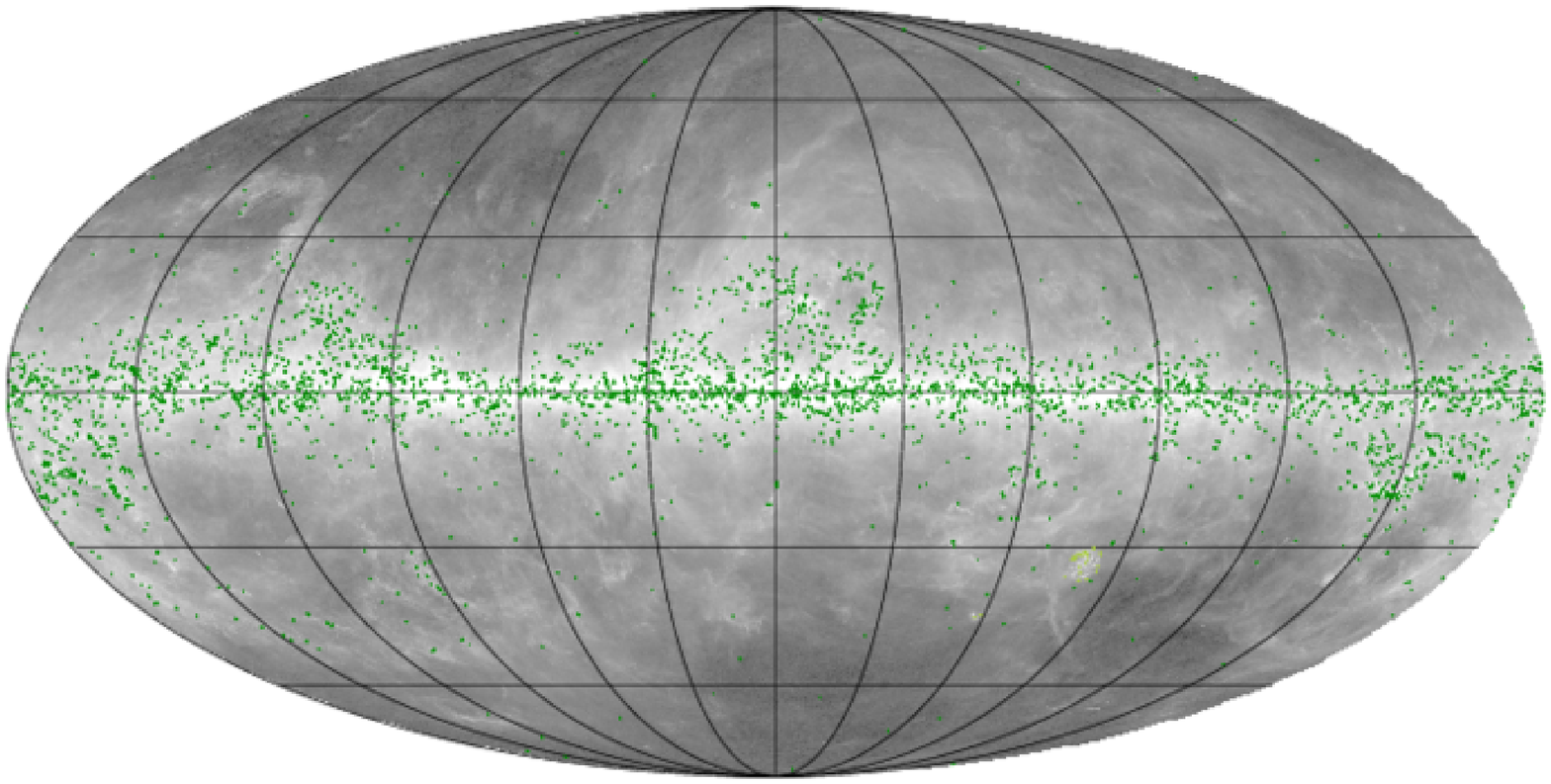} \\
\includegraphics[width=8cm, angle=90, viewport=35 20 420 730]{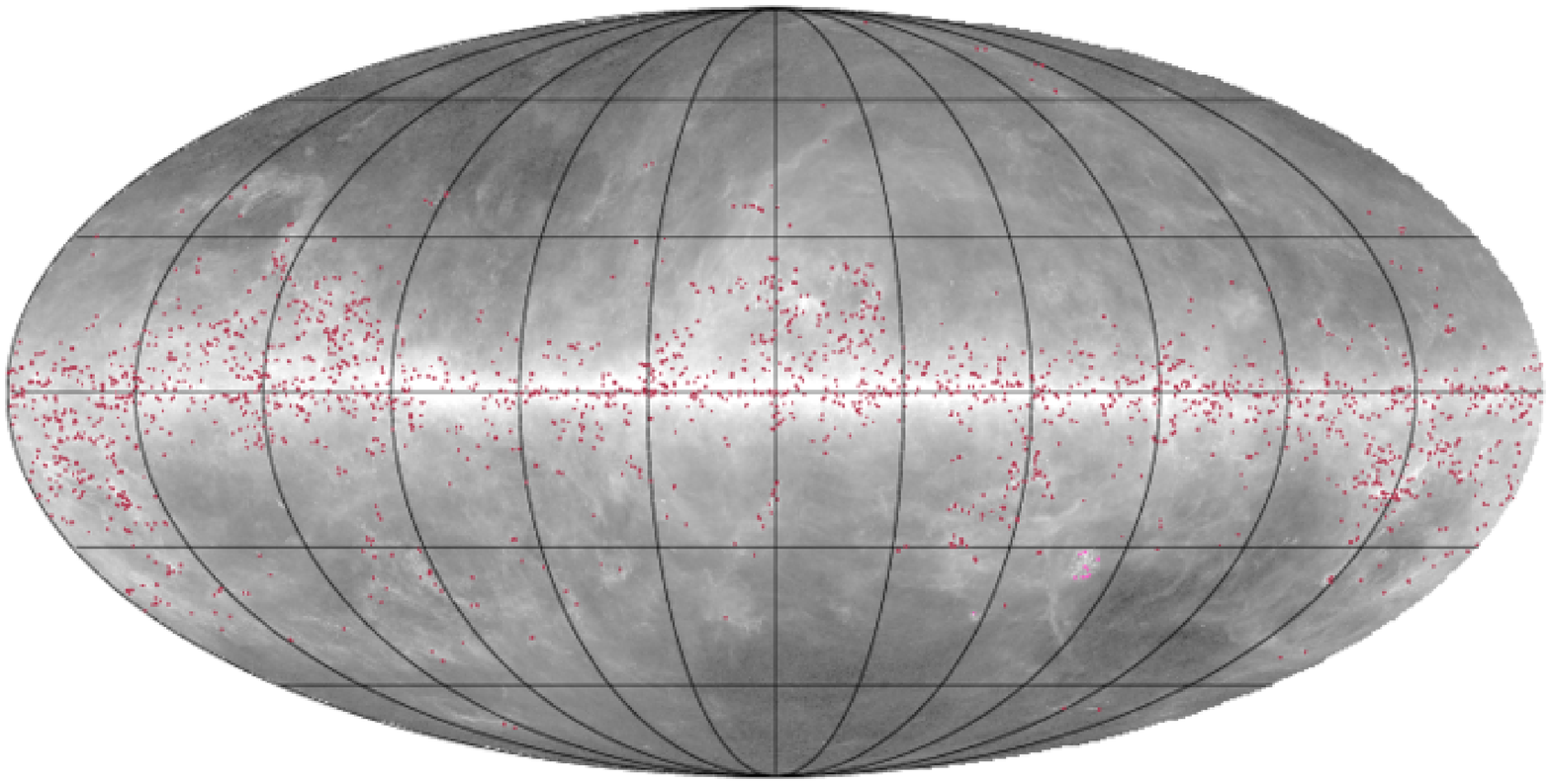} 
\end{tabular}
\caption{All-sky distribution of the PGCC sources according to their {\asciifamily FLUX\_QUALITY} category: "{\it Reliable flux densities}" (FQ=1, {\it top panel}), 
"{\it Missing 3\,THz flux density}" (FQ=2, {\it middle panel}) and "{\it Detection only}" (FQ=3, {\it bottom panel}). 
Sources located in the LMC and SMC are also shown in light colours.}
\label{fig:allsky_fluxquality}
\end{figure*}

In this section, we describe the final selection of sources, 
starting from the source list generated by the detection algorithm. In particular, we discuss how we have increased the reliability of the 
catalogue by rejecting spurious sources and extragalactic contaminants. Finally, we provide details on the catalogue 
content.

\subsection{Selection based on photometric quality}
\label{sec:photometry_quality}

We have seen that the detection algorithm described in Sect.~\ref{sec:detection_method} has been applied to the combined {\Planck} and IRIS 3\,THz data. 
After rejection of the spurious detections obtained in the proximity of the gaps in the IRAS map, we are left with 13832 
sources. As a first validation step, we have used the quality of the photometric measurements discussed in Sect~\ref{sec:photometry} 
to identify additional spurious sources. Following this procedure, 428 detections ($\sim 3$\% of the total) are rejected 
because of highly inaccurate photometry. The typical case of rejected sources at this stage is that of sources with a negative flux density 
estimate, which can be caused by the presence of stripes in the data. The 13404 remaining sources have been divided into three categories, 
according to the quality of their flux density values (see {\asciifamily FLUX\_QUALITY} flag). The three categories are:

\begin{description}
\item[category 1: "{\it Reliable flux densities}"]: sources with flux densities at S/N$ > 1$ in both {\Planck} (857, 545, and 353\,GHz) and IRIS 3\,THz bands, allowing a full characterization
of their colour ratio and their temperature, as required for their validation. These sources represent the highest quality sample of the PGCC catalogue, 
and have {\asciifamily FLUX\_QUALITY}=1. 
\item[category 2: "{\it Missing 3\,THz flux density}"]: sources with flux densities S/N$ > 1$ in all bands except for 
 the IRIS 3\,THz band, where we have obtained only an upper limit. These sources are typically characterized by 
low flux densities and extremely cold temperatures, and have no  detectable counterparts in the infrared. 
They are potentially interesting very cold clump candidates, and have {\asciifamily FLUX\_QUALITY}=2. 
\item[category 3: "{\it Detection only}"]: sources for which the quality of the elliptical Gaussian fit is very poor, thus no 
reliable flux density estimate can be obtained. These sources are likely extended or embedded in a complex environment. 
They have {\asciifamily FLUX\_QUALITY}=3. 
\end{description}

We provide the number of sources in each category in the first row of Table~\ref{tab:pgcc_numbers}. 
After removal of the contaminants, such as sources affected by the presence of nearby hot sources 
(see Sect.~\ref{sec:nearby_hot_sources}) or extragalactic objects  (see Sect.~\ref{sec:extragal}), 
we obtain the final numbers of sources shown in the last row of the Table. The all-sky distributions of the sources 
in each {\asciifamily FLUX\_QUALITY} (hereafter FQ) category are shown in Fig.~\ref{fig:allsky_fluxquality}.

\subsection{Blending issue}
\label{sec:deblending}
 
A total number of 1757 sources are affected by blending, i.e.,  their
flux density estimates have been compromised by the presence of a nearby and partly overlapping source.
In this case, the flag {\asciifamily FLUX\_BLENDING} is raised, and an 
approximate estimation of the contamination level is performed. This 
is expressed in terms of a relative bias ({\asciifamily FLUX\_BLENDING\_BIAS}) of the original flux density estimate in each band, 
when flux density estimates are available for both involved sources. The median value of this bias is around $-$37\%, although it varies greatly from one source to  another.
This bias is only indicative and cannot be used to correct the flux density estimates.  A more accurate estimate of these flux densities 
could be obtained by performing a detailed analysis on individual sources, which, in particular, should take into account the local background fluctuations, hence 
the relative contribution of each source component to the integrated flux density.
A visual inspection to the ID cards of the blended sources may help to get an idea of the complexity of each case.
For each source impacted by blending, we also provide the catalogue index of the companion source ({\asciifamily FLUX\_BLENDING\_IDX})  with
the angular distance to its centroid ({\asciifamily FLUX\_BLENDING\_ANG\_DIST}).

\subsection{Extragalactic contamination}
\label{sec:extragal}

The goal of the PGCC catalogue is to contain a selection as large as possible of Galactic cold clump candidates. Since
 the {\tt CoCoCoDeT} detection algorithm is applied to all-sky maps, it is possible to have contamination from extragalactic objects. 
Hereafter we describe the three independent methods that we have used to identify and reject this type of contaminants. 

The first step consists of a colour-colour selection of 'radio-type' objects, characterized by a  flat  spectral energy distribution  (SED) 
in the submillimetre and millimetre wavelength range. This kind of objects may have been detected by our algorithm because of the flattening of their SED 
around 857 and 545\,GHz, which tends to mimic a cold black body spectrum. In this case, 
we have used the 353\,GHz to 545\,GHz flux density ratio to discriminate between radio-emitting and other type of objects. 
We found 26 objects with a ratio $S_{353}/S_{545}>0.9$, typical of an extremely flat or increasing SED in the millimetre domain.

The second step consists in cross-correlating the PGCC catalogue with extragalactic catalogues, such as: the Messier \citep{Messier1981} catalogue, the NGC \citep{Dreyer1888} 
and IC \citep{Dreyer1895} catalogues of nearby galaxies, and the 3C \citep{Edge1959,Bennett1962} and 4C \citep{Pilkington1965,Gower1967} catalogues of quasars. 
The cross-correlation has been performed using a 5{\arcmin} radius, leading to 66 found associations between a cold clump and an extragalactic object.

In the last step, we have searched for possible optical counterparts in the Digitized Sky Survey (DSS) data \citep{Djorgovski2003}.
The whole sample of cold clump candidates has been visually inspected to look for extended and/or bright sources in DSS images located close to or 
at the cold clump coordinates. After selecting some 800 sources with potential DSS counterparts, for each of these we have carefully examined ancillary \citet{Dame2001} CO data, 
extinction maps obtained with the NICER algorithm \citep{Lombardi2009} and {\it cold residual} maps. In addition, we have also searched for possible counterparts 
in the {\Planck} Low Frequency Instrument (LFI) data. The combination of all these data sets has allowed us to identify 43 sources that 
are likely of extragalactic origin. 

Finally, we have merged the three samples of extragalactic contaminants and rejected 114 unique objects from the initial source list. 

The {\tt CoCoCoDeT} algorithm has also detected sources in the Large Magellanic Cloud (LMC) and the Small Magellanic Cloud (SMC), which have not been 
rejected from the catalogue. Because of the proximity of these two galaxies, the {\Planck} resolution and 
sensitivity allow  the detection of  individual cold clumps, forming a potential very interesting sample. 
Hence 51 {\Planck} clumps falling inside a radius of $4\pdeg09$ centred on the Galactic coordinates  (279.03, -33.60), as defined by \citet{Staveley2003},  
are flagged ({\asciifamily XFLAG\_LMC}) as part of the LMC, 
and 3 others falling inside a radius of $2\pdeg38$ centred on the Galactic coordinates (302.67, -44.46), following \citet{Stanimirovic1999},  
are flagged ({\asciifamily XFLAG\_SMC}) as part of the SMC. 
Remarkably, follow-up observations of the LMC and SMC {\Planck} clump candidates have confirmed the nature of these objects, as discussed in Appendix~\ref{sec:lmc_smc}.

\subsection{Nearby hot source contamination}
\label{sec:nearby_hot_sources}

The detection algorithm {\tt CoCoCoDeT} \citep{Montier2010} used to extract the 
cold clumps from the {\Planck} and IRIS data
is designed to detect sources colder than a median background estimated in the neighbourhood of the source.
As stressed in Sect.~\ref{sec:detection_method}, this method allows us to detect cold regions embedded in a warm background, but 
it can also yield detections of extended envelopes of warm sources that appear colder than their environment but are not intrinsically cold.  
In some extreme cases, a hot source can
cause an overestimation of the background temperature in its proximity and thus lead to spurious detections.

In order to investigate this type of contamination, we use the {\it cold residual} maps as an indicator of the warm background around
the source and look for negative contiguous pixels. By definition, where the {\it cold residual} is positive, the relative temperature is colder than the background, 
and the other way round, i.e.,  where the residual is negative, the temperature is higher. Thus we can build a list of hot 
point sources by using the same detection algorithm as for the cold clumps, but this time by applying it to the reverse of the 
{\it cold residual} maps. We then compute the minimum distance between any cold clump candidate 
and  hot source detections in a 15{\arcmin} radius from the cold clump coordinates centre. 
This yields 2464 cold-hot associations with distances ranging from 2{\parcm}6 to 15{\arcmin}.

We emphasize that the presence of hot sources in the proximity of cold sources does not lead systematically to spurious detections. 
This kind of association is expected in star formation regions, as both pre- and proto-stellar cores often reside in the same 
molecular cloud. In fact, the formation of cold and compact condensations may even be triggered by nearby star formation. 
For this reason, we only reject cold clump detections that are associated with a hot source located inside a 5{\arcmin} radius from the 
cold clump coordinate centre (48 sources) and we flag the other  2416 cases while providing the distance between the 
centre of the cold clump and the hot source ({\asciifamily NEARBY\_HOT\_SOURCE}).

\subsection{Description of the catalogue content}

The final PGCC catalogue counts 13188 Galactic sources and 54 sources located 
in the LMC and SMC, divided into three categories of flux density quality ({\asciifamily FLUX\_QUALITY} flag).
The number of sources in each category is given in Table~\ref{tab:pgcc_numbers}. The Table also provides 
the number of sources that are flagged due to the presence of nearby hot sources ({\asciifamily NEARBY\_HOT\_SOURCE} flag)
or because they are located in the LMC and SMC.
The all-sky distribution of the PGCC sources, overlaid on the 857\,GHz {\Planck} data, is shown in Fig.~\ref{fig:allsky}, while 
Fig.~\ref{fig:allsky_fluxquality} shows the distribution per flux density quality flag. 
The columns in the catalogue and their meaning is given in Tables~\ref{tab:pgcc_listing_1} and \ref{tab:pgcc_listing_2} in Appendix~\ref{sec:cat_content}. 
The distance estimates and  related physical properties are described in Sects.~\ref{sec:distance} and \ref{sec:physical_properties}. 
The PGCC catalogue is available online \footnote{ESA PGCC link}, together with 30'$\times$30' cutouts, respectively, from the IRIS 3\,THz, the {\Planck} frequency map, and
the {\Planck} {\it cold residual} maps.

\begin{table}[t]
\caption{Description of the PGCC content, providing the total amounts of initial, rejected, final and flagged sources, and split in each category of the flux density quality.}
\label{tab:pgcc_numbers}
\nointerlineskip
\setbox\tablebox=\vbox{
\newdimen\digitwidth 
\setbox0=\hbox{\rm 0} 
\digitwidth=\wd0 
\catcode`*=\active 
\def*{\kern\digitwidth} 
\newdimen\signwidth 
\setbox0=\hbox{+} 
\signwidth=\wd0 
\catcode`!=\active 
\def!{\kern\signwidth} 
\newdimen\pointwidth 
\setbox0=\hbox{.} 
\pointwidth=\wd0 
\catcode`?=\active 
\def?{\kern\pointwidth} 
\halign{\hbox to 1.3 in{#\leaderfil}\tabskip=1.0em&
\hfil#\hfil&
\hfil#\hfil&
\hfil#\hfil&
\hfil#\hfil\tabskip=0pt\cr
\noalign{\doubleline}
\omit  & \multicolumn{3}{c}{{\asciifamily FLUX\_QUALITY}} & Total \cr
\omit & 1 & 2 & 3 & \cr
\noalign{\vskip 3pt\hrule\vskip 4pt}
 Initial & 7062 & 3833 & 2509 & 13404 \cr
 Extragalactic sources & **46 & **40 & **28 & **114 \cr
 LMC / SMC &  **19 &  **25 &  **10 &   ***54 \cr
Nearby hot sources &  ***4 &  **13 &  **31 &   ***48 \cr
\noalign{\vskip 3pt\hrule\vskip 4pt}
 Final & 6993 & 3755 & 2440 & 13188 \cr 
\noalign{\vskip 3pt\hrule\vskip 4pt}
 Flag nearby hot sources &  *758 &  1025 &  *633 &   *2416 \cr 
 Flag blending &  *726 &  *528 &  *503 &   *1757 \cr 
 With distance estimate & 2940  &    1686      &    *948   &     *5574 \cr
 \noalign{\vskip 3pt\hrule\vskip 4pt}
}}
\endPlancktable
\end{table}

\section{Quality assessment}
\label{sec:statistical_validation}

\subsection{MCQA simulations}

We have carried out a Monte Carlo Quality Assessment (MCQA) to quantify the performance of  
the detection algorithm applied to the combined IRIS 3\,THz and {\Planck} data set. 
To this end, we have generated all-sky simulations by injecting artificial sources in the IRIS and {\Planck} maps,  
and then applied the {\tt CoCoCoDeT} algorithm described in Sect.~\ref{sec:detection_method}. 
In total, 150\,000 sources have been injected over the whole sky, divided into 15 sky realizations. 
Each source is characterized by a temperature, a fixed emissivity spectral index $\beta$=2,  
a flux density at 857\,GHz, and an elliptical Gaussian profile  
(major and minor axis, ellipticity and position angle). The simulated sources are randomly distributed across the sky, using a uniform spatial distribution, 
at a minimum distance of 12{\arcmin} from the true {\Planck} cold clump centre coordinates.

The synthetic temperatures range from 6 to 20\,K, while the synthetic flux densities at 857\,GHz follow 
 a  uniform distribution in logarithmic scale between 1.5\,Jy and 500\,Jy, indicating that we inject more faint sources that are effectively detected. 
Temperatures and flux densities are 
independently assigned to a value, i.e.,  no functional relation is assumed between these two quantities. 
The ellipticity varies between 1 and 2, and $\theta$ ranges from 5{\arcmin} to 7{\arcmin}. 

We note that these simulations do not intend to accurately reproduce the  {\Planck} cold clump properties. Their 
goal is rather to cover entirely the physically acceptable parameter space, in order 
to allow us to recover the {\tt CoCoCoDeT} transfer function. 

The detection and extraction procedure, described in Sects.~\ref{sec:detection_method} and \ref{sec:photometry},
is then applied to the simulated data set, and a catalogue of detected sources, with corresponding 
flux densities and {\asciifamily FLUX\_QUALITY} flag, is built.

\subsection{Completeness}
\label{sec:completeness}

\begin{figure}[t]
\center
\begin{tabular}{c}
\psfrag{legend}{{\tiny $S_{857}^{IN}>15\, \mathrm{Jy}$}}
\includegraphics[width=8.5cm]{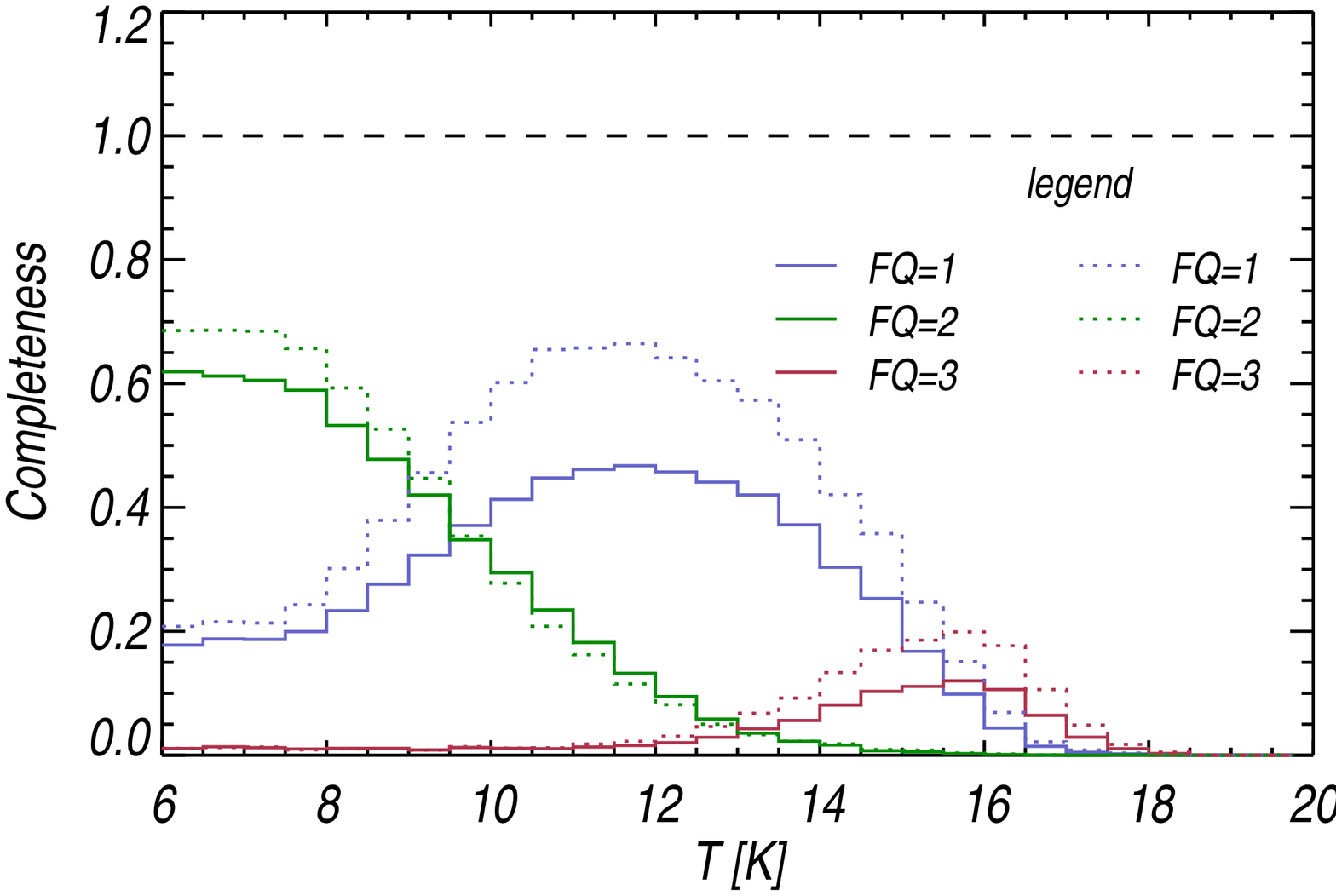} \\
\psfrag{legend}{{\tiny $S_{857}^{IN}>15\, \mathrm{Jy}$}}
\includegraphics[width=8.5cm]{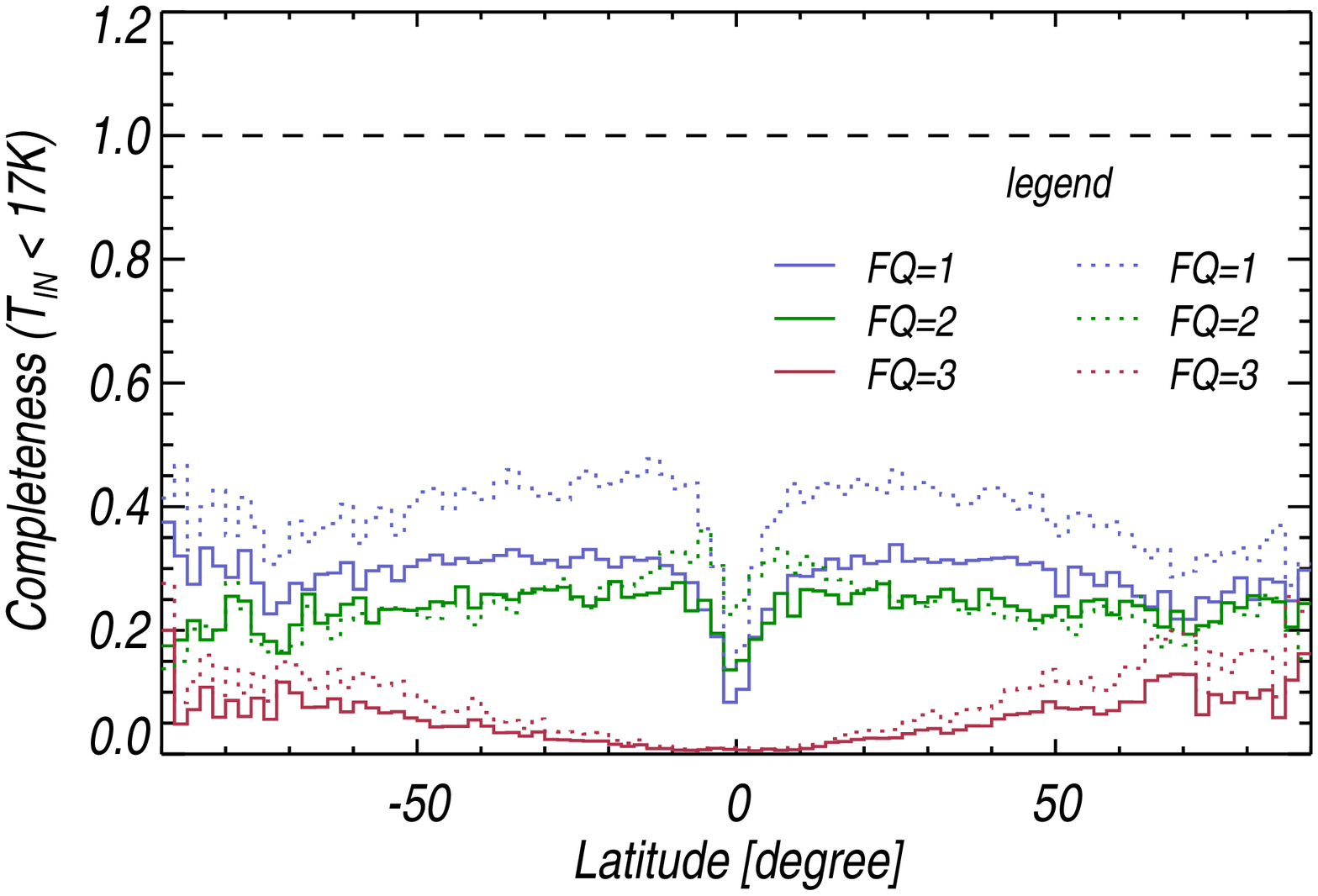} \\
\psfrag{-----xtitle-----}{$M \, [M_{\odot}]$}
\includegraphics[width=8.5cm]{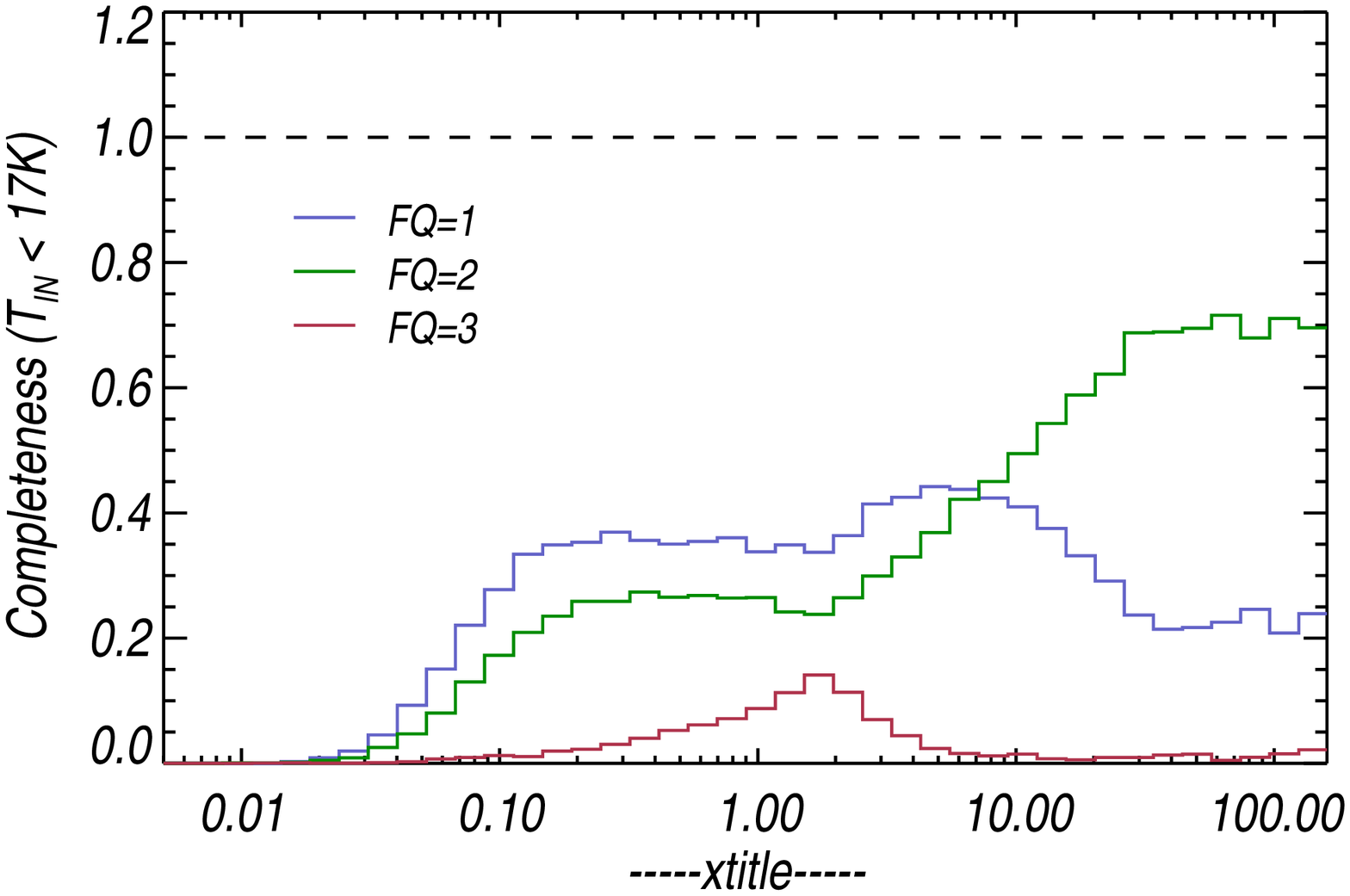} 
\end{tabular}
\caption{
Completeness as a function of the input temperature ({\it top panel}), latitude ({\it middle panel}), and 
mass ({\it bottom panel}) of the injected sources. For the latitude and mass cases, the input temperature was lower than 17\,K. In 
addition, when considering the dependence of completeness on mass, all the sources are assumed to be at 100\,pc from the Sun. 
Each panel shows the distributions obtained from selecting only the sources in a 
given {\asciifamily FLUX\_QUALITY} category. In particular, FQ=1, 2 and 3 correspond to the 
blue, green and pink curves, respectively. Finally, the dotted line in each panel denotes the completeness for the 
sample of sources with input flux density $S_{857} > 15$\,Jy. } 
\label{fig:mcqa_completeness}
\end{figure}

\begin{figure*}
\hspace{-0.5cm}
\begin{tabular}{cc}
\includegraphics[width=8.5cm]{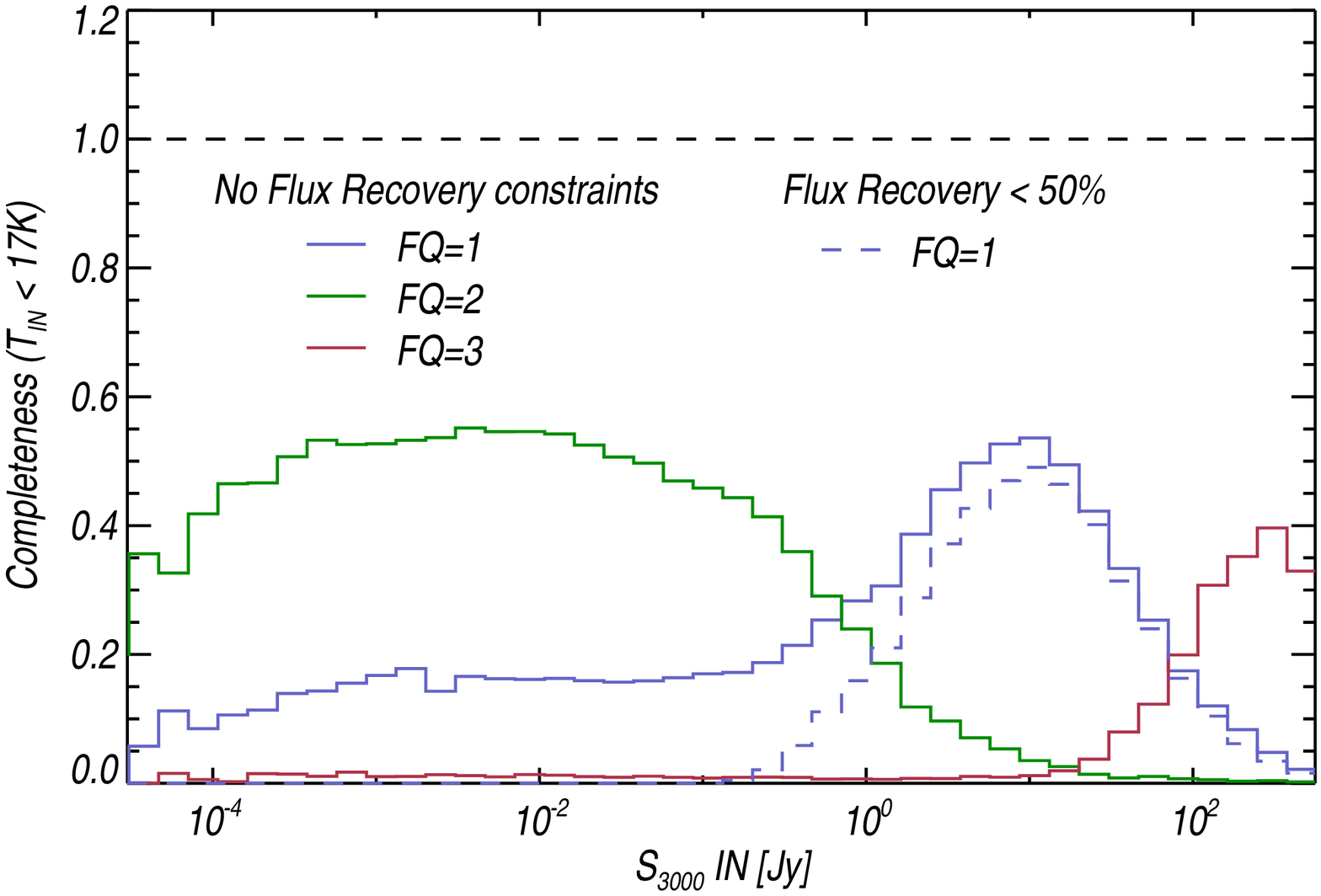} &
\includegraphics[width=8.5cm]{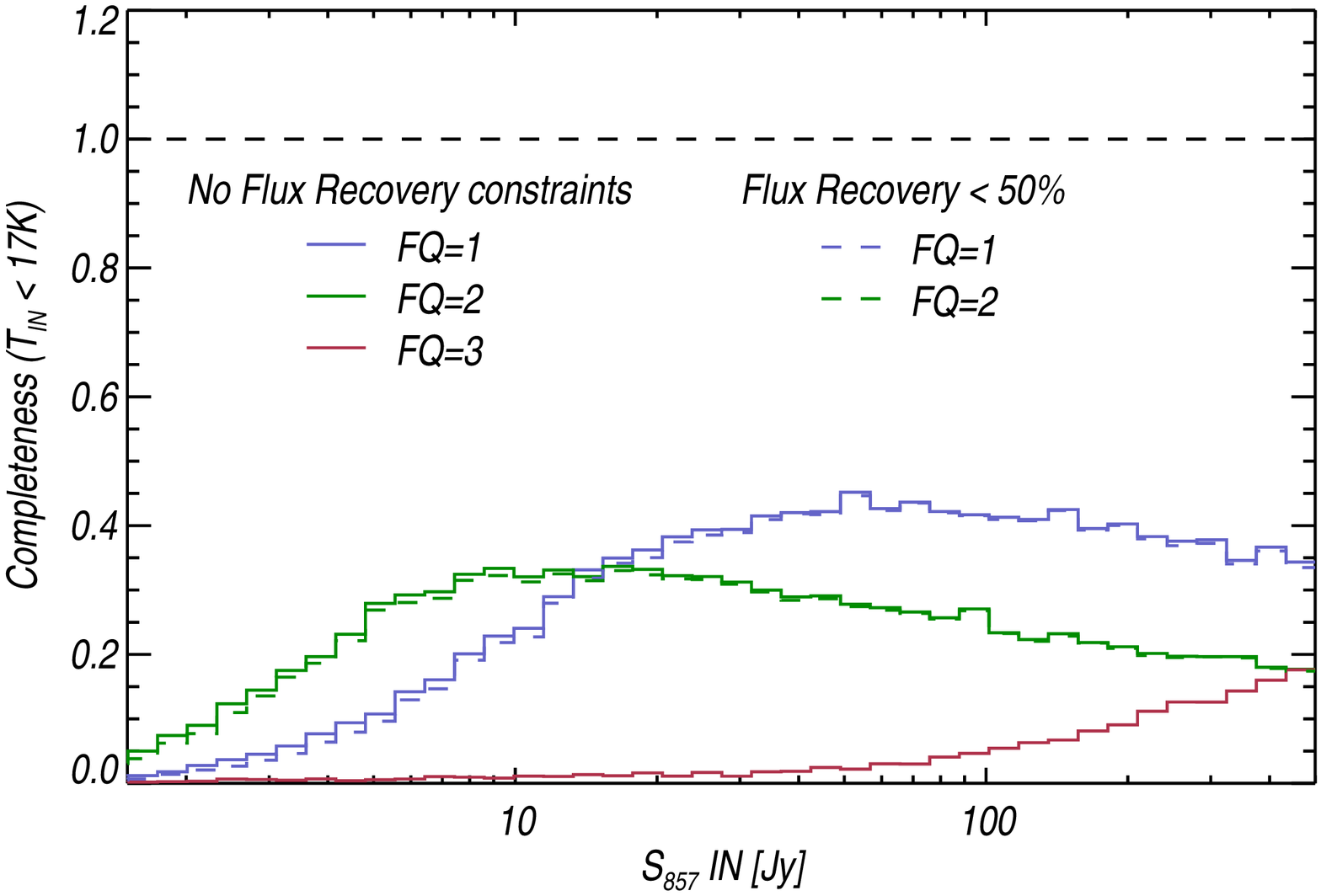} \\
\includegraphics[width=8.5cm]{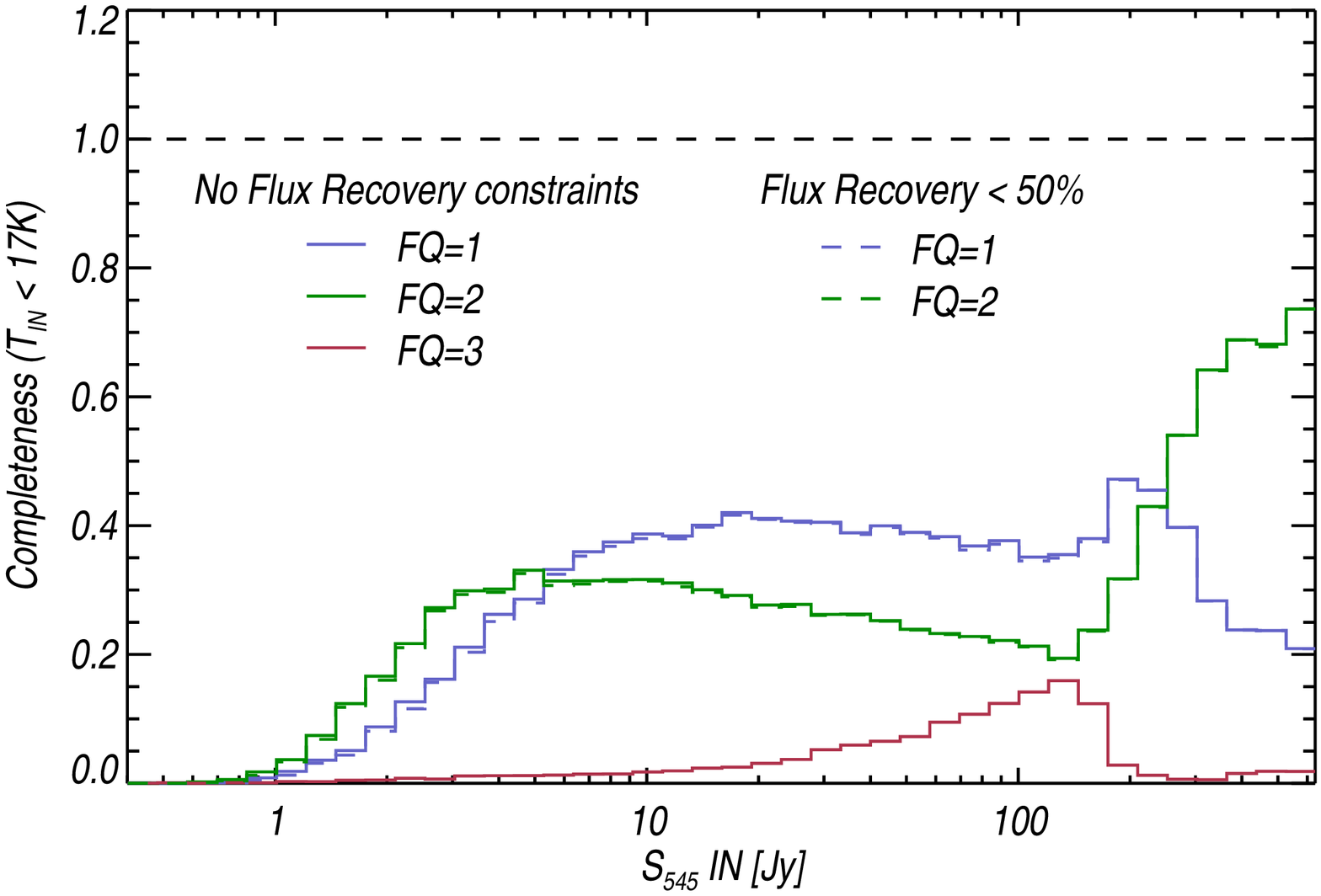} &
\includegraphics[width=8.5cm]{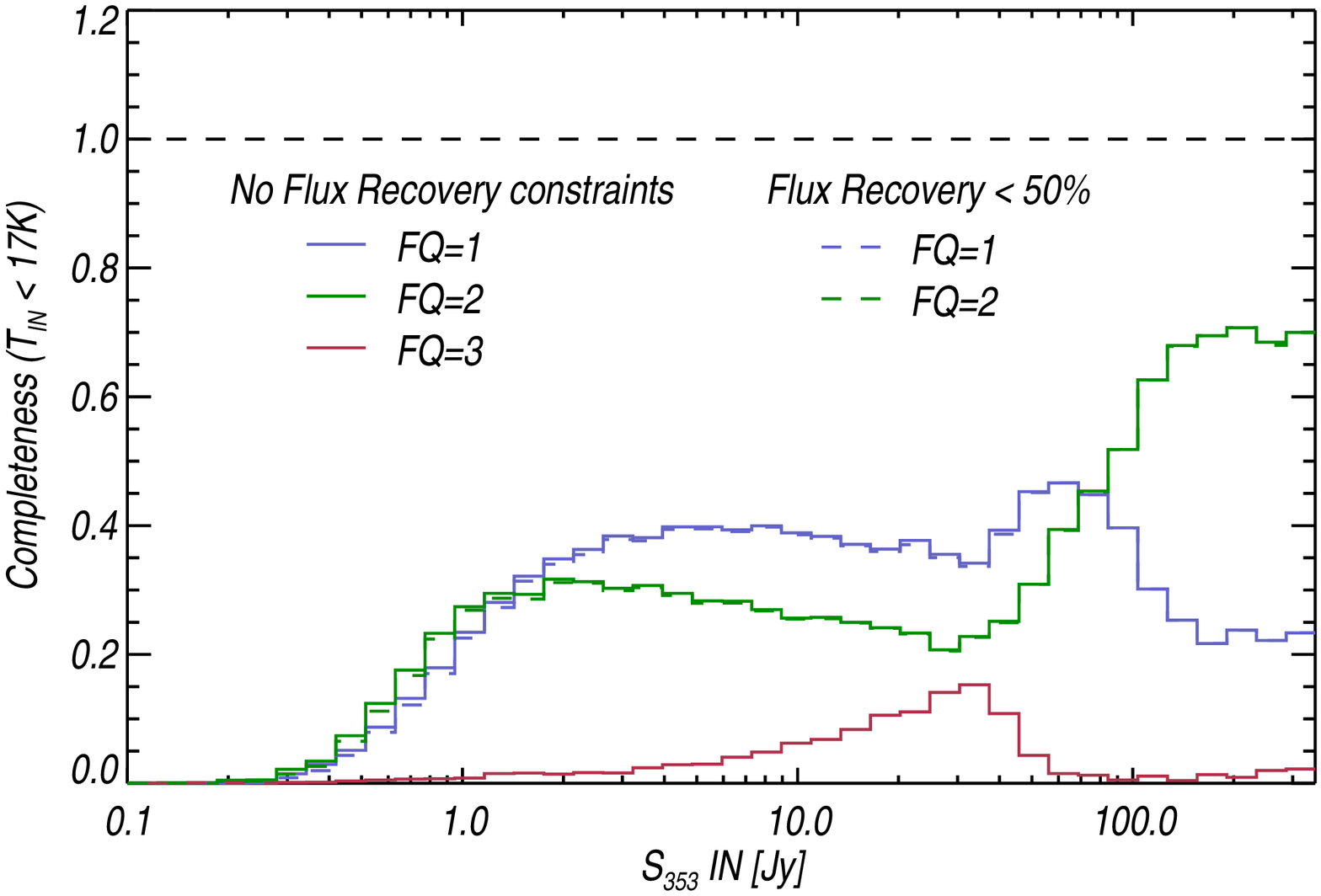} 
\end{tabular}
\caption{Completeness  (computed on a restricted sample of sources with injected temperature below 17\,K) 
as a function of the injected flux density in the IRAS 3\,THz ({\it top left panel}) and {\Planck} upper frequency bands, 
857\,GHz ({\it top right panel}), 545\,GHz ({\it bottom left panel}), and 353\,GHz ({\it bottom right panel}). 
It is shown per-category of flux density quality, i.e.,  FQ=1, 2 and 3 categories
(blue, green and pink, respectively). The completeness is also computed on a more restricted sample 
of sources with injected temperature below 17\,K and with a flux density accuracy better than 50\% ({\it dashed lines}).}
\label{fig:mcqa_completeness_flux}
\end{figure*}

Completeness is defined as the ratio of the number of detected sources to the total number
of injected sources. Using the Monte Carlo analysis described in the previous section, we have investigated 
whether the completeness of the catalogue generated with the {\tt CoCoCoDeT} algorithm depends on temperature,  
Galactic latitude, flux density and mass.

The {\it top panel} of Fig.~\ref{fig:mcqa_completeness} shows the relation between catalogue completeness and temperature of the injected 
sources. The completeness is about 60\% ({\it black solid line}) for temperatures lower than 10\,K, while it drops below 1\% for temperatures larger than 17\,K. 
In addition, the completeness increases to almost 90\%, for temperatures below 10\,K  if we consider only sources with 
input flux densities $S_{857}>15$\,Jy (dotted line). This result confirms that {\\t CoCoCoDeT} is a method optimized 
to detect cold sources embedded in a warm environment, while rejecting warm sources. We also note that  
sources flagged with FQ=3, (i.e.,  {\it Detection only}) are found in correspondence of 
relatively warm temperatures (between 12 and 18\,K), indicating that, as discussed in 
Sect.~\ref{sec:photometry_quality}, they might not be cold clumps. 
On the  other hand, the completeness of the sources with FQ=2  (i.e.,  {\it Missing 3\,THz flux}) 
increases towards lower temperatures, further suggesting that they are probably very cold. Notice that the 
completeness of sources with very low temperature (close to 6\,K) is still about 60\%, which means that, if these sources indeed exist, our algorithm is 
able to detect them. 
Finally, the completeness of the most reliable set of sources (FQ=1, blue) appears to peak (at 40\%) around 12\,K, 
ranging from 17\,K to  6\,K, i.e.,  the floor of the temperature distribution of the injected sources. 
Because the detection efficiency drops to zero beyond 17\,K, in the following we limit the discussion to 
simulated sources with temperatures lower than this threshold. 

The {\it middle panel} of Fig.~\ref{fig:mcqa_completeness} illustrates the completeness as a function of Galactic latitude. 
Outside the Galactic plane, the completeness cumulated over all FQ categories remains quite constant and around 
60\%, while it drops to 30\% for $|b| < 10^{\circ}$. This effect is expected and due to confusion in the 
Galactic plane. Sources with FQ=1 and 2 present a similar behaviour. On the contrary, the simulated 
sources with FQ=3 are mainly detected outside the plane, as observed for the real 
PGCC sources in this flux category (see {\it bottom panel} of Fig.~\ref{fig:allsky_fluxquality}). 

We have also explored the dependence of the completeness on the  injected 
flux density in the IRIS 3\,THz and {\Planck} bands. The result is in Fig.~\ref{fig:mcqa_completeness_flux}. 
The completeness of the FQ=3 sources increases with flux density, especially at 
the two highest frequencies. This is indeed the behaviour we expect from relatively warm sources. 
Conversely, the completeness of the FQ=2 sources, which
are presumably very cold, 
peaks at bright flux densities in the two lower frequencies, and at faint flux densities (below 1\,Jy) at 3\,THz, 
 where it has not been possible to get any output flux density estimates.
 The completeness at 3\,THz of the FQ=1 sources
 that, by definition, have S/N\,$ > 1$, is 
about 20\% below the IRIS sensitivity limit (1\,Jy), and may appear inconsistent with it.
This comes from the fact that completeness
in a given band is here defined based on the input source flux density rather than on the output one. If we estimate the completeness 
by using only the sources for which the recovered flux has an accuracy of 50\% or more ({\it dashed line}), we obtain 
a result in agreement to within 1$\sigma$ with the IRIS sensitivity limit, i.e., it drops to 0\% for flux densities below 0.5\,Jy. 
Interestingly, the completeness drops to 0\% for $S_{545}$ and $S_{353}$ below 1\,Jy and 0.3\,Jy, respectively, for all FQ categories: 
these two bands define the detection limit of our catalogue.

This flux density limit can be converted into a mass detection limit if we assume that all the sources are located at 100\,pc from the Sun. 
The completeness as a function of mass is shown in the {\it bottom panel} of Fig.~\ref{fig:mcqa_completeness}. 
The catalogue appears very incomplete for sources with a mass below 0.1\,$\mathrm{M}_{\odot}$, which 
explains the lower cutoff of the PGCC mass distribution in Fig.~\ref{fig:distribution_physical_properties}. 
At the same time, for sources with FQ=1 the catalogue is complete at the 20\% -- 40\% level across the entire mass range, meaning 
that no mass selection is introduced by {\tt CoCoCoDeT}. Therefore, the only (significant) bias introduced in the PGCC mass distribution 
originates from the availability of distance estimates. Finally, the simulated FQ=2 detections are characterized by a higher degree of completeness for 
relatively high mass values, as also observed also for the PGCC sources.

\begin{figure}
\center
\vspace{-0.5cm}
\includegraphics[width=8.5cm, viewport= 0 0 600 600]{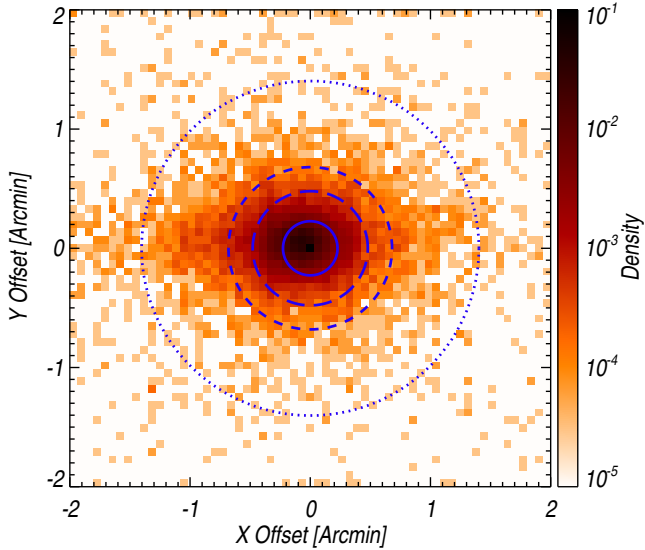} 
\caption{Density distribution of the positional offsets computed from 100~000 Monte Carlo realizations. 
Circles show the cumulative distributions at 68\% (solid line), 90\% (long-dashed line), 
95\% (dashed line) and 99\% (dotted line).}
\label{fig:mcqa_position_accuracy}
\end{figure}

\begin{figure}
\center
\includegraphics[width=8.5cm, viewport=-20 50 610 1163]{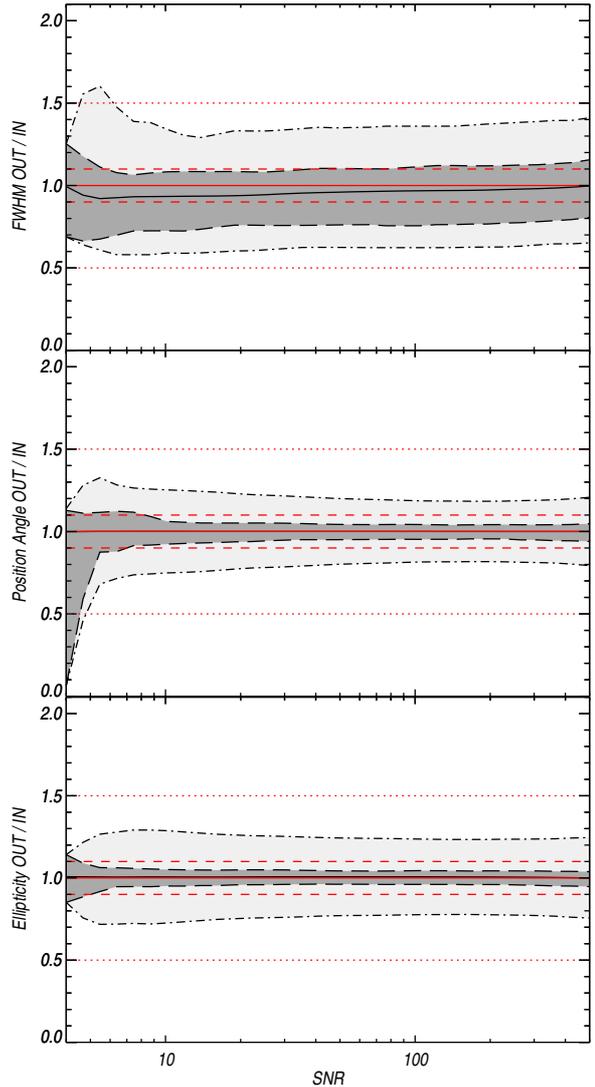}
\caption{Ratio of the recovered (OUT) to injected (IN) geometric parameters for the simulated sources detected with 
FQ=1 and 2 : $\theta$ ({\it top}), position angle ({\it middle}) and ellipticity ({\it bottom}). 
The ratio is given as a function of the detection S/N estimated from the {\it cold residual} maps.
The light and dark grey shaded regions (denoted by the dot-dashed and long-dashed lines, respectively) 
highlight the behaviour of 95\% and 68\% of the sources, respectively, in each S/N bin.
The median of the ratio distributions are shown with a solid line. 
The 0\%, 10\% and 50\% levels of uncertainty are overlaid using a red solid, dashed and dotted line, respectively.}
\label{fig:mcqa_geometry_accuracy}
\end{figure}

\subsection{Geometric accuracy}
\label{sec:geometry_accuracy}

By {\it geometry}, we mean the ensemble of parameters 
(e.g.,  centroid, ellipticity, position angle and equivalent full width half maximum) that describes 
the location, size and orientation of the source. 
The accuracy of these parameters is crucial for accurate photometric measurements, as described in Sect.~\ref{sec:photometry}.

Figure~\ref{fig:mcqa_position_accuracy} summarizes the catalogue positional accuracy, which we define as the offset between the input and
the recovered centroid of the synthetic sources. The cumulative distributions are shown at four confidence levels: 
68\%, 90\%, 95\% and 99\%. 68\% of the sources have a recovered centroid with a 0{\parcm}2 uncertainty, and 
95\% with a 0{\parcm}8 uncertainty. The median of the position offset distribution is 10{\arcsec}. Thus, despite background 
confusion, our catalogue appears to contain accurate source coordinates. 

We have also checked the accuracy for the other geometric parameters. For this purpose, we have investigated the relation 
between the ratio of the recovered to injected quantity (OUT/IN) and the S/N.
The result is illustrated in Fig.~\ref{fig:mcqa_geometry_accuracy}. 

For each bin in S/N, 
we have computed the cumulative OUT/IN distributions and corresponding median (solid line): in Fig.~\ref{fig:mcqa_geometry_accuracy}, 
each panel shows 68\% (light shaded contours) and 95\% (dark shaded contours) of the sources at a given frequency. The uncertainty 
on the recovered position angle (Fig.~\ref{fig:mcqa_geometry_accuracy}, {\it middle panel}) 
and ellipticity (Fig.~\ref{fig:mcqa_geometry_accuracy}, {\it bottom panel}) is very good, and below 10\% for 68\% of the sources at S/N~$ > 6$. 
When we include 95\% of the sources at S/N~$ > 6$, the uncertainty varies between 20 and 30\%. 
Note that for S/N~$> 10$, the uncertainty on the recovered parameters remains fairly constant up to very high S/N. 
The reconstructed size of the sources, $\theta$, (Fig.~\ref{fig:mcqa_geometry_accuracy}, {\it top panel}) 
appears more uncertain at all S/N, with
an uncertainty of about $\pm$15 to 20\% for 68\% of the data
at S/N~$ > 6$, and $\pm$30\% of uncertainty
when we include 95\% of the data. More importantly the recovered $\theta$ is systematically underestimated by 
about 10\% compared to the injected one, due to background. 
This effect has a direct impact on the photometric accuracy, as discussed in the following section.

\subsection{Photometric accuracy}
\label{sec:photometry_accuracy}

 \begin{figure}
\center
\includegraphics[width=8.5cm, viewport= 50 80 600 1320]{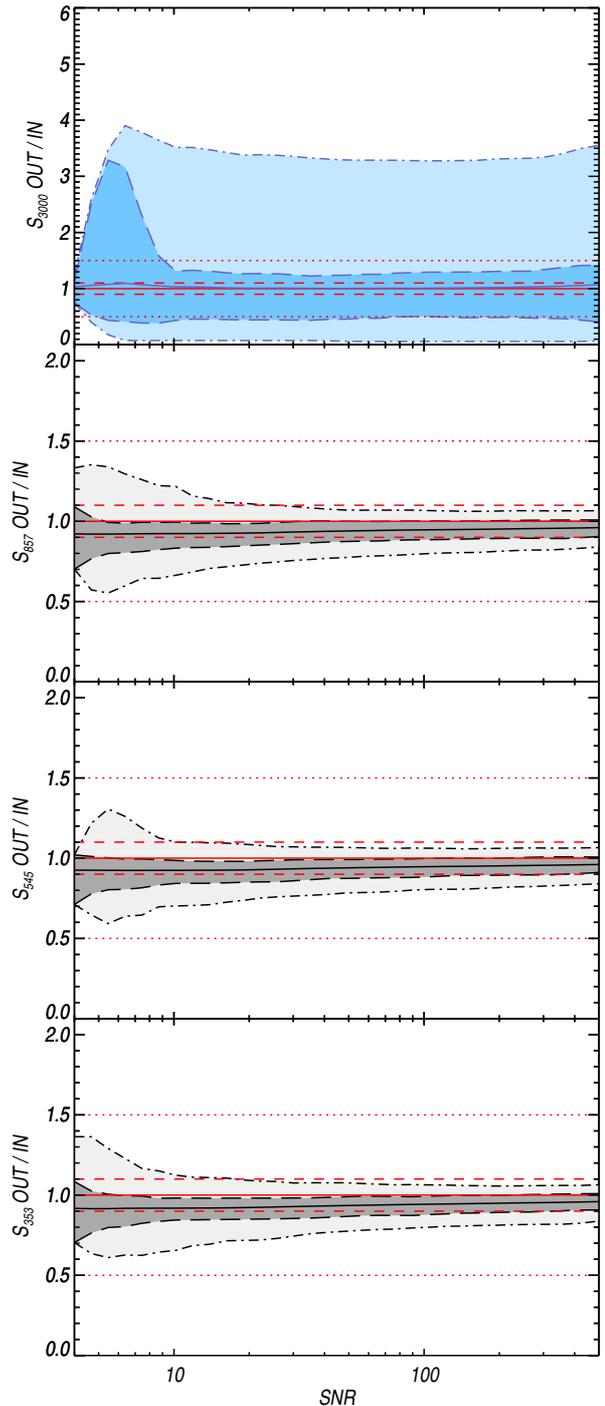}
\caption{Ratio of the recovered (OUT) to injected (IN) flux densities for the simulated sources detected with 
FQ=1 and 2. The ratio is given as a function of the detection S/N estimated from the {\it cold residual} maps.
The light and dark grey shaded regions  (denoted by the {\it dot-dashed} and {\it long-dashed lines}, respectively) 
highlight the behaviour of 95\% and 68\% of the sources, respectively, in each S/N bin.
Note that at 3\,THz ({\it top panel}) we only consider sources with FQ=1, and the corresponding 
contours and shaded regions are highlighted in colour.
The median of the ratio distributions are shown with a {\it solid line}. 
The 0\%, 10\% and 50\% uncertainty levels are overlaid using a red {\it solid}, {\it dashed} and {\it dotted line}, respectively.}
\label{fig:mcqa_accuracy_flux}
\end{figure}

Following a procedure similar to the one applied in the case of the geometric parameters, we have estimated 
the accuracy of the photometric measurements. Accordingly, we have computed the ratio of the 
recovered (OUT) to injected (IN) flux densities and analysed its behaviour as a function of S/N. 
We have limited the analysis to sources with FQ=1 at 3\,THz and 
FQ=1 and 2 in the {\Planck} bands. 

The poorest performance appears to be in the 3\,THz band. The S/N has to be as large as 15 for the accuracy to reach a level of about 
50\% for 68\% of the data, while the uncertainty is higher by a factor 4 if we include 95\% of the distribution. 
At lower S/N ($<15$), the OUT/IN ratio can reach a factor of 10, while, by definition of  
FQ=1, it should be below 2, because of the  $S_{3000}^{OUT}$/$\sigma_{S_{3000}^{OUT}}>1$ contraint. Thus, 
the uncertainty at 3\,THz  is severely underestimated during the photometry measurement.
 
If we compute the ratio of the difference between the recovered and injected flux density to flux density uncertainty, 
($S_{3000}^{OUT}$ - $S_{3000}^{IN}$)/$\sigma_{S_{3000}^{OUT}}$, we find a mean value of 5. 
We interpret this result as due to the fact that, at this frequency, the performance of the photometric measurements   
is dominated by modeling uncertainties, such as the removal of the warm background. 
However, this effect does not introduce any bias on the flux density measurements at 3\,THz.

The photometric accuracy of the photometry in the {\Planck} bands (second, third and fourth panel from the top in Fig.~\ref{fig:mcqa_accuracy_flux}) 
is much better than at 3\,THz. Indeed the uncertainty goes down to less than 10\% for 68\% of the sources at S/N~$>6$, once the data are bias-corrected. 
In fact, the flux densities in the {\Planck} bands are systematically underestimated, i.e.,  by 10\% at low S/N and by 5\% at high S/N, as
a consequence of the underestimation of $\theta$ (see Sect.~\ref{sec:geometry_accuracy}),  a well-known effect also studied in the framework of both the
ERCSC \citet{planck2011-1.10} and the PCCS \citet{planck2013-p05, planck2014-a35}. 
This bias is about at the same level as the flux density uncertainty, thus it is included in the 1$\sigma$  uncertainty.


\section{Distance estimates}
\label{sec:distance}

In this section we describe the derivation of the distance estimates for the PGCC sources which have been obtained by using 
four different methods:
i) cross-checking with kinematic distance estimates already available, ii) using the optical or near-infrared extinction due to the PGCC sources as an indicator of their distance, 
iii) associations with known molecular complexes, iv) estimates from the literature.

\subsection{Kinematic}

The \citet{Simon2006a} and \citet{Jackson2008} catalogues of infrared dark clouds (IRDC) provide 
kinematic distances for 497 objects. These distances are obtained by combining 
the gas observed radial velocity with a Galactic rotation curve, in the assumption of gas circular motion. 
Accordingly, an observed radial velocity at a given longitude corresponds to a unique 
Galactocentric distance solution while, at least in the inner Galaxy, two heliocentric 
distances are allowed. By cross-correlating the PGCC sources with the IRDC catalogues in a 5{\arcmin} radius, 
we have found 92 associations, mainly located along the Galactic plane. To these sources we have assigned 
the distance flag {\asciifamily DIST\_KINEMATIC}. We note that, when two heliocentric solutions are available, 
we always choose the near solution. An arbitrary 25\% uncertainty on these distance estimates is adopted. 
In the following, we will refer to kinematic distances as method [1].

\subsection{Optical extinction}

Distances derived from optical extinction are based on processing of two
independent Sloan Digital Sky Survey (SDSS) photometry-based data sets 
containing the computed distances and interstellar reddening to each star. 
The first data set, based on SDSS DR7 photometry and covering 1467 PGCC sources, 
is that of \citet{Berry2012} who fit reddening and
the stellar locus colours of \citet{Covey2007} to the observed photometry. 
The second data set, from \citet{McGehee2015}, makes use of the SDSS DR9 catalogue.
This targets a total of 1769 cold clumps and computes the reddening from
$g-i$ colour excess, where the intrinsic 
stellar $g-i$ colour is derived from the 
reddening invariant indices defined by \citet{McGehee2005}.

Distance moduli values and  uncertainties for each cold clump
are inferred from analysis of 100 Monte
Carlo runs using stars within a 10{\arcmin} radius 
of the catalog position. For each realization the distance moduli and 
$E(B-V)$ reddening of each star are randomly varied assuming normal 
distributions of $N(0,0.282^2)$ and $N(0,0.072^2)$, respectively,
These variance values, which are similar for both data sets, 
were computed by propagating the observed stellar locus width in low
extinction  $(A_r < 0.05$) regions and the stated photometric  uncertainties through
the relations for $E(B-V)$ and $m-M$.

Each of these profiles are processed by a Canny edge detection filter 
\citep{Canny1986} with the location of the cold clump set by the distance 
modulus for which the edge detection signal is maximized. A Gaussian sigma of 
0.3 magnitudes in $m-M$ is used in the Canny filter for smoothing and 
noise reduction. 
Implicit in this approach is the assumption 
that there is only a single interstellar cloud along the line-of-sight.
For each PGCC source we assign a value and  uncertainty to the
distance modulus based on the mode and standard deviation of the distribution.
This distribution is
obtained via kernel density estimation [KDE] on a 0.01 magnitude grid 
using the values returned from the Monte Carlo realizations and with the
bandwidth of the KDE set by the normal distribution approximation.

We adopted the following prescriptions of \citet{McGehee2015} to build a final selection of robust distance estimates:
i) distance estimates obtained with the M dwarf based technique \citep{McGehee2015} towards sources with an extinction $E(B-V)< 0.4$
are rejected, and ii) distance modulus estimates with  uncertainties larger than 1.0130 and 0.7317, for the \citet{Berry2012} or  \citet{McGehee2015} methods, respectively, 
are also rejected. Furthermore we have performed a sanity check on the altitude of the sources, rejecting those
with an altitude above or below 2 times the Galactic scale height, 
which has been recently estimated by \citet{Jones2011} at 119$\pm$15\,pc.

Two sets of distance estimates are finally provided using optical extinction with SDSS data, 
depending on the SDSS data version. Hence 1083 sources have been assigned a distance estimate based on the DR7 SDSS data version using the \citet{Berry2012} method [2]
(distance flag {\asciifamily DIST\_OPT\_EXT\_SDSS\_DR7}) and  191 sources based on the DR9 SDSS data version using the  \citet{McGehee2015} method [3]
(distance flag {\asciifamily DIST\_OPT\_EXT\_SDSS\_DR9}).

\subsection{Near-infrared extinction}

By comparing observed stellar colours to the
predictions of the Besançon Galactic model \citep{Robin2003, Robin2012}, we have attempted to infer the most probable three-dimensional 
extinction distribution along the line-of-sight. The line-of-sight extinction is parametrised using a number of points, 
each described by a distance and an extinction. These parameters are probed using a Markov Chain Monte Carlo (MCMC) method, 
based on the Metropolis Hastings algorithm with $10^5$ iterations. Modeled stars are reddened using linear interpolation 
between points and comparison with observations is performed on the colour distribution using a Kolmogorov Smirnov test.

Stars are chosen from the Two Micron All Sky Survey \citep[2MASS, ][]{Skrutskie2006} point source catalogue that lie 
within the ellipse defining the cold clump, so in the location of the cloud along the line-of-sight should be 
detectable as a sharp rise in extinction. The resultant extinction vs distance profile is then analysed to detect the presence of any clouds. 
The dust density with respect to distance is calculated via the derivative of the extinction distance relation 
and the diffuse extinction is estimated from the continuum. Any peaks in dust density 3 $\sigma$ over the diffuse are flagged. 
If a line-of-sight contains more than one cloud, the one with the highest extinction is chosen and the presence of 
a second cloud is flagged. Only lines-of-sight with a single detected cloud have been included in the present PGCC catalogue.

The principle of the method is similar to that described by \citet{Marshall2006} 
and \citet{Marshall2009}. The motivation for the change in our case is to provide more robust estimates when 
stellar density is low, as well as to more fully characterize the uncertainty via the MCMC exploration of the parameter space.

The former version of this method was first applied in \citet{Marshall2009} on the \citet{Simon2006a} {\it MSX\/} catalogue of IRDCs, providing 
1218 distance estimates. After cross-correlating the PGCC sources with this IRDC catalogue using a 5{\arcmin} radius, 
we have found 182 associations, leading to the distance estimates  of the method [4] ({\asciifamily DIST\_NIR\_EXT\_IRDC} field).
We have adopted an uncertainty for the distance estimate of 25\% following \citet{Marshall2009} prescriptions.
The improved algorithm has been then applied to the PGCC sources using 2MASS data. After performing the sanity check on the altitude (as for the optical extinction methods), 
and including 787 upper limits (marked as negative estimates), this has led to 2810 estimates (method [5], {\asciifamily DIST\_NIR\_EXT} field). 
The uncertainty on these new set of distance estimates is provided individually by the algorithm.

\subsection{Molecular complexes}
\label{sec:mol_comp}

A simple inspection of the all-sky distribution of cold clumps (see Fig.~\ref{fig:allsky}) suggests that
it follows the distribution of known molecular complexes  at intermediate latitude, as it is illustrated in Appendix~\ref{sec:co}. 
Many of these molecular complexes have distance estimates in the
literature. To assign the distance of a complex to a particular
cold clump, we have used the all-sky CO {\Planck} map. In particular, we have checked the presence of a given PGCC source inside 
a molecular cloud, by using a mask generated from the CO map. This method has been applied to 11 molecular complexes listed in
Table~\ref{tab:molecular_complexes} and located outside the Galactic plane, which allows us to  
minimize the effect of confusion. Following this procedure, we have obtained 1895 distance estimates (method [6], {\asciifamily DIST\_MOLECULAR\_COMPLEX} field) 
with associated uncertainties.

\subsection{{\it Herschel} follow-up}

As mentioned in the introduction and further discussed in Sect.~\ref{sec:ancillary_validation} we have performed a high angular resolution follow-up 
with the PACS and SPIRE instruments on-board {\it Herschel}, in the framework of the {\it Herschel} key-programme Galactic Cold Cores (HKP-GCC). 
 This follow-up programme has targeted  349 PGCC sources, for which we have 
obtained distance estimates from the literature \citep{Montillaud2015}, the most reliable of which (228 sources) are 
reported in the final catalogued and flagged as {\asciifamily DIST\_HKP\_GCC} field. In the following, 
we refer to this method as method [7].

\begin{table}
\caption{Molecular complexes used to assign a distance estimate to the PGCC sources. 
For more details see Sect.~\ref{sec:mol_comp}.}
\label{tab:molecular_complexes}
\nointerlineskip
\setbox\tablebox=\vbox{
\newdimen\digitwidth 
\setbox0=\hbox{\rm 0} 
\digitwidth=\wd0 
\catcode`*=\active 
\def*{\kern\digitwidth} 
\newdimen\signwidth 
\setbox0=\hbox{+} 
\signwidth=\wd0 
\catcode`!=\active 
\def!{\kern\signwidth} 
\newdimen\pointwidth 
\setbox0=\hbox{.} 
\pointwidth=\wd0 
\catcode`?=\active 
\def?{\kern\pointwidth} 
\halign{\hbox to 0.9 in{#\leaderfil}\tabskip=1.0em&
\hfil#\hfil&
\hfil#\hfil&
\hfil#\hfil&
\hfil#\hfil&
\hfil#\hfil\tabskip=0pt\cr
\noalign{\doubleline}
\omit\hfil Name\hfil& $l$& $b$& Area& Distance  & No. \cr
\omit& [deg]& [deg]& [$\rm{deg}^2$]& [pc]&  \cr
\noalign{\vskip 3pt\hrule\vskip 4pt}
Aquila Serpens& *28&   !*3& *30& 260*$\pm$*55 $^{[1,2]}$ & *51 \cr
 Polaris Flare& 123&   !24& 134& 380*$\pm$*40 $^{[3]}$* & *68 \cr
Camelopardalis& 148&   !20& 159& 200*$\pm$*30 $^{[3]}$*  & *19 \cr
    Ursa Major& 148&   !35& *44& 350*$\pm$*35 $^{[3]}$*  & *22 \cr
        Taurus& 177& $-$15& 440& 140*$\pm$*15 $^{[2]}$* & 384 \cr
Taurus Perseus& 163& $-$15& 440& 230*$\pm$*20 $^{[2]}$* &  224 \cr
 $\lambda$ Ori& 196& $-$13& 113& 400*$\pm$*40 $^{[4]}$* & *70 \cr
         Orion& 212& *$-$9& 443& 450*$\pm$*50 $^{[2]}$* & 333 \cr
    Chamaeleon& 300& $-$16& *27& 170*$\pm$*15 $^{[5]}$* &  114 \cr
     Ophiuchus& 355&   !17& 422& 150*$\pm$**5 $^{[2]}$* & 316 \cr
      Hercules& *45&   *!9& *35& 300*$\pm$*75 $^{[6]}$* & *19 \cr
\noalign{\vskip 3pt\hrule\vskip 4pt}
} }
\endPlancktable
\tablenote  {{\rm }} {\tiny $^{[1]}$ \citet{Bontemps2010b}};  {\tiny $^{[2]}$ \citet{Loinard2013}};  
 {\tiny $^{[3]}$ \citet{Schlafly2014}}; {\tiny $^{[4]}$ \citet{Murdin1977}}; 
 {\tiny $^{[5]}$ \citet{Bertout1999}}; {\tiny $^{[6]}$ \citet{Andersson1991}}. \par
\end{table}

\begin{figure*}
\center
\includegraphics[width=8cm, angle=90, viewport=35 20 420 730]{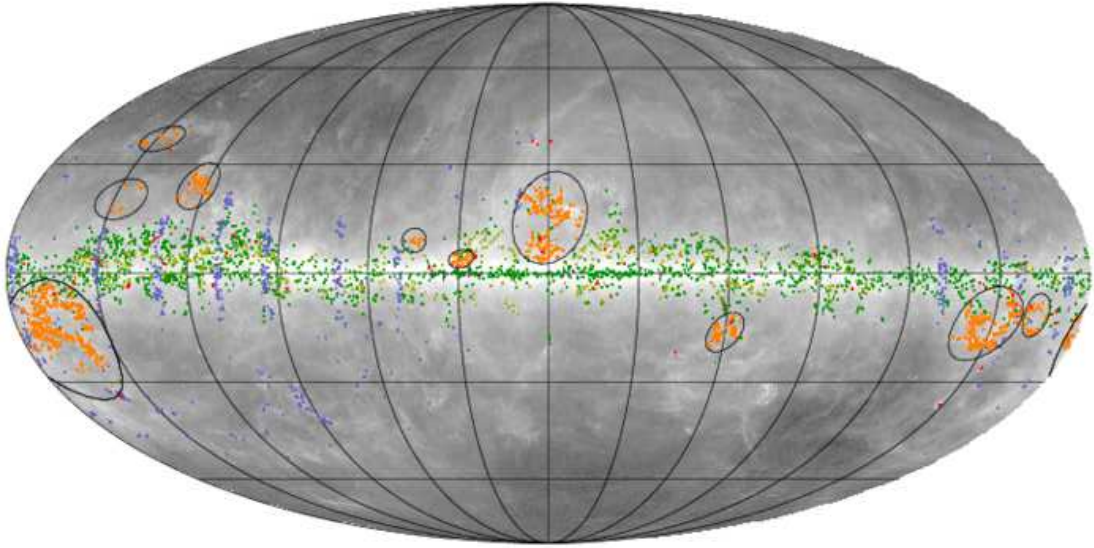} 
\caption{All-sky distribution of the  4655 PGCC sources for which a distance estimate with a {\asciifamily DIST\_QUALITY} flag equal to 1 or 2 is available.
The various types of distance estimates are defined as follows : kinematic (purple), 
optical extinction (blue), near-infrared extinction (green), molecular complex association (orange), and
{\it Herschel} HKP-GCC (red). We also show the distribution of  the 664 sources with an upper-limit estimate  ({\asciifamily DIST\_QUALITY}=4) 
provided by the near-infrared extinction method in light green. The regions covered by the molecular complexes are shown as black contours (see Table~\ref{tab:molecular_complexes}).}
\label{fig:allsky_distances}
\end{figure*}

\subsection{Combined distance estimates}
\label{sec:distance_combination}

We rank all methods by increasing order 
of confidence level, starting with the kinematic estimates ([1]), which several authors indicate as less reliable 
than extinction estimates \citep{Foster2012}, due to the distance ambiguity in the inner Galaxy. 
The estimates derived from optical extinction are ordered as [2] then [3], according to the SDSS data release version (DR7 to DR9). 
The estimates obtained from the near-infrared extinction come next, starting from those obtained towards IRDCs ([4]), and then considering the ones 
from the improved algorithm ([5]). The association with molecular complexes ([6]) provides  
consistent estimates for sources belonging to the same cloud. Finally, the distance 
estimates derived from the analysis of the {\it Herschel} observations appear as the most reliable ([6]). 

In the final catalogue we have assigned to each source a unique distance value ({\asciifamily DIST}) , corresponding to the distance estimate with the highest 
confidence level among the available estimates. This approach allows us to avoid assigning to a source an average distance computed from individual 
estimates obtained from very different methods. Importantly,  we have checked the internal consistency among the available estimates for a given clump 
using the following relation:
\begin{equation}
X_{D} = \sqrt{\frac{1}{C_n^2} \sum_{i \ne j} \left( \frac{| D_i - D_j |}{\sigma_i + \sigma_j} \right)^2} < 1 \, ,
\end{equation}
where $n$ is the number of available distances, $D_i$ and $D_j$ are the distance estimates and their corresponding standard deviation
$\sigma_i$ and $\sigma_j$, and $C_n^2$ is the number of combinations of pairs between the $n$ estimates.  
A value $X_{D}<1$ indicates that the mean distance between distance estimates is compatible with a 1 $\sigma$  uncertainty on each estimate. 
We have used the $X_D$ parameter to assign a {\asciifamily DIST\_QUALITY} flag to each source. In particular, $X_D$ can take up the following values: 
 
\begin{description}
\item[0. "{\it No estimate}"]: no distance estimate available;
\item[1. "{\it Consistent}"]: few estimates are available and consistent within 1$\sigma$ ($X_D \le 1$);
\item[2. "{\it Single}"]:  one single estimate;
\item[3. "{\it Unconsistent}"]: few estimates are available, but not consistent within 1$\sigma$ ($X_D > 1$);
\item[4. "{\it Upper limit}"]: only an upper limit is available.
\end{description}

\begin{figure}
\center
\includegraphics[width=8.5cm]{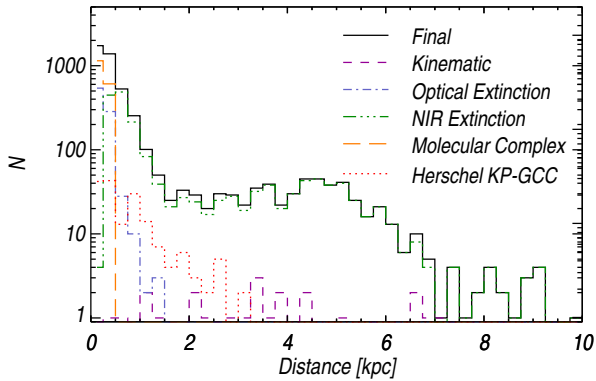} 
\caption{Distance distribution  per type: kinematic, optical extinction, near-infrared extinction, molecular complex association, and
{\it Herschel} HKP-GCC.}
\label{fig:distribution_distance}
\end{figure}

\begin{figure}
\vspace{-0.18cm}
\center
\includegraphics[width=8cm,viewport=10 10 600 550]{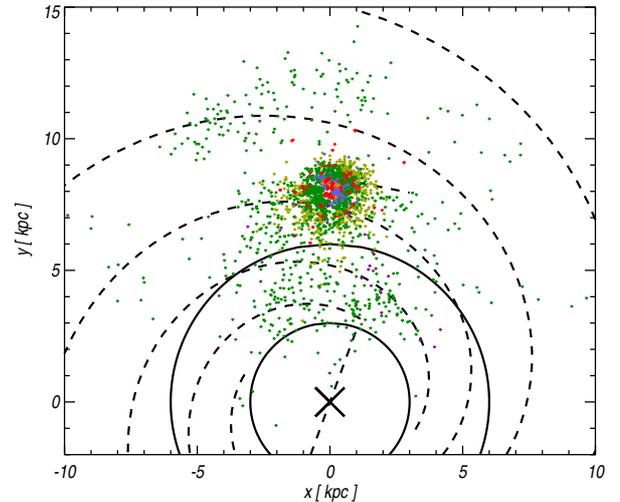} 
\caption{Distribution of the PGCC sources as seen from the North Galactic pole. 
Only distance estimates with a {\asciifamily DIST\_QUALITY} flag equal 1 and 2 are plotted. Different methods have been used to 
derive the distances: kinematic (purple), optical extinction (blue), near-infrared extinction (green), molecular complex association (orange), and
{\it Herschel} HKP-GCC (red). We also show the distribution of the distance upper-limits ({\asciifamily DIST\_QUALITY}=4) 
provided by the near-infrared extinction method (light green).  The red dashed circle shows the 1\,kpc radius around the Sun.  Black {\it dashed
lines} represent the spiral arms and the local bar.
The black circles, centred on the cross, provide  an indication of the location of the molecular ring.} 
\label{fig:gal_distribution_distance}
\end{figure}

Following the procedure described above, we have obtained a total of 5574 distance estimates distributed across the whole sky 
with an associated {\asciifamily DIST\_QUALITY} flag. The method selected for each distance estimate 
({\asciifamily DIST}) is specified with the {\asciifamily DIST\_OPTION} field, which ranges from 0 to 7 (see Table~\ref{tab:distance_estimates}). 
The statistical properties of the distance distribution are shown in Table~\ref{tab:distance_estimates}.
We emphasize that these distance estimates have been compiled without taking into account the quality of the photometric measurements.
In total, there are 4655 PGCC sources with  {\asciifamily DIST\_QUALITY} flag equal to 1 and 2. 
 Of these, only 2489 sources 
have reliable flux densities in all bands (FQ=1), while 1378 sources have reliable flux densities in the {\Planck} 
bands, but not at 3\,THz (FQ=2). For more details see Table~\ref{tab:dist_flux_quality}. 

The cross-correlation between the {\asciifamily DIST\_QUALITY}  and {\asciifamily FLUX\_QUALITY} flags is shown 
in Table~\ref{tab:dist_flux_quality}, and reveals that only 2489 sources have  both  reliable distance estimates {\asciifamily DIST\_QUALITY}=1 or 2) 
and reliable flux densities in all bands (FQ=1), while 1378 sources have reliable flux densities in {\Planck} bands 
 but not at 3\,THz (FQ=2). 

The spatial distribution of the sources having a distance estimate (see Fig.~\ref{fig:allsky_distances}) strongly depends on the adopted 
distance method. For example, sources with distances derived from optical extinction (blue) mainly follow
the SDSS sky-coverage, while distances obtained from associations with molecular cloud complexes present 
a patchy distribution, mirroring that of the host clouds (orange). The near-infrared extinction estimates (green) are mostly 
concentrated towards the inner Galactic plane where
the density of stars is sufficient to provide high extinction contrast, while the near-infrared extinction upper-limit estimates (light green) 
are spread at larger latitudes for the same reason.

A similar type of consideration applies to the analysis of the statistical distribution of all distance estimates (see Fig.~\ref{fig:distribution_distance}). 
The near-infrared extinction and kinematic methods allows us to probe distant regions (from 1\,kpc to 9\,kpc) across the Galactic plane. 
On the contrary, the optical extinction and molecular complex associations methods are applicable in the nearby Galaxy only (up to 1\,kpc and 0.5\,kpc, respectively). 
As seen from the North Galactic pole (Fig.~\ref{fig:gal_distribution_distance}), the complementarity of the different methods shows clearly. 
 About 88\% of the sources with a reliable distance estimate ({\asciifamily DIST\_QUALITY}=1 or 2) lie within 2\,kpc from the sun. Therefore, the PGCC catalogue mainly
probes the solar vicinity. It is interesting to notice that the distribution of the PGCC sources at larger distance follows at first order the Galactic arms and the molecular ring.
This is especially significant for the Perseus arm towards the outer Galaxy and for the Scutum-Centaurus and Norma arms in the inner Galaxy.
We conclude this section by emphasizing that, due to the variety of distance estimators, any statistical analysis involving distances or the related quantities, will be 
affected by severe biases which, given the fact that the catalogue is not flux density complete, are very hard to quantify.

\begin{table}
\caption{Number of sources with a distance estimate for each of the seven methods, 
before (column 2) and after (column 3) the combination process described in Sect.~\ref{sec:distance_combination}, 
in which only one distance is selected among the ones available.
The repartition between methods of the final distance estimates provided in the catalogue ({\asciifamily DIST} field) 
is shown in columns 4 to 7 for each category of the distance quality ({\asciifamily DIST\_QUALITY} flag) :
{\it Consitent} (1), {\it Single} (2), {\it Unconsistent} (3), {\it Upper limit} (4). }
\label{tab:distance_estimates}
\nointerlineskip
\setbox\tablebox=\vbox{
\newdimen\digitwidth 
\setbox0=\hbox{\rm 0} 
\digitwidth=\wd0 
\catcode`*=\active 
\def*{\kern\digitwidth} 
\newdimen\signwidth 
\setbox0=\hbox{+} 
\signwidth=\wd0 
\catcode`!=\active 
\def!{\kern\signwidth} 
\newdimen\pointwidth 
\setbox0=\hbox{.} 
\pointwidth=\wd0 
\catcode`?=\active 
\def?{\kern\pointwidth} 
\halign{\hbox to 1. in{#\leaderfil}\tabskip=1.0em&
\hfil#\hfil&
\hfil#\hfil&
\hfil#\hfil&
\hfil#\hfil&
\hfil#\hfil&
\hfil#\hfil\tabskip=0pt\cr
\noalign{\doubleline}
\omit \hfil Method \hfil & \multicolumn{2}{c}{No} & \multicolumn{4}{c}{ {\asciifamily DIST\_QUALITY}} \cr
\omit \hfil & Init. & Final & 1 & 2 & 3 & 4 \cr
\noalign{\vskip 3pt\hrule\vskip 4pt}
$[1]$  Kinematic  	& **92  	& **23  	& **- & **23 & **- & **- \cr
$[2]$ Opt. Ext. DR7  	& 1083 	& *719 	& **-  & *717  & **2  & **- \cr
$[3]$ Opt. Ext. DR9  	& *191 	& *173 	& *63 & **89 & *21 & **- \cr
$[4]$  NIR Ext. IRDC &  *182 	&  *106 	& *19 & **81  & **6 & **- \cr
$[5]$  NIR Ext. 		&  2810 	&  2491 	& 138 & 1601 & *88 & 664 \cr
$[6]$  Complexes  	& 1895 	& 1834 	& 177 & 1576 & *81 & **- \cr
$[7]$  HKP-GCC 	& *228 	& *228 	& *67 & *104 & *57 & **- \cr
\noalign{\vskip 3pt\hrule\vskip 4pt}
Total & & 5574 & 464 & 4191 & 255 & 664 \cr
\noalign{\vskip 3pt\hrule\vskip 4pt}
}}
\endPlancktable
\end{table}

\begin{table}
\caption{Number of sources with a distance estimate in each category of the {\asciifamily DIST\_QUALITY}  and  {\asciifamily FLUX\_QUALITY} flags.}
\label{tab:dist_flux_quality}
\nointerlineskip
\setbox\tablebox=\vbox{
\newdimen\digitwidth 
\setbox0=\hbox{\rm 0} 
\digitwidth=\wd0 
\catcode`*=\active 
\def*{\kern\digitwidth} 
\newdimen\signwidth 
\setbox0=\hbox{+} 
\signwidth=\wd0 
\catcode`!=\active 
\def!{\kern\signwidth} 
\newdimen\pointwidth 
\setbox0=\hbox{.} 
\pointwidth=\wd0 
\catcode`?=\active 
\def?{\kern\pointwidth} 
\halign{\hbox to 0.7 in{\hfil#\hfil}\tabskip=1.0em&
\hfil#\hfil&
\hfil#\hfil&
\hfil#\hfil\tabskip=0pt\cr
\noalign{\doubleline}
{\asciifamily DIST\_QUALITY} & \multicolumn{3}{c}{ {\asciifamily FLUX\_QUALITY}} \cr
 & 1 & 2 & 3 \cr
\noalign{\vskip 3pt\hrule\vskip 4pt}
 1 & *240 & *146 & *78  \cr
2 & 2249 & 1232 & 710  \cr
3 & *116 & **88 & *51 \cr
4 & *335 & *220 & 109 \cr
\noalign{\vskip 3pt\hrule\vskip 4pt}
}}
\endPlancktable
\end{table}

\section{Physical parameters}
\label{sec:physical_properties}

For the {\Planck} Galactic cold clumps we  have derived: temperature, column density and, when distance estimates are available, size, mass, mean density and luminosity.
The propagated uncertainties on the computed quantities are obtained, for each clump, from $10^6$
Monte-Carlo simulations and are provided in two different fashions: i) as 1$\sigma$ standard dispersion; ii) as defined 
by the lower and upper limits of the 68\%, 95\% and 99\% confidence intervals of the Monte Carlo distributions. The latter 
takes into account the non gaussian behaviour of the same Monte-Carlo distributions. 

Figure~\ref{fig:distribution_physical_properties} illustrates the temperature, column density, size, mass, mean density and luminosity distributions  
for the PGCC sources with {\asciifamily FLUX\_QUALITY} equal to 1 and 2. We recall that 
sources with FQ=1  have reliable flux densities in both the IRIS 3\,THz and {\Planck} bands, 
which allows the derivation of reliable temperatures and emissivity index, while sources with FQ=2 are likely faint and cold, and 
their temperatures estimated are obtained using only the three {\Planck} bands and a fixed emissivity spectral index of 2. 
In the case of physical quantities for which a distance estimate is needed, we also require that {\asciifamily DIST\_QUALITY} is either 1 or 2. 

Additionally, we investigated the impact of the individual uncertainties estimated above on the overall source catalogue distribution for each 
physical quantity. To this end, we have again used a Monte-Carlo approach. In this case, for each quantity, we build 10\,000 synthetic
samples with the same number of sources as the original sample. Then, starting 
from the computed values, we add random noise based on the uncertainty associated with each source. At this stage, we generate 
the 1$\sigma$ upper and lower contours for each of the 10\,000 samples, where the 1$\sigma$ limit is estimated based on the median 
of the distribution. Fig.~\ref{fig:distribution_physical_properties} shows, as expected, that the contours follow reasonably well the distribution of the  
computed values when the uncertainties are relatively small. On the contrary, there is quite a significant departure from these 
when the uncertainties are large.

\subsection{Temperature and emissivity}
\label{sec:temperature}

The source temperatures and the local warm background temperatures have been estimated using the flux density measurements and 
their corresponding uncertainties  in the IRIS 3\,THz band and the {\Planck} 857, 545, and 353\,GHz channels (see Sect.~\ref{sec:photometry}). 
The fits of the SEDs have been performed assuming that the emission can be described as a modified black body:
\begin{equation}
F_{\nu} = F_{\nu,0} B_{\nu}(T) (\nu/\nu_0)^{\beta} \, ,
\label{eq:mbb}
\end{equation}
where $F_{\nu}$ are the observed flux densities, $\beta$ is the emissivity spectral index, $T$ is the fitted colour temperature, 
and $F_{\nu,0}$ is the fitted flux density at a reference wavelength $\nu_0$. In Eq.~\ref{eq:mbb} we have assumed that the observed 
emission is optically thin at frequencies $\nu \le 3$\,THz, that 
the emissivity spectral index is constant within the fitted wavelength range, and that the source is isothermal. 
The first two assumptions likely hold, however, the sources are not necessarily isothermal. In particular, 
the temperature is expected to vary along the line-of-sight due to radiative transfer effects, with lower values in the inner, denser part of the clump than  
the averaged colour temperature obtained from Eq.~\ref{eq:mbb}. The derived column density and mass can then be underestimated up to a factor 3 \citep{Ysard2012}. 
For each source, two independent temperature estimates have been obtained, i.e.,  by fixing the emissivity spectral index, $\beta$, to 2.0, 
or by letting $\beta$ be a free parameter. These estimates
and their uncertainties have been calculated with a MCMC approach, using a Bayesian formulation with 
flat priors distributions for the temperature, emissivity spectral index (when this is a free parameter), and the amplitude of the fitted
modified black body \citep[see][and references therein]{Juvela2013}. The allowed parameter ranges are:
$5.0 < T < 30$\,K, $0.5 < \beta < 5.0$ and $F_{\nu,0}>0.0$. 

The MCMC calculations have been made in chunks of 10$^7$ iterations.  After reaching convergence (following the initial burn-in phase), 
the parameters have been estimated from the final $5\times 10^7$ steps. The parameters values quoted in the catalogue 
correspond to the mean and the marginalised 68\% confidence intervals calculated from the MCMC samples.   
 The uncertainty of the MCMC temperature estimate caused by the finite length of the MCMC chain is always negligible.

For sources with FQ=1, we have compared the temperature and emissivity spectral index obtained from $\chi^2$ fits 
and from the MCMC results, and these are on average within 1\%.  For sources with the lowest S/N, the $\chi^2$ method tends 
to give higher $T$ values and lower $\beta$ values than the MCMC technique. This is consistent with the
behaviour found in simulations where, when $\beta$ is treated as a free parameter, the joint probability distribution of $T$
and $\beta$ is asymmetric and presents a long tail towards low temperatures and high values of the emissivity spectral index 
\citep{Juvela2012b,Shetty2009b}. The standard deviations of the MCMC vs $\chi^2$ estimates are $\sigma(T)=0.3\,\rm{K}$ and
$\sigma(\beta)=0.08$. Again, these numbers reflect mostly the behaviour of sources with the lowest S/N measurements. 
In general, the difference between the MCMC and $\chi^2$ estimates are not significant when compared to the 
uncertainties. The MCMC results have also been compared to $\chi^2$ fits performed over a regular grid with a step $\Delta T=0.2\,\rm{K}$
and $\Delta \beta=0.05$, where, at each grid position, we fit only the amplitude of the model spectrum. The results are consistent 
within the  uncertainties, confirming that MCMC has correctly localised the absolute minimum across the entire allowed parameter space. 

\begin{figure*}
\vspace{-0.7cm}
\center
\begin{tabular}{cc}
\includegraphics[width=8.5cm]{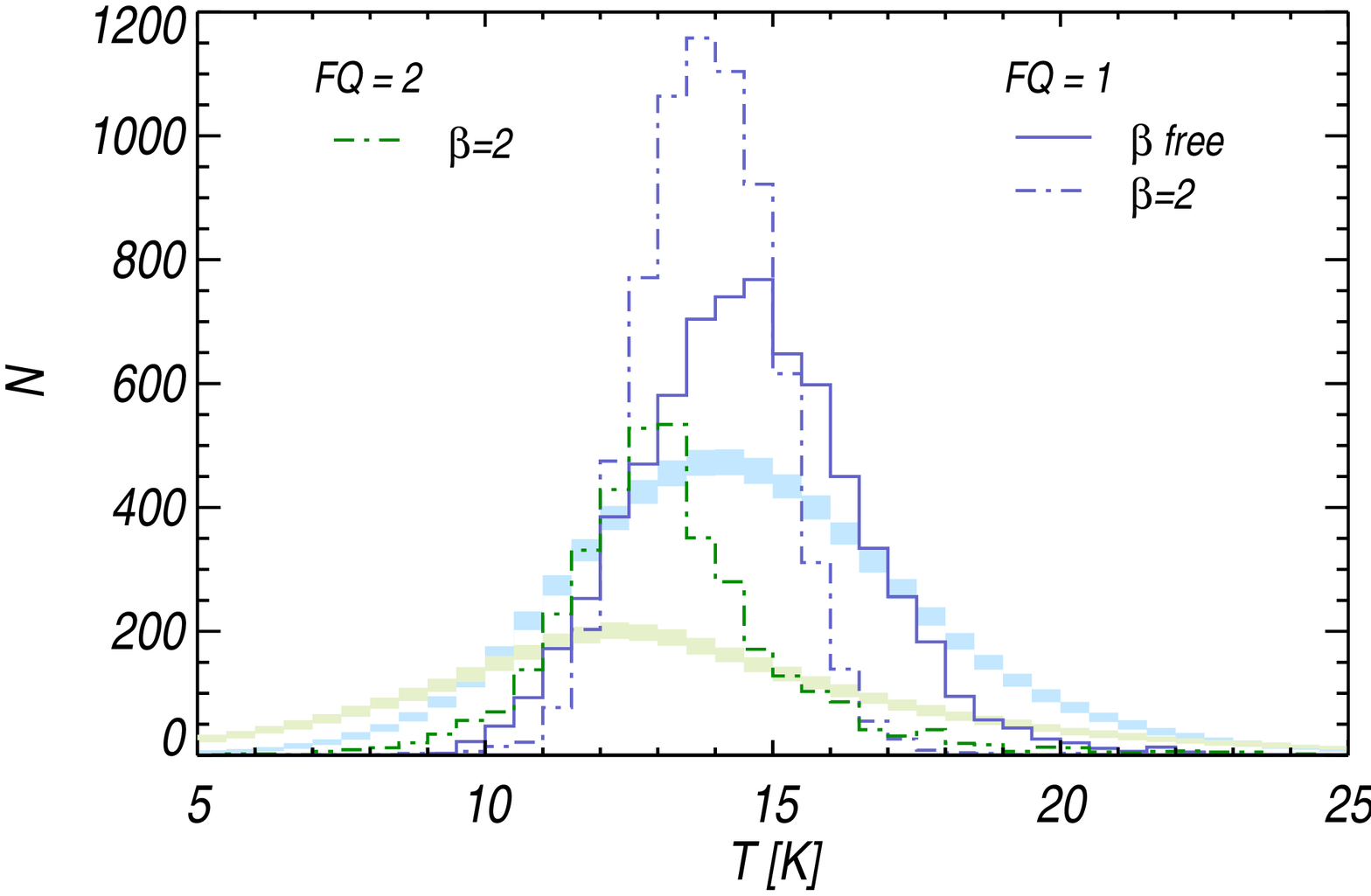} &
\includegraphics[width=8.5cm]{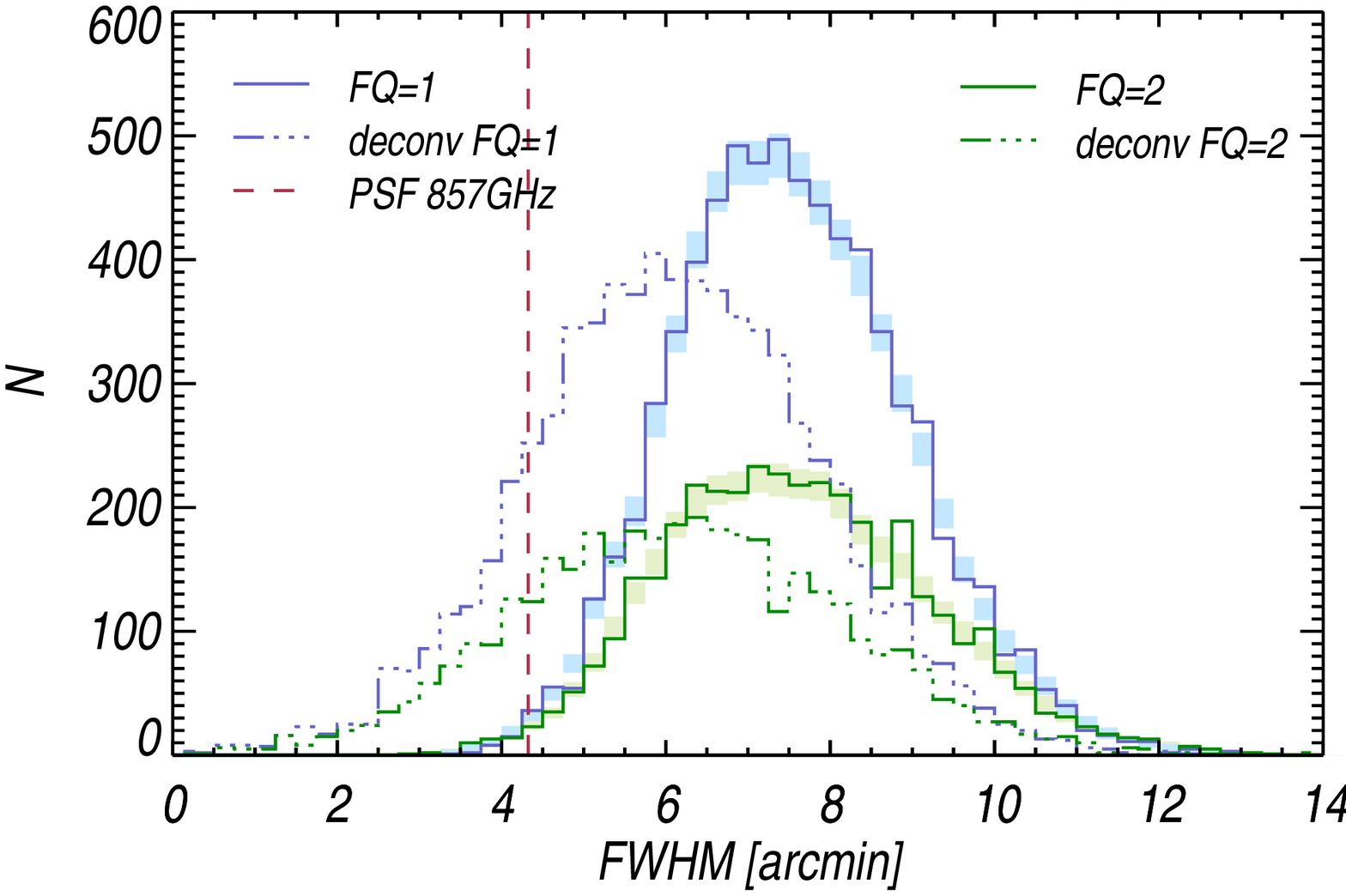} \\
\includegraphics[width=8.5cm]{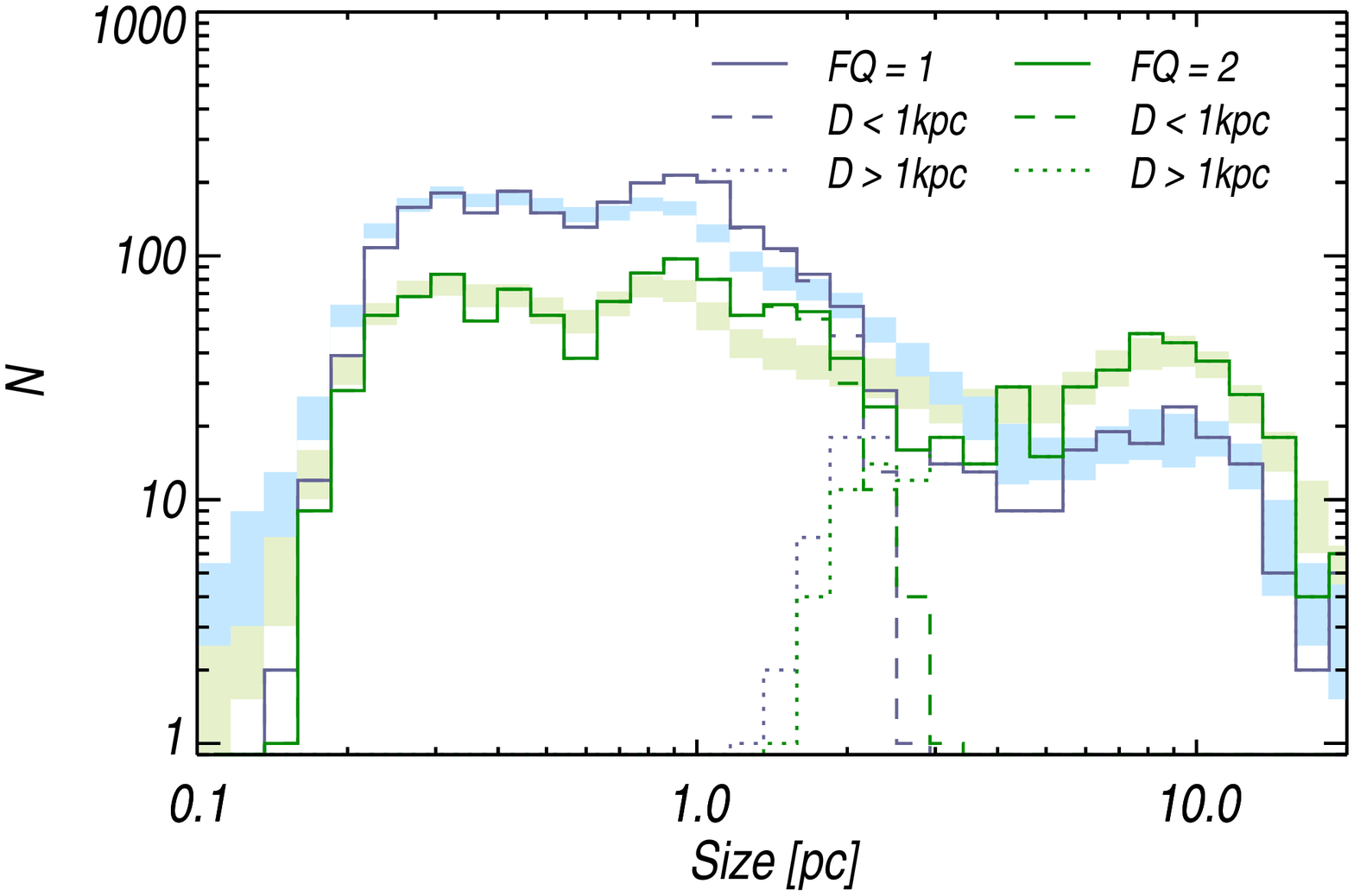} &
\includegraphics[width=8.5cm]{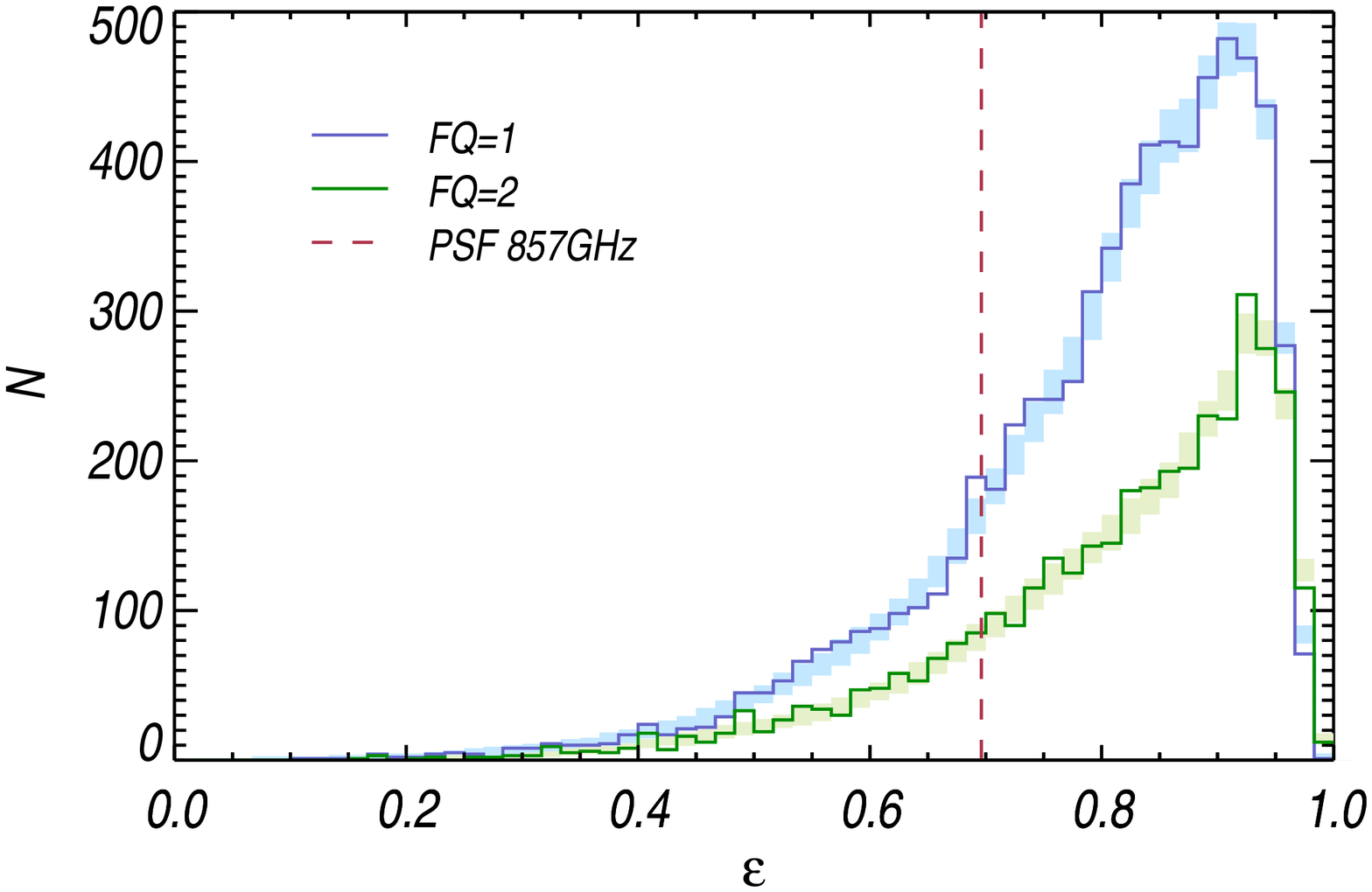} \\
\includegraphics[width=8.5cm]{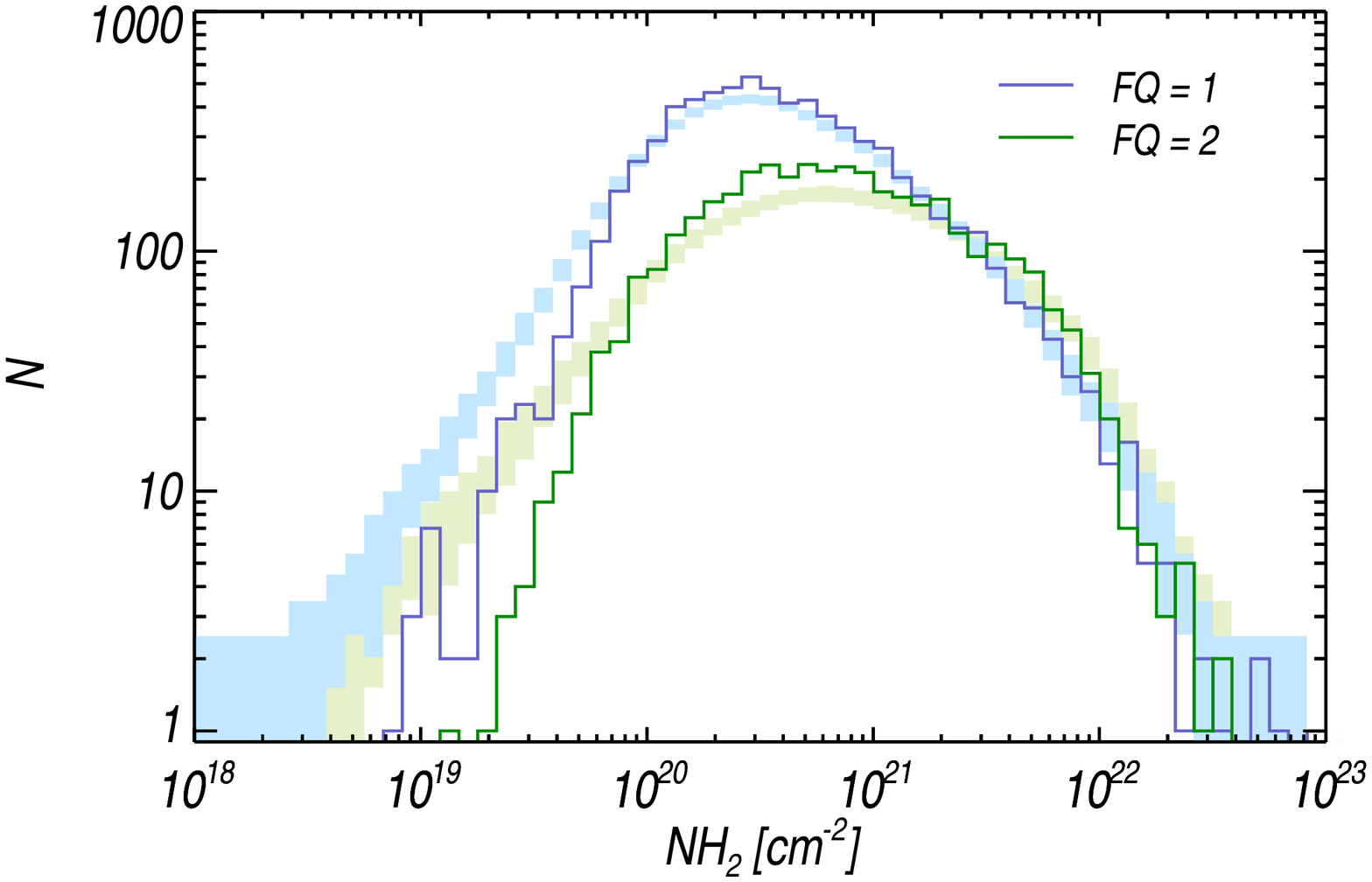} &
\psfrag{xtitle}{$M \, [M_{\odot}]$}
\includegraphics[width=8.5cm]{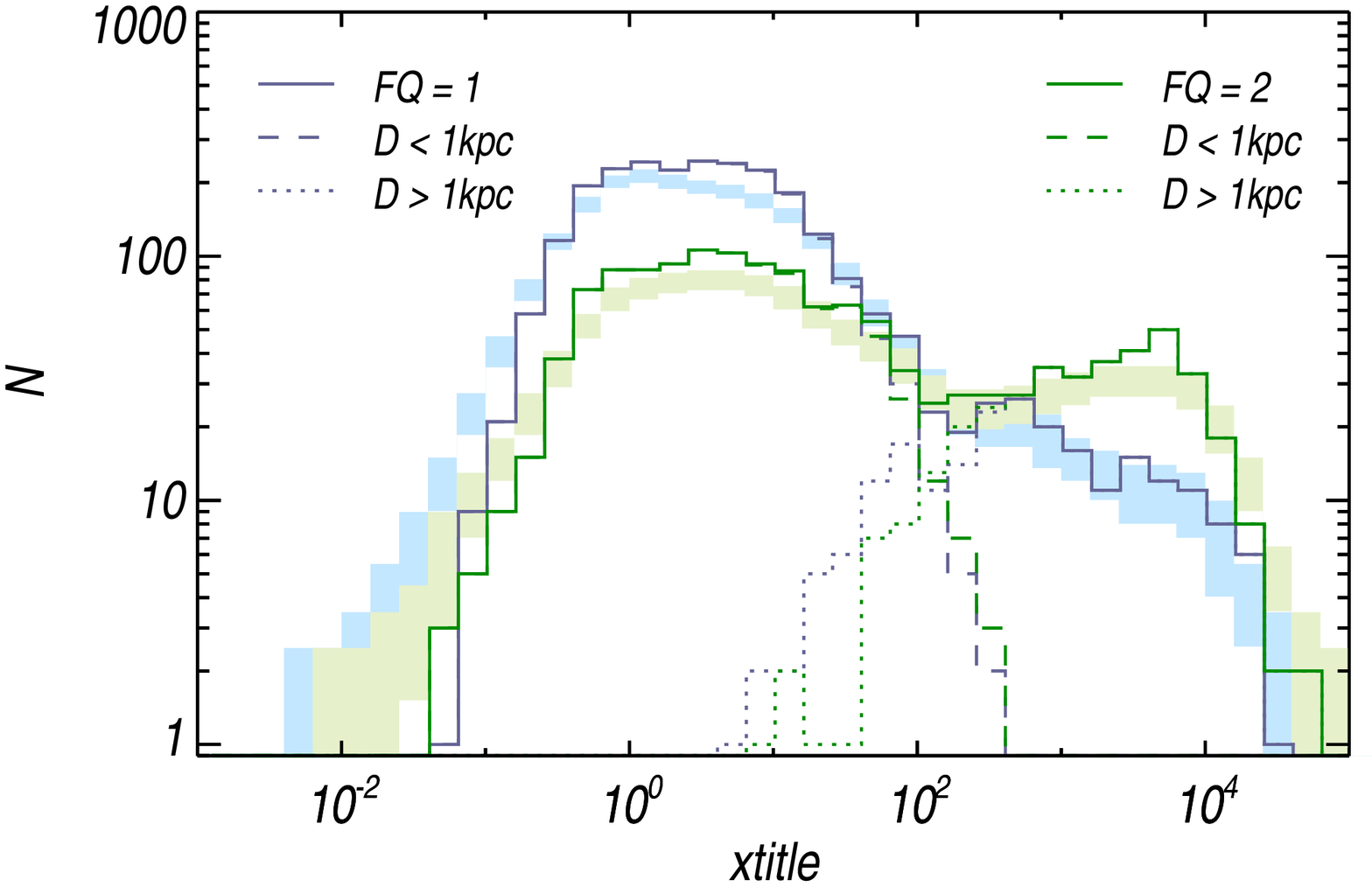} \\
\psfrag{xtitle}{$L \, [L_{\odot}]$}
\includegraphics[width=8.5cm]{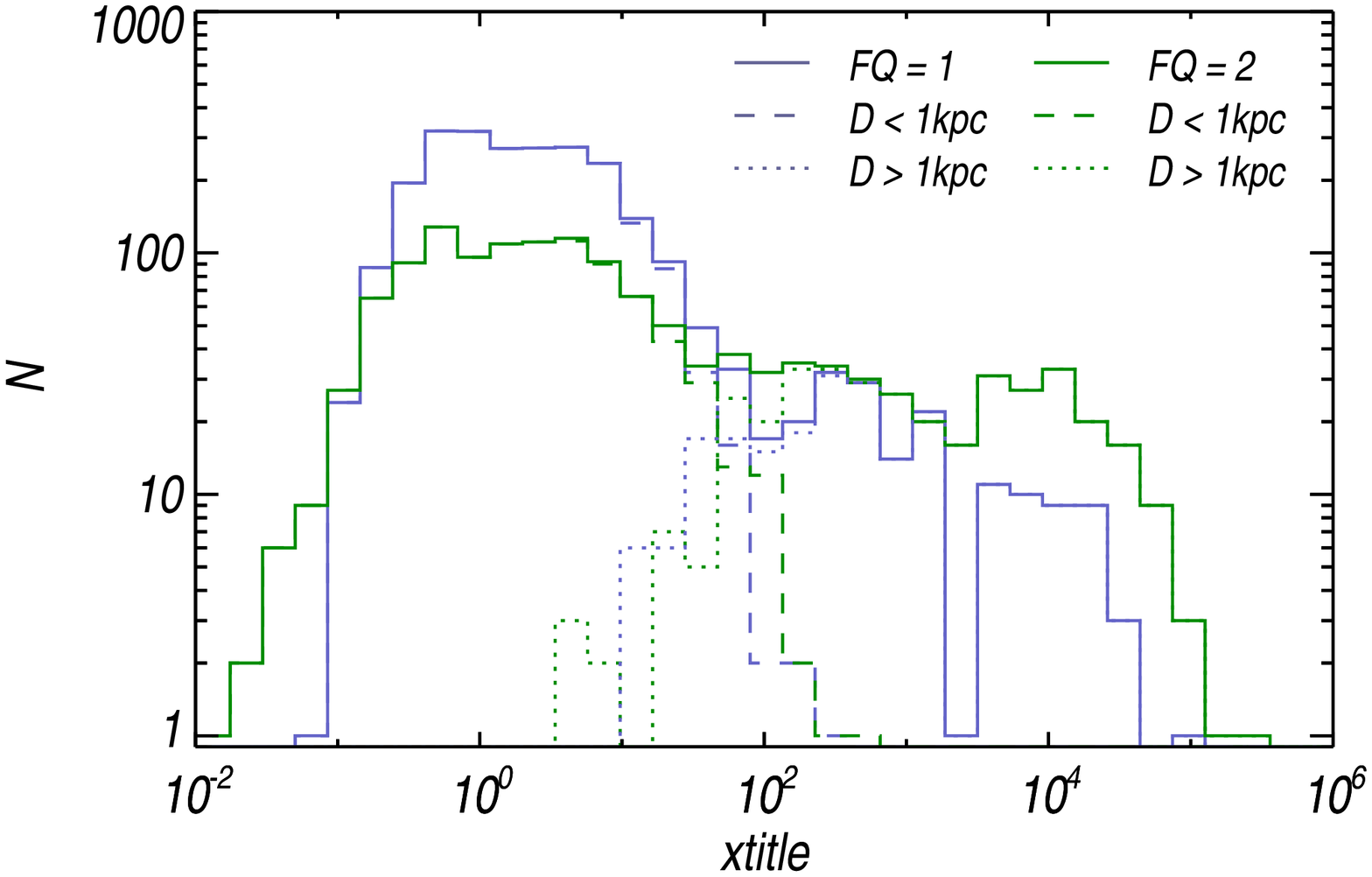}  &
\psfrag{-----xtitle-----}{$n_{\mathrm{H}_2} \, [cm^{-3}]$}
\includegraphics[width=8.5cm]{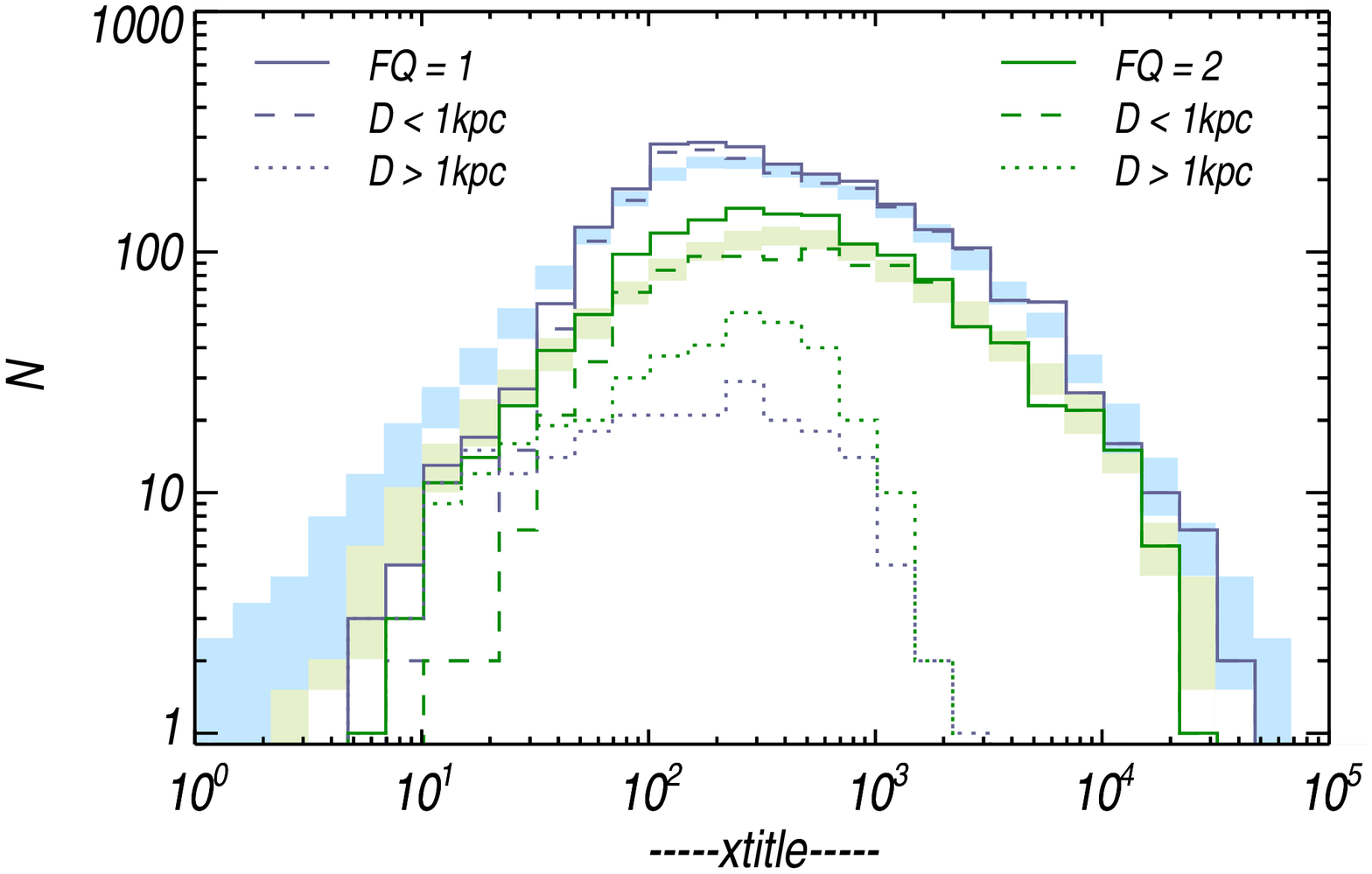} \\
\end{tabular}
\caption{Distribution of the physical properties of the PGCC objects with FQ=1 (blue) and FQ=2 (green).
The coloured shaded regions  provide an estimate of the impact of the individual uncertainties on the statistical distribution at a 1$\sigma$
dispersion level around the mean value obtained from 10\,000 MC realizations (see Sect.~\ref{sec:physical_properties} for more details). 
 The temperature distribution is shown for two temperature estimates, i.e.,  using
 a free emissivity spectral index $\beta$ ({\it solid line}) or a fixed $\beta$=2 ({\it dot-dashed line}).
 The distribution of the intrinsic size, after deconvolution, is shown on the {\it top right panel} in {\it dot-dot-dot-dashed line}.
 The distributions of the physical size, mass, mean density and luminosity of the clumps are provided for
 sources with reliable distance estimates ({\asciifamily DIST\_QUALITY}=1 or 2).
These distributions are also shown separately for sources at solar distances smaller or larger than 1\,kpc from the Sun, in {\it dashed} and {\it dotted lines}, respectively. 
 }
\label{fig:distribution_physical_properties}
\end{figure*}

 We stress that the temperature estimates of the sources correspond to the temperature of the clump-only after removal of the warm background, 
while the colour temperature is usually associated with the total emission on the line-of-sight.
The resulting clump temperature distribution is shown in Fig.~\ref{fig:distribution_physical_properties} for 
the two categories of sources, with FQ=1 and 2.
When the emissivity spectral index $\beta$ is allowed to vary ({\it solid lines}), the temperature of 
the {\Planck} cold clump candidates with  FQ=1 ranges from 8.6 to 30\,K, 
 and actually from 10.5 to 19.9\,K when excluding the 2\% extreme percentiles, 
with a peak of the distribution at about 14.5\,K. Likewise, when we take $\beta$=2, the distribution is narrower
 (ranging from 6 to 22.5\,K,  and from 11.1 to 16.8\,K when excluding the 2\% extreme percentiles)  but still peaks 
around 14\,K. These values are about 1\,K higher than the temperature estimates derived in \citet{planck2011-7.7b}. 
This can be ascribed to the change in calibration of the {\Planck} high frequency bands (857 and 545\,GHz) with respect 
to the data used for the {\Planck} Early Papers and the {\Planck} Results 2013 \citep[see][]{planck2013-p03f}. 
The temperature distribution of sources with FQ=2 has a peak at lower  
temperatures, around 13\,K,  and spans a range of lower temperatures, 
down to 5.8\,K, while it ranges from 8.6 to 22.3\,K when excluding the 2\% extreme percentiles. 
This is consistent with the aforementioned hypothesis  that these sources are colder than sources with FQ=1.

If we compare this  actual  distribution with the expected temperature  completeness of Fig.~\ref{fig:mcqa_completeness}, we notice that 
the temperature distribution of the PGCC sample with FQ=1 drops to zero below 9\,K, while the
expected detection efficiency is still about 15\% below 8\,K for the same category. Considering now sources with FQ=2, 
 about 0.5\% of the sample exhibits temperature below 8\,K, but no sources below 5.8\,K, where the expected completeness reaches almost 45\% for the same category.
The lower temperature limits at 8.6 and 5.8\,K for sources with FQ=1 and 2, respectively, 
appear therefore as physical thresholds, rather than a bias introduced by the detection algorithm.
However, it is important to keep in mind that colder sources (below 6\,K) may still exist on smaller angular scales.

\subsection{Size}
\label{sec:size}

The geometric parameters of the PGCC sources are derived from the photometric measurements by means of an elliptical Gaussian
fit to the 857\,GHz {\it cold residual} map with an elliptical Gaussian. From the fit, we can 
estimate: the source centroid, its major and minor axis, ellipticity and position angle. In the catalogue, these 
quantities are available only for sources with FQ=1 or 2. 

The distribution of $\theta$, defined as the geometric mean of the major and minor FWHM,  
$ \theta = \sqrt{\theta_{\rm{maj}} \theta_{\rm{min}}} $, is shown in Fig.~\ref{fig:distribution_physical_properties}. 
The PGCC sources have an average $\theta$ of 7{\arcm}5 (red {\it dashed line}), thus appearing slightly extended with respect to 
the effective beam, $\theta_{PSF}$, at 857\,GHz \citep[4.325$\pm$0{\parcm}055, ][]{planck2013-p01}. 
We have also computed the intrinsic size, $\theta_i$, of the sources by deconvolution of the observed 
diameter from the effective beam, and yielding an average value of 5{\parcm}6. We have obtained an average $\theta_i/\theta_{PSF}$ ratio of 1.4, 
which is consistent with other similar observations, e.g.,  by BLAST \citep[1.1,][]{Netterfield2009} and Bolocam \citep[1.5,][]{Enoch2007}. 
This effect might have to do with the hierarchical structure of the interstellar medium, and to the fact 
that cold clumps are likely to be characterized by the presence of extended envelopes. 
Note that, statistically speaking, PGCC sources with FQ=1 or 2 have comparable $\theta$ distributions.

The ellipticity of the sources is defined as: 
\begin{equation}
\varepsilon = \sqrt{1 - \left( \frac{\theta_{\rm{min}}}{\theta_{\rm{maj}}}   \right)^2}\, .
\end{equation}
As shown in Fig.~\ref{fig:distribution_physical_properties}, PGCC clumps exhibit a distribution of the ellipticity peaking around 0.9 with 
a median value of 0.83, larger than the average ellipticity of the effective beam at 857\,GHz, which is about
0.70 \citep{planck2013-p01}. While the uncertainty on the major and minor FWHM determination 
can lead to an artificial elongation of the sources, we estimate that a 45\% level of uncertainty is required to produce a median value of the ellipticity of 0.83 instead of 0.7, 
whereas the effective accuracy of the recovered ellipticity is about 10\% (see Sect. \ref{sec:geometry_accuracy}). 
Therefore, the observed elongation of the PGCC sources cannot be explained by noise only.
Furthermore, we have verified that the position angle of the PGCC sources is not correlated with the {\Planck} scanning strategy. 
Indeed, for the previous catalogue version, particularly the ECC \citep{planck2011-1.10}, 
a statistical analysis revealed that the orientation of the sources was slightly spatially correlated with the scan direction. Such an effect was 
induced by the instrument transfer function, which at the time was not fully characterized. This issue has been resolved in the 
public release of {\Planck} data \citep[see][]{planck2013-p03c} used to build the present catalogue. 
About 40\% of sources have $0.87<\varepsilon<0.95$, which translates to axial ratios
between 2 and 3. PGCC clumps appear therefore clearly elongated, which may be due to their association with filamentary structures.

For the sources with a reliable distance estimate ({\asciifamily DIST\_QUALITY}=1 or 2), i.e., 
2489  and 1378 sources with FQ=1 and 2, respectively, we have also derived their linear size. 
A few things need to be noted regarding the linear size distribution shown in Fig.~\ref{fig:distribution_physical_properties}. 
First of all, this distribution is limited to objects for which a distance estimate is available, and these constitute, as discussed in 
Sect.~\ref{sec:distance}, a very heterogeneous sample of sources. Secondly, this distribution is strongly influenced by the {\Planck} 
angular resolution (5{\arcmin}), therefore small ($<1$\,pc) and compact clumps  are likely local objects ($\rm{D}<1$\,kpc), while 
larger objects (a few pc) are likely intrinsically extended structures located at large distances ($\rm{D}>1$\,kpc).

\subsection{Column density}
\label{sec:nh}

The column density of the  {\Planck} cold clumps  has been evaluated following the prescription in \citet{planck2011-7.7b}:
\begin{equation}
N_{\mathrm{H}_2} = \frac{S'_{\nu} / \Omega }{ \mu m_{\mathrm{H}}\,  B_{\nu}(T) \,  \kappa_{\nu}} \, ,
\end{equation}
where $S'_{\nu}$ is the flux density per beam at the frequency $\nu$, which is integrated over the solid angle  of the clump defined by
$\Omega$=$\pi \theta_{\mathrm{min}} \theta_{\mathrm{maj}} /4$,
 $\mu$=2.33 is the mean molecular weight, $m_{\mathrm{H}}$ is the mass of the atomic hydrogen, and 
 $\kappa_{\nu}$ is the dust opacity. We have adopted the dust opacity from \citet{Beckwith1990}, $\kappa_{\nu}$=$0.1(\nu/1\rm{THz})^{\beta}\, \rm{cm^{2}g^{-1}}$, 
which is appropriate for the case of dense clouds at intermediate densities. The column density is computed at 857\,GHz, 
close to the 1\,THz reference of \citet{Beckwith1990}, which allows us to minimize the impact of assuming a fixed emissivity spectral index, $\beta$=2.
Changing the emissivity spectral index in the range 1--3 yields up to 15\% variations on the emissivity estimate, which is negligible compared to the intrinsic uncertainty
of the emissivity. 
This is also the {\Planck}-HFI band where the S/N is the highest, and where dust emission remains optically thin. 
Note that $S'_{857}$, integrated over $\Omega$, is half the flux density provided in the catalogue,  $S_{857}$, which is the flux density integrated over the whole clump.

The peak column density is defined by: 
\begin{equation}
N^{\mathrm{peak}}_{\mathrm{H}_2} = \frac{I_{\nu}}{ \mu m_{\mathrm{H}}\,  B_{\nu}(T) \,  \kappa_{\nu}} \, ,
\end{equation}
where $I_{\nu}$ is the surface brightness measured at the peak, and it can be derived from $N_{\mathrm{H}_2}$ 
by multiplying by 1.38.  We stress that these two estimates of the column density are derived for the {\Planck} clumps only after removal of the warm component, 
which is not what is typically found in the literature. 

The $N_{\mathrm{H}_2}$ distribution is shown in Fig.~\ref{fig:distribution_physical_properties}. This ranges 
from 6.8$\times$$10^{18}$ to 1.2$\times$$10^{23}$\,$\rm{cm^{-2}}$, with a median of 3.4$\times$$10^{20}$\,$\rm{cm^{-2}}$ and 
6.3$\times$$10^{20}$\,$\rm{cm^{-2}}$ for sources with FQ=1 and 2, respectively. 

Sources with FQ=2 have slightly larger column densities, which is expected if these sources effectively 
correspond to colder and denser objects. About 80\% of the sources have a column density between
1.4$\times$$10^{20}$ and 3.7$\times$$10^{21}$\,$\rm{cm^{-2}}$. We emphasize that these column densities 
are averaged over the size of the clump which, at the {\Planck} resolution, is likely
inducing a bias at low values, because of  beam dilution effects. Indeed, as shown in 
\citet{planck2011-7.7a}, the {\it Herschel} higher angular resolution
observations of a sample of PGCC sources have revealed complex sub-structures, characterized by lower temperatures 
and higher densities compared to the parent clump.

\subsection{Mass and mean density}
\label{sec:mass}

For sources with a reliable distance estimate, the mass of the clump is given by:
\begin{equation}
M = \frac{S_{\nu}\,  D^2}{ \kappa_{\nu}\,   B_{\nu}(T)}\, ,
\end{equation}
where $S_{\nu}$ is the flux density measured at 857\,GHz, integrated over the solid angle $\Omega$ defined in Sect.~\ref{sec:nh}, 
$D$ is the distance, $\kappa_{\nu}$ is the 
dust opacity defined in Sect.~\ref{sec:nh}, and $B_{\nu}(T)$ is the Planck function
for a dust temperature $T$.

The mass distribution shown in Fig.~\ref{fig:distribution_physical_properties}, ranges from
a few $10^{-2}$ to almost $10^{5}$ $M_{\odot}$, probing a large variety of objects, from cores to giant molecular clouds.
We stress that this mass distribution, as the linear size distribution discussed in Sect.~\ref{sec:size}, is biased by our distance 
sample, which we know being highly heterogeneous (see Sect.~\ref{sec:distance}). This becomes clear when we partition the mass 
distribution according to distance, in particular by separating sources located closer ({\it dashed line}) or further ({\it dotted line}) than 1\,kpc from the Sun. 
Furthermore, except for less than 40\% of the cases where the distance estimates are obtained from molecular complex association providing highly reliable estimates, 
the uncertainty on the computed mass is mainly dominated by the uncertainty on the distance, which is
about two to three times larger than the uncertainties on the temperature and flux density involved in the calculation. 
Hence about 84\% of sources have mass estimates with a S/N between 1 and 2, 
and 16 \% with a S/N above 2. Note also that we have almost reached the theoretical sensitivity limit of {\Planck}
 to low-mass cold cores, as it has been derived in Sect.~\ref{sec:completeness},  which is about 0.03\,$\mathrm{M}_{\odot}$
for a cold source located at 100\,pc and having a column density of $10^{20}$\,$\mathrm{cm}^{-2}$.

Finally, we have obtained the mean density of the clump, computed as $n_{\mathrm{H}_2}$=$M / V$, 
where $V$ is the volume of the clump, modeled as a sphere of diameter equal to the physical size of the clump derived in Sect.~\ref{sec:size}. 
Figure~\ref{fig:distribution_physical_properties} shows that the mean density ranges from 5.3 to $3.5\times10^4$\,$\mathrm{cm}^{-3}$, again 
spanning a wide range of object categories. 
While the smallest estimates are not consistent with typical density of large clouds \cite[i.e., $\sim10^2\,\rm{cm}^{-3}$, ][]{Williams2000, Blitz1993}, 
a closer look at the correlation with the size of the objects reveals that these values can be explained by a dilution effect in the {\Planck} beam, as discussed in Sect.~\ref{sec:matrix_correlation}.

\begin{figure}
\center
\psfrag{---------------xtitle---------------}{$log(L/M) \, [log(L_{\odot}/M_{\odot})]$}
\includegraphics[width=8.5cm]{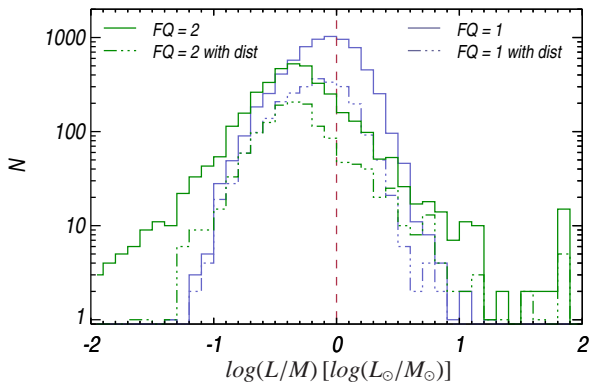} 
\caption{ Distribution of the luminosity to mass ratio of the PGCC clumps for sources with FQ=1(blue line) and FQ=2 (green line).
The {\it dot-dot-dot-dashed line} denotes the distribution for the sub-sample of sources 
with a reliable distance estimate.}
\label{fig:distribution_mass_luminosity}
\end{figure}

\begin{table*}
\caption{Statistical description of the physical properties of the PGCC clumps with FQ=1. 
The statistics of the quantities denoted with $\star$ has been computed on the sub-sample of sources having a distance estimate with 
{\asciifamily DIST\_QUALITY}=1 or 2.}
\label{tab:physical_properties_1}
\nointerlineskip
\setbox\tablebox=\vbox{
\newdimen\digitwidth 
\setbox0=\hbox{\rm 0} 
\digitwidth=\wd0 
\catcode`*=\active 
\def*{\kern\digitwidth} 
\newdimen\signwidth 
\setbox0=\hbox{+} 
\signwidth=\wd0 
\catcode`!=\active 
\def!{\kern\signwidth} 
\newdimen\pointwidth 
\setbox0=\hbox{.} 
\pointwidth=\wd0 
\catcode`?=\active 
\def?{\kern\pointwidth} 
\halign{\hbox to 0.8in{#\leaderfil}\tabskip=1.0em&
\hfil#\hfil&
\hfil#\hfil&
\hfil#\hfil&
\hfil#\hfil&
\hfil#\hfil&
\hfil#\hfil&
\hfil#\hfil&
\hfil#\hfil&
\hfil#\hfil\tabskip=0pt\cr
\noalign{\doubleline}
\omit\hfil Percentile\hfil& $T_{\rm c}$ & FWHM & size$^{\star}$ & $\varepsilon$ & $N_{{\rm H}_2}$ & D$^{\star}$ & M$^{\star}$ & $n_{\mathrm{H}_2}$$^{\star}$ & L$^{\star}$ \cr
\omit\hfil  \hfil& [K] & [{\arcmin}]& [pc] & & [${\rm cm}^{-2}$] & [kpc]  &  [$\rm{M}_{\odot}$] & [cm$^{-3}$]&  [$\rm{L}_{\odot}$] \cr
\noalign{\vskip 3pt\hrule\vskip 4pt}
min. 		&  *8.6 	& *3.4 	& *0.14 	&   0.10 	&  6.8$\times$$10^{18}$  	& *0.07 	&  5.1$\times$$10^{-2}$ 	& 5.3$\times$$10^{0}$   	&  7.6$\times$$10^{-2}$   \cr
first 1\% 	& 10.5 	& *4.6 	&  *0.20 	&  0.36 	&   3.2$\times$$10^{19}$ 	&  *0.12 	& 1.4$\times$$10^{-1}$	& 1.6$\times$$10^{1}$ 	& 1.4$\times$$10^{-1}$ \cr
first 10\% 	& 12.1	& *5.8	&   *0.27	&   0.61	&   9.8$\times$$10^{19}$	&  *0.14 	&  4.5$\times$$10^{-1}$	&  6.8$\times$$10^{1}$	&  3.6$\times$$10^{-1}$\cr
median	&  14.5 	&  *7.5 	& *0.71 	&  0.83 	&   3.4$\times$$10^{20}$	&   *0.33 	& 3.3$\times$$10^{0}$?  	&  3.0$\times$$10^{2}$  	&  2.1$\times$$10^{0}$?  \cr
last 10\%  & 17.0	&  *9.3 	&   *1.91 	&   0.94 	&  1.8$\times$$10^{21}$	&  *0.92 	&   5.9$\times$$10^{1}$? 	&   2.4$\times$$10^{3}$ 	&   6.9$\times$$10^{1}$? \cr
 last 1\%  	& 19.9 	& 11.0 	&  11.72 	&   0.97 	& 8.6$\times$$10^{21}$ 	&   *5.51 	&  6.8$\times$$10^{3}$?  	&  1.3$\times$$10^{4}$   	&   7.5$\times$$10^{3}$?  \cr
max. 	& 30.0 	& 13.1 	& 25.16  	&   0.98	& 8.8$\times$$10^{22}$	&   10.48  	& 2.6$\times$$10^{4}$?	& 3.5$\times$$10^{4}$  	&  7.6$\times$$10^{4}$? \cr
\noalign{\vskip 3pt\hrule\vskip 4pt}
}}
\endPlancktable
\end{table*}

\begin{table*}
\caption{Same as Table~\ref{tab:physical_properties_1} with FQ=2. 
The statistics of the quantities denoted with $\star$ has been computed on the sub-sample of sources having a distance estimate with 
{\asciifamily DIST\_QUALITY}=1 or 2. We recall that for this class of sources, we only provide an upper limit of the temperature.}
\label{tab:physical_properties_2}
\nointerlineskip
\setbox\tablebox=\vbox{
\newdimen\digitwidth 
\setbox0=\hbox{\rm 0} 
\digitwidth=\wd0 
\catcode`*=\active 
\def*{\kern\digitwidth} 
\newdimen\signwidth 
\setbox0=\hbox{+} 
\signwidth=\wd0 
\catcode`!=\active 
\def!{\kern\signwidth} 
\newdimen\pointwidth 
\setbox0=\hbox{.} 
\pointwidth=\wd0 
\catcode`?=\active 
\def?{\kern\pointwidth} 
\halign{\hbox to 0.8in{#\leaderfil}\tabskip=1.0em&
\hfil#\hfil&
\hfil#\hfil&
\hfil#\hfil&
\hfil#\hfil&
\hfil#\hfil&
\hfil#\hfil&
\hfil#\hfil&
\hfil#\hfil&
\hfil#\hfil\tabskip=0pt\cr
\noalign{\doubleline}
\omit\hfil Percentile\hfil& $T_{\rm c}$ & FWHM & size$^{\star}$ & $\varepsilon$ & $N_{{\rm H}_2}$ & D$^{\star}$ & M$^{\star}$ & $n_{\mathrm{H}_2}$$^{\star}$ & L$^{\star}$ \cr
\omit\hfil  \hfil& [K] & [{\arcmin}]& [pc] & & [${\rm cm}^{-2}$] & [kpc]  &  [$\rm{M}_{\odot}$] & [cm$^{-3}$]&  [$\rm{L}_{\odot}$] \cr
\noalign{\vskip 3pt\hrule\vskip 4pt}
min. 		&  *5.8 	& *2.8 	& *0.15 	&   0.16 	&  1.4$\times$$10^{19}$  	& *0.07 	&  5.0$\times$$10^{-2}$ 	& 6.4$\times$$10^{0}$   	&  9.2$\times$$10^{-3}$   \cr
first 1\% 	& 8.6 	& *4.2 	&  *0.19 	&  0.37 	&   4.9$\times$$10^{19}$ 	&  *0.12 	& 1.4$\times$$10^{-1}$	& 1.4$\times$$10^{1}$ 	& 6.4$\times$$10^{-2}$ \cr
first 10\% 	& 11.0	& *5.6	&   *0.28	&   0.61	&   1.4$\times$$10^{20}$	&  *0.14 	&  6.1$\times$$10^{-1}$	&  6.4$\times$$10^{1}$	&  2.7$\times$$10^{-1}$\cr
median	&  13.0 	&  *7.5 	& *0.95 	&  0.84 	&   6.3$\times$$10^{20}$	&   *0.41 	& 8.7$\times$$10^{0}$?  	&  3.5$\times$$10^{2}$  	&  3.7$\times$$10^{0}$?  \cr
last 10\%  & 15.7	&  *9.7 	&   *8.55 	&   0.95 	&  3.7$\times$$10^{21}$	&  *4.38 	&   3.1$\times$$10^{3}$? 	&   2.6$\times$$10^{3}$ 	&   3.3$\times$$10^{3}$? \cr
 last 1\%  	& 22.2 	& 11.7 	&  15.69 	&   0.98 	& 1.0$\times$$10^{22}$ 	&   *6.90 	&  1.4$\times$$10^{4}$?  	&  1.2$\times$$10^{4}$   	&   4.6$\times$$10^{4}$?  \cr
max. 	& 30.0 	& 14.5 	& 30.56  	&   0.98	& 1.2$\times$$10^{23}$	&   10.16  	& 6.3$\times$$10^{4}$?	& 2.4$\times$$10^{4}$  	&  2.9$\times$$10^{6}$? \cr
\noalign{\vskip 3pt\hrule\vskip 4pt}
}}
\endPlancktable
\end{table*}

\begin{figure*}
\vspace{2.5cm}
  \begin{center}
\psfrag{xtitle1}{$T_{\rm{d}} \, [\rm{K}]$}
\psfrag{-----xtitle2-----}{$N_{\rm{H}_2} \, [\rm{cm}^{-2}]$}
\psfrag{xtitle3}{$D \, [\rm{kpc}]$}
\psfrag{---xtitle4---}{$\rm{size} \, [\rm{pc}]$}
\psfrag{--xtitle5--}{$M \, [\rm{M}_{\odot}]$}
    \includegraphics[width=8.9cm, angle=90, viewport=-60 -50 600 500]{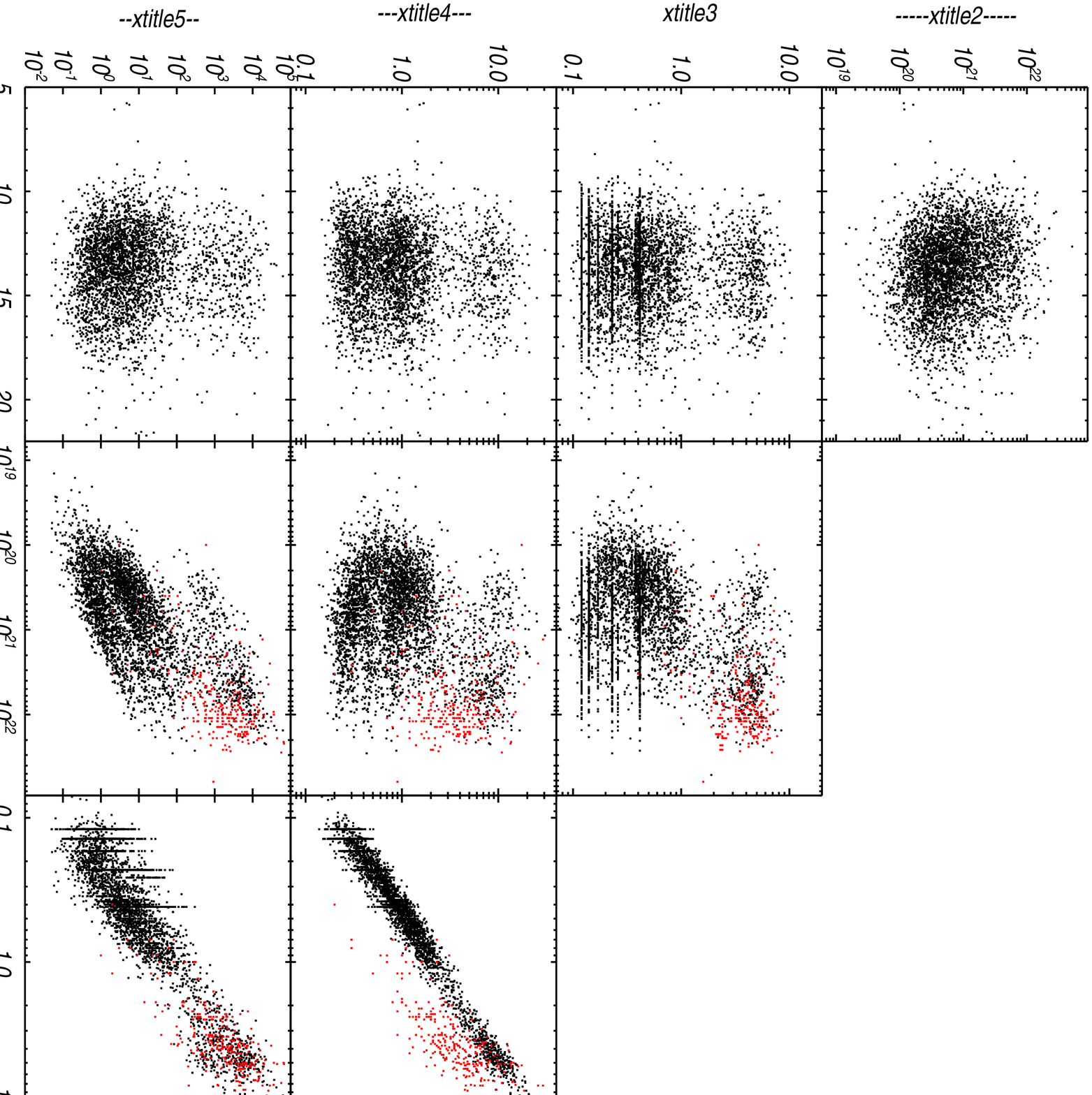}
     \caption{Correlation plots for: temperature, column density, distance, size and mass. The black dots represent 
3867 PGCC sources with a robust distance and flux density estimate. The red dots denote the IRDCs from \citet{Simon2006b}.between the following physical properties of the 3867 PGCC sources for which a robust distance estimate is available combined with a reliable flux density: temperature, column density, distance, size and mass. It is compared to the physical properties of IRDCs from \citet{Simon2006b} in red.}
     \label{fig:matrix_correlation}
  \end{center}
\end{figure*}

\subsection{Luminosity}
\label{sec:luminosity}

We have computed the  bolometric luminosity of the cold clumps as: 
\begin{equation}
L = 4 \pi D^2 \int_{\nu} S'_{\nu} d\nu \, ,
\end{equation}
where $D$ is the distance and $S'_{\nu}$ is the flux density of the clump modeled with a modified black body function 
with temperature $T$ and emissivity spectral index $\beta$, as described in Sect.~\ref{sec:temperature}, 
and normalized to $S_{\nu}$ defined in Sect.~\ref{sec:nh}. 
The integral is taken over the frequency range $300\,\rm{GHz}<\nu<10$\,THz, which extends beyond the IRAS and {\Planck} spectral coverage. 
Any emission at shorter wavelengths is not included in this calculation.
From Fig.~\ref{fig:distribution_physical_properties} we note that the luminosity 
distributions for sources in the FQ=1 and 2 categories are very heterogeneous, ranging from $10^{-2}$ to $10^6$\,$\mathrm{L}_{\odot}$.

Figure~\ref{fig:distribution_mass_luminosity} illustrates the $L$/$M$ distribution. This quantity presents the advantage of not depending on distance, thus it 
can be derived for all the sources in the catalogue. It is often used as an indicator of the evolutionary stage of cold sources. 
About 73\% and 83\% of the sample with FQ=1 and 2, respectively, have a $L/M<1$, 
which is typical of objects with low degree of evolution. Interestingly, we find the same 
proportion when we analyse the $L/M$ distribution ({\it dotted lines}) for objects with a reliable distance estimate, 
suggesting that the sample of sources with distance estimates is statistically representative of the entire sample.

\subsection{Correlation between physical properties}
\label{sec:matrix_correlation}

\begin{table*}
\caption{Result of the positional cross-correlation of the PGCC catalogue with the other {\Planck} internal catalogues.}
\label{tab:xflag}
\nointerlineskip
\setbox\tablebox=\vbox{
\newdimen\digitwidth 
\setbox0=\hbox{\rm 0} 
\digitwidth=\wd0 
\catcode`*=\active 
\def*{\kern\digitwidth} 
\newdimen\signwidth 
\setbox0=\hbox{+} 
\signwidth=\wd0 
\catcode`!=\active 
\def!{\kern\signwidth} 
\newdimen\pointwidth 
\setbox0=\hbox{.} 
\pointwidth=\wd0 
\catcode`?=\active 
\def?{\kern\pointwidth} 
\halign{\hbox to 1.5in{#\leaderfil}\tabskip=2.0em&
\hfil#\hfil&
\hfil#\hfil&
\hfil#\hfil&
\hfil#\hfil&
\hfil#\hfil&
\hfil#\hfil\tabskip=0pt\cr
\noalign{\doubleline}
\omit\hfil  &  &  & \multicolumn{3}{c}{PGCC  {\asciifamily FLUX\_QUALITY}} & \cr
\omit\hfil Catalogue\hfil &  option & N & 1 & 2 & 3 &  Total \cr
\omit\hfil  &  &  & (6993) & (3755) & (2440) &  (13188) \cr
\noalign{\vskip 3pt\hrule\vskip 4pt}
ECC & & **915 & *622 & *237 & *55 & *892 \cr
\noalign{\vskip 3pt\hrule\vskip 4pt}
PCCS 857 & zone 0!* & *4891 & **55 & ***9 & ***5 & **69 \cr
\omit\hfil  & zone 1$-$3 & 43290 & 5361 & 2511 & 1725 & 9597 \cr
\noalign{\vskip 3pt\hrule\vskip 4pt}

PCCS 545 & zone 0!* & *1694 & **67 & **15 & **12 & **94 \cr
\omit\hfil  & zone 1$-$3 & 31068 & 5010 & 2412 & 1596 & 9018 \cr
\noalign{\vskip 3pt\hrule\vskip 4pt}

PCCS 353 & zone 0!* & *1344 & **59 & **17 & ***8 & **84 \cr
\omit\hfil  & zone 1$-$3 & 22665 & 4645 & 2227 & 1393 & 8265 \cr
\noalign{\vskip 3pt\hrule\vskip 4pt}

PCCS 217 & zone 0!* & *2135 & *168 & **15 & **27 & *210 \cr
\omit\hfil  & zone 1$-$3 & 16842 & 3963 & 1836 & 1129 & 6928 \cr
\noalign{\vskip 3pt\hrule\vskip 4pt}

PCCS 143 & zone 0!* & *2160 & *106 & **12 & ***9 & *127 \cr
\omit\hfil  & zone 1$-$3 & *4139 & *959 & *748 & *320 & 2027 \cr
\noalign{\vskip 3pt\hrule\vskip 4pt}

PCCS 100 & zone 0!* & *1742 & *119 & **26 & **20 & *165 \cr
\omit\hfil  & zone 1$-$3 & *2487 & *545 & *478 & *225 & 1248 \cr
\noalign{\vskip 3pt\hrule\vskip 4pt}

PCCS 70 & zone 0!* & *1101 & **17 & **23 & **14 & **54 \cr
\omit\hfil  & zone 1$-$3 & **195 & ***6 & ***8 & ***6 & **20 \cr
\noalign{\vskip 3pt\hrule\vskip 4pt}

PCCS 44 & zone 0!* & **830 & ***6 & **17 & **13 & **36 \cr
\omit\hfil  & zone 1$-$3 & **104 & ***- & ***5 & ***1 & ***6 \cr
\noalign{\vskip 3pt\hrule\vskip 4pt}

PCCS 30 & zone 0!* & *1435 & **16 & **19 & **18 & **53 \cr
\omit\hfil  & zone 1$-$3 & **125 & ***2 & ***5 & ***6 & **13 \cr
\noalign{\vskip 3pt\hrule\vskip 4pt}

PCCS 857x545x357 & zone 0!* & **648 & **27 & ***8 & ***2 & **37 \cr
\omit\hfil  & zone 1$-$3 & 20534 & 3946 & 1890 & 1165 & 7001 \cr
\noalign{\vskip 3pt\hrule\vskip 4pt}

PSZ & & *1653 & **31 & **16 & **18 & **65 \cr
\noalign{\vskip 3pt\hrule\vskip 4pt}

PH$z$ & & *1261 & ***2 & **13 & ***- & **15 \cr
\noalign{\vskip 3pt\hrule\vskip 4pt}
}}
\endPlancktable
\end{table*}

A summary of the statistical properties of the PGCC clod clumps physical quantities is 
provided in Table~\ref{tab:physical_properties_1}, for sources with FQ=1, and in 
Table~\ref{tab:physical_properties_2}, for sources with FQ=2.
For each quantity the 1\%, 10\%  lower and higher percentiles and the minimum, median and maximum values are reported.
We stress that there is no systematic correlation between physical properties for the objects in the lowest and highest percentile bins. 
Indeed, the existence of a potential correlation between the physical properties of the clumps is explored in 
Fig.~\ref{fig:matrix_correlation}. Only sources with robust distance estimates ({\asciifamily DIST\_QUALITY}=1 or 2) and reliable flux 
densities ({\asciifamily FLUX\_QUALITY}=1 or 2) have been considered, amounting to a total of 3867 clumps. 
The locus of the \citet{Simon2006b} IRDCs properties is shown as red points,  including local thermodynamic equilibrium (LTE) mass estimates.

As expected, the clumps size and mass are strongly correlated with the distance estimate, 
while the FWHM of the PGCC sources is mainly constrained by the 5{\arcmin} {\Planck} beam, ranging from 5{\parcm}6 to 9{\parcm}7 for 80\% of the sources.
Artefacts due to the distance association with molecular clouds are visible as straight lines in all the scatter plots including the distance. 
Comparing the PGCC distributions (black) to the IRDC population (red), it appears that the PGCC 
Galactic cold clumps correspond statistically to smaller objects located in the solar neighbourhood and detected at all latitudes, while the 
IRDCs are preferentially detected towards the Galactic plane and represent  distant structures. 
Furthermore, IRDCs appear more massive and dense than the PGCC sources.

It is also interesting to notice that the population of PGCC at very low column density ($<10^{20}\,\rm{cm}^2$) corresponds to sources with an intermediate linear size, from 0.5 to 2\,pc.
This is fully consistent with the dilution of a dense core of 0.1\,pc with a column density of $10^{22}\,\rm{cm}^2$ at a distance of 500\,pc in the {\Planck} beam (5{\arcmin}).
 Similarly, we may explain the very low density estimates (a few $\rm{cm}^{-3}$), which are not expected even for large clouds.

\section{Ancillary Validation}
\label{sec:ancillary_validation}

\subsection{Identification with {\Planck} internal catalogues}

We have performed a positional cross-correlation of the PGCC catalogue with the other 
{\Planck} internal catalogues, using a 5{\arcmin} radius. In particular, we have considered: 
the Early Cold Cores catalogue (ECC), the Planck Catalogue of Compact Sources (PCCS), 
the Planck Catalogue of Sunyaev-Zeldovich sources (PSZ), and the Planck Catalogue of  High-redshift source candidates (PH$z$). 
The results are summarized in Table~\ref{tab:xflag}.

\subsubsection{ECC}

The Early Cold Cores catalogue contains 915 sources that were detected with high S/N 
by the {\Planck} Early Mission using the same colour-detection method used for the PGCC catalogue. 
We find matches  in positions for 892 ECC sources. The missing 23 ECC sources have a strong signal in all three detection maps. 
However, we have deblended the detection maps using the local maxima of the detection signal 
\citep[see][Sect.~5.2]{Montier2010}, and adopted the positions of these local maxima for the coordinates of the detections at each frequency.  
The individual frequency catalogues have then been merged to create the final catalogue. 
This last step of band merging has been improved in the final version of the catalogue to ensure a higher degree of compactness of the detected clump 
across the frequencies. The missing 23 ECC sources do no longer satisfy the compactness criterion.  
This does not mean that these ECC sources are spurious, rather it suggests that they are slightly more extended compared 
to the rest of the sources in the catalogue, hence they are discarded.

\subsubsection{PCCS}

The Planck Catalogue of Compact Sources \citep[PCCS, ][]{planck2014-a35} contains 
frequency catalogues in each {\Planck} band.The PCCS detections are obtained from individual frequency maps to which a Mexican Hat Wavelet filter 
has been applied. This procedure allows removal of extended emission and noise. Because 
of this data treatment, the S/N of PCCS sources depends on the frequency and on the area of the sky.
The detections in each band are divided into two categories depending on the region of the sky: 
zone 0 corresponds to regions where the reliability of the sources has been quantified, while zones 1 to 3 correspond to filaments, Galactic regions or 
filaments in Galactic regions, respectively. The second category of PCCS sources is considered to have a lower reliability than the zone 0 sources.

The number of retrieved matches is given in Table~\ref{tab:xflag}. In the Table we provide  the 
total number of matches (last column), as well as by {\asciifamily FLUX\_QUALITY} flag (second, third and fourth 
columns). This is given for each sub-category of the PCCS catalogue: zone 0 and zones 1$-$3.
Since the PCCS catalogue contains warm sources that are rejected from the PGCC catalogue, it is not surprising that
we do not find PGCC matches for all the PCCS detections. However, an important fraction of the PGCC sources 
have a counterpart in the PCCS catalogue, especially in the highest frequency channels, up to 75\% in the 857\,GHz band when including the
zones 1$-$3 sources. Notice also that the proportion of matched sources in the low frequency channels becomes extremely small (131 in total) and mainly 
consists of zone-0  sources, which could represent a population of extragalactic radio contaminants that has not been fully  
rejected. More interestingly, the PGCC detection algorithm requires a detection in all 
three frequency {\it cold residual} maps, while the PCCS only requires single frequency detections. 
There are 21\,182 PCCS sources that are simultaneously detected at 857, 545, and 353\,GHz. 
Of these, 7038 have a match in the PGCC catalogue in a 5{\arcmin} radius. 
Thus the  PGCC catalogue contains $\sim 45$\% of new sources not already identified in the PCCS catalogue
in the three upper {\Planck} frequencies as cold sources.

\subsubsection{PSZ}

The  Planck Catalogue of  Sunyaev-Zeldovich sources \citep[PSZ, ][]{planck2014-a36} 
contains 1653 detections.  These are exclusively extragalactic sources, so the overlap between 
the PSZ and PGCC catalogues is  small, i.e., 65 sources. Most of these sources exhibit cold temperature, except one which is warm.
Since the {\tt CoCoCoDeT} algorithm is suited to detect sources whose SED peaks 
around 857\,GHz, it is not expected to be sensitive to the SZ effect from galaxy clusters, which is characterized by a peak at 353\,GHz and almost no emission 
contribution at 857\,GHz.
Hence it is likely that this sample of 65 matched sources, lying at the limit of the Galactic mask used when building the PSZ catalogue,
consists of Galactic cold clumps at intermediate to high latitudes, 
therefore representing a contaminant for the PSZ catalogue.

\subsubsection{PH$z$}

The Planck Catalogue of high-$z$ source candidates \citep[PH$z$, ][]{planck2015_PHz} is a catalogue of 
sources detected at high latitude, in the 26\% cleanest fraction of the sky, and consists of a sample of high-redshift candidates
 identified by their 'red' colours in the {\Planck} highest  frequency bands. 
 By cross-correlating the PGCC and PH$z$ catalogues, we 
found 15 common sources. This result was partly expected, due to the similar spectral behaviour of Galactic cold objects and 
extragalactic red sources. Nevertheless, at the time of writing the nature of these matches is uncertain and requires further investigation.

\subsection{Crosscheck with follow-up observations}
\subsubsection{\textit{Herschel} imaging}

A sub-sample of the PGCC sources has been further investigated in the {\it Herschel}
Open Time Key Programme Galactic Cold Cores (hereafter, HKP-GCC). {\it Herschel} PACS and SPIRE
instruments were used to observe 115 fields at five wavelengths, from  
100\,$\mu$m to 500\,$\mu$m, with an angular 
resolution from 12$\arcsec$ to $37\arcsec$. The fields were
selected based on an early version of the PGCC catalogue, and target 
349 individual {\Planck} clumps, spanning a wide range in S/N, latitude and temperature.
 
The sensitivity and resolution of the {\it Herschel} instruments \citep{Pilbratt2010, Poglitsch2010,
Griffin2010} enable detailed studies of the structure of the clumps and their interplay with their parent clouds. First results of the {\it Herschel} 
follow-up have been presented in \citet{Juvela2010, Juvela2011, Juvela2012a} and in \citet{planck2011-7.7a}.

For the purpose of target selection, the {\Planck} cold clumps have been  
binned according to their Galactic coordinates, their estimated dust
colour temperature, and their mass. The bins are identified by the following boundary values: 
$l$= 0, 60, 120, and 180 degrees; $|b|=$1, 5, 10, and 90
degrees; $T_{\rm dust}$=6, 9, 11, and 14\,K; $M=$0, 0.01, 2.0, 500, 10$^6$\,$\rm{M}_{\sun}$. 
Here $T_{\rm dust}$ is the clump cold dust temperature. After creating the bins, 
the targets have been selected using a Monte Carlo technique, in which we have uniformly sampled 
the sources in the various bins. This procedure allows us to cover entirely the parameter space, including 
sources at high latitudes and with extreme mass values. 
The $M$=0 bin has been reserved for sources with no distance
estimate. 

Noteworthy, the selection of the PGCC sources 
had to avoid areas of the sky covered by other {\it Herschel} Key Programmes, such as 
the Galactic plane for $|b|<1^{\degr}$, which was targeted by 
the Hi-GAL programme \citep{Molinari2010}, and several nearby clouds
that are included in the Gould Belt \citep{Andre2010} and HOBYS
\citep[e.g., ][]{Motte2010} programmes.

The final selection of 115 fields includes 16 fields at Galactic latitudes
above 20 degrees. The median peak column density in 
these fields is $N_{\rm{H}_2}=1.5\times 10^{21}$\,cm$^{-2}$ \citep[for details on the column density 
calculation see][]{Juvela2012a}. In each followed-up field, the {\Planck} clump coordinates clearly identify a coherent structure in the 
{\it Herschel} surface brightness maps. This is the case even when the 
surface brightness is below $5\,\rm{MJy}\,\rm{sr}^{-1}$ at $250\,\mu$m, suggesting that the {\tt CoCoCoDeT} 
algorithm is able to reliably extract very low column density features from the {\Planck} data.  

\begin{figure}
\center
\psfrag{-----ytitle-----}{$N_{\rm{H}_2} \, [\mathrm{cm}^{-2}]$}
\psfrag{----------bar----------}{$alog_{10} \, dN / dT / dN_{\rm{H}_2}$}
 \includegraphics[width=9cm,viewport=0 0 650 550]{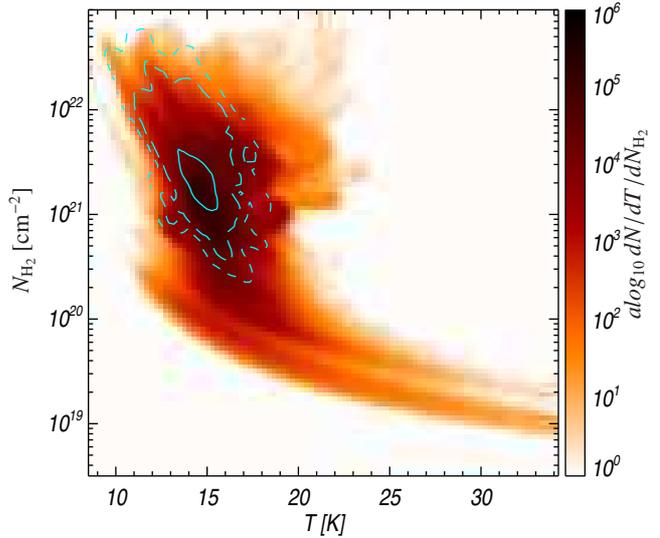} 
   \caption{
   Temperature vs. column density two-dimensional distribution for all PGCC cold clumps 
followed-up by {\it Herschel} in the HKP-GCC programme. The plotted data points are identified as IN and OUT: IN points are 
located inside the {\Planck} clump solid angle, while the OUT points are located outside.  
The distribution of the OUT points is shown in colour, while the distribution of the 
IN points is denoted by cyan contours at three different levels : 10, 50 and 90\% of the maximum.}
     \label{fig:hist2d_herschel_nht}
\end{figure}

The {\it Herschel} data have been used to generate column density and colour
temperature maps \citep{Montillaud2015}.  Figure~\ref{fig:hist2d_herschel_nht} shows 
the pixel-to-pixel two-dimensional distribution of column density vs temperature. 
Pixels located within 1$\sigma$ of the elliptical Gaussian fit of the PGCC sources are 
defined as IN pixels, and are shown with cyan contours in the Figure. They exhibit a much 
narrower distribution with respect to pixels located outside the PGCC sources, defined as OUT pixels. 
The IN pixels are mostly found in correspondance of the coldest and densest regions of the 
two-dimensional diagram, while the OUT pixels are also in regions of low column density and relatively 
high temperature. This result highlights the dense and cold nature of the PGCC sources.

Based on the {\it Herschel} follow-up, \citet{Montillaud2015} have also built 
a catalogue of compact sources. After removing extragalactic contaminants, 
pre-stellar candidates have been identified with a multi-wavelength analysis. 
\citet{Montillaud2015} have performed a detailed analysis aimed at classifying the 
{\it Herschel} sources according to their evolutionary stage, i.e.,  from starless cores to Young Stellar Objects (YSO). 
For each {\it Herschel} field, the fraction of YSOs and starless sources falling inside the 3-$\sigma$ elliptical Gaussian contour of the PGCC clumps are derived, providing
a unique information of the evolutionary stage of the PGCC clumps observed as a whole with {\Planck}.
A preliminary analysis of the YSO and starless populations in the {\it Herschel} fields around the 349 PGCC sources has already shown that 
{\Planck} clumps contain a mixture of the two populations in various proportions:  we find a number of {\Planck} clumps 
with only one source that is a starless core ($f_{\rm{starless}}$=1), or a YSO ($f_{\rm{YSO}}$=1), but more often sources composed by a mixture of YSOs and starless cores with 
fractions ranging from 10\% to 50\%. Again this illustrates the variety of sources covered by the PGCC catalogue, probing the ISM in extremely various evolutionary stages.

\subsubsection{Gas tracers}
\label{sec:molecular_tracers}

In order to characterize the gas content of the {\Planck} cold
clumps, several ground-based follow-ups of dense medium tracers have been performed.  

\citet{Wu2012} carried out a large survey of the $^{12}$CO,  $^{13}$CO
and C$^{18}$O $J$=1$\rightarrow$0 transition, targeting 674 ECC clumps 
with the 13.7 m telescope of the Purple Mountain Observatory. All the clumps 
(except for one) have $^{12}$CO and $^{13}$CO detections, and 
$68\%$ of them have C$^{18}$O emission. The three line peak velocities are found to coincide, suggesting that the {\Planck} clumps are quite cold and quiescent. 
The comparison between the derived excitation temperature and the 
dust temperature shows that dust and gas are well coupled in $95\%$ of
the clumps. Ten of the ECC clumps were also mapped, evidencing sub-structures, such as cores, and 
filamentary/elongated morphologies.

\section{Conclusion}
\label{sec:conclusion}

The highest frequency bands of the {\Planck}-HFI instrument provide an extremely powerful tracer of Galactic cold dust. 
By combining data from the {\Planck} three upper bands with IRAS 3\,THz data, we have conducted a multi-frequency compact source 
detection, and generated the Planck Catalogue of Galactic Cold Clumps (PGCC). A first version of this catalogue 
was released in 2011, i.e.,  the Early Cold Core catalogue (ECC), together with the Early Release Compact Source Catalogue (ERCSC). 
At that time, 915 sources, selected for their low temperature and high S/N, were made publicly available to the astronomical community. 
With the present work, we are releasing the whole PGCC catalogue, containing 13188 Galactic sources and 54 
 cold sources located in the LMC and SMC.

We have applied the {\tt CoCoCoDeT} algorithm \citep{Montier2010} to the {\Planck} 857, 545, and 353\,GHz maps and to the 
IRAS  3\,THz data. The combined use of these maps allow the separation of the cold and warm emission components, hence the identification of sources colder than their local 
environment. Through a dedicated analysis, we have removed all possible extragalactic contaminants. In particular, we have cross-correlated 
the PGCC catalogue with publicly available extragalactic catalogues, leading to the rejection of less than one hundred sources, mostly at high 
Galactic latitude. In parallel, we have also discarded detections, which have turned out to be contaminated by the presence of 
nearby hot sources. Noteworthy, the final catalogue contains 54 sources located in the LMC and SMC. These sources have been 
kept in the catalogue because the proximity of these dwarf galaxies makes it possible to detect individual clouds with {\Planck}. 

The PGCC sources have been divided into three categories, depending on the quality of their flux density estimates.   
6993 sources have accurate photometry in both the {\Planck} and IRAS bands. These sources correspond to
FQ=1 and are the most reliable in the catalogue. An amount of 3755 sources have accurate flux densities in the 
{\Planck} bands but not in the IRAS 3\,THz band. These sources correspond to FQ=2, and are likely so cold that 
their emission at 3\,THz falls below the IRAS detection limit. The last category comprises 2440 sources, and for these  
no accurate flux density has been measured in at least two bands. These sources correspond to FQ=3 and might be intrinsically very faint
or still deeply embedded. Despite the poor photometry, they are considered real detections. 

We have combined seven independent methods to assign a distance estimate to 5574 sources. 
In the PGCC catalogue, for each source we quote all the available distances derived from the different methods, 
however, only the most reliable estimate is used to compute other source physical quantities such as mass and luminosity. 
Distance estimates from different methods have been compared and validated. Accordingly, we have assigned, to each clump, 
a {\asciifamily DIST\_QUALITY} flag: 464 sources have consistent distance estimates ({\asciifamily DIST\_QUALITY}=1); 
4191 sources have only one estimate ({\asciifamily DIST\_QUALITY}=2); 255 have incompatible estimates ({\asciifamily DIST\_QUALITY}=3); 
664 sources have only distance upper limits ({\asciifamily DIST\_QUALITY}=4). More detailed on this analysis can be 
found in the on-line version of the PGCC catalogue. The 4655 sources with an accurate distance estimate ({\asciifamily DIST\_QUALITY}=1 or 2) 
are mainly located in the solar neighbourhood, with about 85\% of sources at less than 1\,kpc, and 91.4\% within 3\,kpc from the Sun.  

We have obtained the temperature for 10748 sources, using a free
 (for sources with FQ=1) or fixed ($\beta$=2, for sources with FQ=2) 
emissivity spectral index. The catalogue temperature distribution confirms that the PGCC clumps are not only colder than their local environment (as by construction of the catalogue) but,  
more importantly, that they are intrinsically cold sources, with a median between 13 and 14.5\,K, depending on the quality of the flux density measurements. 
The minimum temperature of the sources in the catalogue is 5.8\,K, reached for sources with FQ=2. 
It is important to emphasize that this value is not a threshold artificially 
induced by our detection method, which in fact has been shown (through a MCQA analysis) to provide a 60\%  completeness level at temperatures as low as  6\,K. 
Therefore, we can confidently state that, at least at the {\Planck} angular resolution, no Galactic source is colder than 5.8\,K. 

From the flux densities, temperature and distance estimates, we have derived other physical properties of the PGCC clumps, namely:  
column density, physical size, mass, mean density and luminosity. The column density of the {\Planck} clumps covers almost five orders of magnitude, reaching a value as low as 
6.8$\times$$10^{18}$\,$\mathrm{cm}^{-2}$, which can be compared to  the sensitivity limit (3$\times$$10^{21}$\,$\mathrm{cm}^{-2}$) of the {\it MSX} absorption measurements 
used to detect IRDCs \citep{Peretto2010}. Hence objects detected in emission by {\Planck} with the {\tt CoCoCoDeT} algorithm may not be detected in absorption by MSX,
meaning that the {\Planck} PGCC sources might represent a larger class of objects than the IRDCs, and 
might include less dense and/or more deeply embedded objects.  
Furthermore, the PGCC sources are characterized by a wide range of sizes and mean densities, which is indicative of a variety of astrophysical 
sources and evolutionary stages. The physical size of the catalogue sources ranges from 0.14\,pc to 30.6\,pc, i.e., from the typical size of a cold core to the one of a giant molecular cloud. 
Similarly the mean density spans four orders of magnitude, from  5.3 to 3.5$\times$$10^{4}$\,$\mathrm{cm}^{-3}$, encompassing the three
categories introduced by \citet{Williams2000}, that are cores, clumps and clouds.

We emphasize that, although we have adopted the term {\it clump} to refer to the generic source in the PGCC catalogue, we are aware that, 
depending on the distance, some of these sources are in fact cores, either pre- or proto-stellar, 
while others are giant molecular clouds. The preliminary {\it Herschel} and gas tracers follow-ups 
have confirmed that the PGCC sources are indeed cold and dense environments, 
but have also shown that they often contain colder sub-structures (e.g.,  cores) and even warm components (e.g., YSOs). 
In the future, other follow-ups of this kind, as well as cross-correlations with already existing ancillary data sets (e.g.,  
{\it Herschel}, {\it WISE} or {\it AKARI}), will become necessary to shed light on the exact nature of the {\Planck} clumps. 

We believe that the PGCC catalogue, covering the whole sky, hence probing wildly different environments, represents a real goldmine for investigations of 
the early phases of star formation. These include, but are not limited to: i) studies of the evolution from molecular clouds to cores and the influence 
of the local conditions; ii) analysis of the {\it extreme} cold sources, such as the most massive clumps or those located at relatively high latitude; 
iii) characterization of the dust emission law in dense regions and the role of the environment. All these topics will be discussed in 
forthcoming publications. 
 
 \begin{acknowledgements}
 The Planck Collaboration acknowledges the support of: ESA; CNES and CNRS/INSU-IN2P3-INP (France); ASI, CNR, and INAF (Italy); NASA and DoE (USA); STFC and UKSA (UK); CSIC, MINECO, JA, and RES (Spain); Tekes, AoF, and CSC (Finland); DLR and MPG (Germany); CSA (Canada); DTU Space (Denmark); SER/SSO (Switzerland); RCN (Norway); SFI (Ireland); FCT/MCTES (Portugal); ERC and PRACE (EU). A description of the Planck Collaboration and a list of its members, indicating which technical or scientific activities they have been involved in, can be found at \url{http://www.cosmos.esa.int/web/planck/planck-collaboration}.
\end{acknowledgements}

\appendix

\section{LMC - SMC}
\label{sec:lmc_smc}

\begin{figure}[b]
\center
\includegraphics[width=8cm, viewport=0 40 520 550]{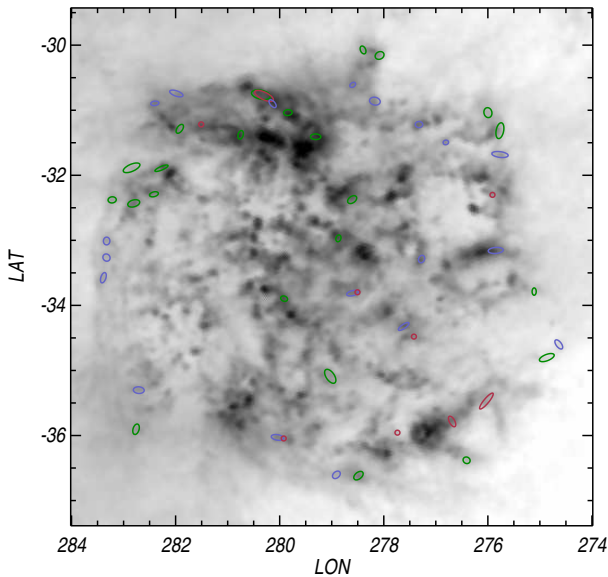} 
\caption{Distribution of the PGCC sources located in the LMC (in Galactic coordinates) for each {\asciifamily FLUX\_QUALITY} category: "{\it Reliable flux densities}" (1, blue), 
"{\it Missing 3\,THz flux density}" (2, green) and "{\it Detection only}" (3, pink).  The grey scale image is the {\Planck} intensity map at 857\,GHz shown in log scale between $10^{-2}$ and 0.5\,MJy$\rm{sr}^{-1}$.} 
\label{fig:lmc}
\end{figure}

\setcounter{figure}{0} \renewcommand{\thefigure}{B.\arabic{figure}} 

\begin{figure*}
\center
\includegraphics[width=8cm, angle=90, viewport=35 20 420 730]{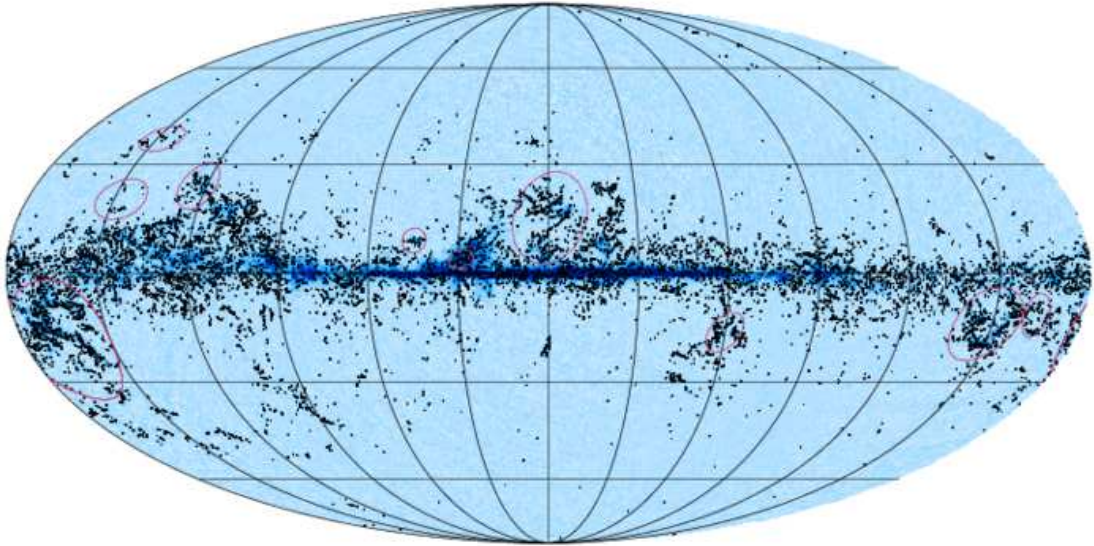} 
\caption{All-sky distribution of the PGCC sources  with FQ=1 or 2, 
displayed over the {\Planck} $^{12}$CO $J$=1$\rightarrow$0 map ranging from -5 to 30\,K.km$\rm{s}^{-1}$. 
The locations of the molecular complexes used in the distance estimate procedure are shown as red circles.}
\label{fig:allsky_co}
\end{figure*}

As noted in Sect.~\ref{sec:extragal}, the PGCC catalogue includes 51
sources located within a $4\pdeg09$ radius of the Large
Magellanic Cloud (LMC), and three sources within a $2\pdeg38$ field
centred on the Small Magellanic Cloud (SMC). Of the 51 sources in the LMC, 
42 sources have FQ=1 or 2, while 9 are only 
considered as poor detections. In the SMC, two sources have a good {\asciifamily FLUX\_QUALITY} flag 
and one is only detected.

At the distance of the
Magellanic Clouds \citep[$D_{\rm{LMC}} = 50.1$\,kpc, $D_{\rm{SMC}} = 61.7$\,kpc,
][]{Walker2012,Hilditch2005}, the working resolution of the {\Planck} and IRAS maps
(5{\arcmin}) corresponds to spatial scales of $\sim80$\,pc, which is
comparable to the characteristic size of giant molecular clouds (GMCs)
in the Milky Way \citep[$\sim50$\,pc, e.g.][]{Blitz1993}. Magellanic
PGCC sources are therefore quite different objects to Galactic PGCC
sources, but they are still of considerable interest for studying the
early phases of star formation. Firstly, the {\Planck} data provide a
census of cold material in the vicinity of the Magellanic Clouds that
is independent of previous observations of HI and CO emission,
the standard tracers of neutral interstellar gas in external
galaxies. Secondly, PGCC sources in the Magellanic Clouds constitute a
useful comparison sample to local GMCs with low levels of star
formation, since the galactic environment hosting the cold molecular
material is quite distinct. The clouds span a small but appreciable range of
sub-solar metallicities \citep[0.2 to 0.5, ][]{Russell1992}, and have
dust-to-gas mass ratios that are $\sim3$ to 10 times lower than the
value in the solar neighborhood \citep[][]{Gordon2014}. They are
also low-mass systems, with shallower gravitational potentials and
lower levels of shear than the normal disc galaxies.

Of the 51 PGCC sources detected towards the LMC, 
34 are located within the field surveyed by NANTEN at 2{\parcm}6 resolution
for CO emission in the LMC \citep{Fukui2008}, while 27 sources
are located inside the region observed by the higher resolution
(45{\arcsec}) MAGMA LMC survey \citep{Wong2011}. All of these 27
sources exhibit CO emission that is well-detected by MAGMA, with peak
integrated $^{12}\rm{CO}$ J=1$\rightarrow$0 intensities brighter than MAGMA's
4$\sigma$ sensitivity limit ($\sim 1.2$\,K.km.$\rm{s}^{-1}$). Dedicated follow-up
observations of the remaining Magellanic PGCC sources with the Mopra
Telescope have detected CO emission associated with a further 15
PGCC sources in the LMC, but none of the three SMC sources \citep{Hughes2015}. 
Figure~\ref{fig:lmc} shows the spatial
distribution of all PGCC sources in the Magellanic Clouds. The
spatial distribution of PGCC sources in the LMC is clearly not random:
only one PGCC source is detected at high stellar surface densities
($\Sigma_{*} > 100\,\rm{Mpc}^2$) even though many CO clouds are detected
there, while there are four PGCC sources that appear to be aligned in
an east-west direction along the southern periphery of the LMC. These
sources were previously noted in the dust property maps obtained by
combining the IRAS and the {\Planck} data as regions of low temperature
and high dust column density.

In the LMC, the {\tt CoCoCoDeT} algorithm therefore seems to be an efficient
tool for identifying cold molecular material. After re-scaling for the
lower dust-to-gas ratio in the LMC and accounting for mismatches
between the intrinsic source size and the {\Planck} and Mopra beam
widths, the catalogued masses of the LMC PGCC sources are in good
agreement with the masses derived from the Mopra CO data. A more detailed
investigation of the spatial distribution and physical nature of the
LMC PGCC sources will be presented in \citet{Hughes2015}.

\section{Correlation with CO map}
\label{sec:co}

In Fig.~\ref{fig:allsky_co} we overlay the all-sky distribution of the PGCC Galactic clumps to the {\Planck} $^{12}$CO $J$=1$\rightarrow$0 
all-sky map \citep[see ][]{planck2013-p03a}.
As expected, for a large majority of the PGCC sources, their location coincides with a $^{12}$CO $J$=1$\rightarrow$0 transition.
This is especially true in the Galactic disc. At high latitude, objects appear less associated 
with a CO signature, although this is likely due to 
the limited sensitivity of the {\Planck} CO map.
A further analysis is required to investigate whether the PGCC sources are associated to CO clumps, or only with diffuse CO emission.

\section{Catalogue content}
\label{sec:cat_content}

\begin{table*}[!hp]
\caption{PGCC catalogue columns.}
\label{tab:pgcc_listing_1}
\nointerlineskip
\setbox\tablebox=\vbox{
\newdimen\digitwidth 
\setbox0=\hbox{\rm 0} 
\digitwidth=\wd0 
\catcode`*=\active 
\def*{\kern\digitwidth} 
\newdimen\signwidth 
\setbox0=\hbox{+} 
\signwidth=\wd0 
\catcode`!=\active 
\def!{\kern\signwidth} 
\newdimen\pointwidth 
\setbox0=\hbox{.} 
\pointwidth=\wd0 
\catcode`?=\active 
\def?{\kern\pointwidth} 
\halign{\hbox to 2.0 in{#\leaderfil}\tabskip=1.0em&
\hfil#\hfil&
#\hfil\tabskip=0pt\cr
\noalign{\doubleline}
\omit \hfil Column Name \hfil & Unit & \hfil Description \cr
\noalign{\vskip 3pt\hrule\vskip 4pt}
\multicolumn{3}{c}{Identification} \cr
\noalign{\vskip 4pt}
 {\asciifamily NAME}  &   & Source Name \cr
 {\asciifamily SNR}  &  &  Maximum S/N over the 857, 545, and 353\,GHz {\Planck} {\it cold residual} maps\cr    
 {\asciifamily SNR\_857}  &  &  S/N of the {\it cold residual} detection at 857\,GHz \cr   
 {\asciifamily SNR\_545}  &   & S/N of the {\it cold residual} detection at 545\,GHz \cr
 {\asciifamily SNR\_353}  &   & S/N of the {\it cold residual} detection at 353\,GHz \cr
\noalign{\vskip 3pt\hrule\vskip 4pt}
\multicolumn{3}{c}{Source position} \cr
\noalign{\vskip 4pt}
 {\asciifamily GLON}  &  [deg] &  Galactic longitude based on morphology fitting \cr
 {\asciifamily GLAT}  &   [deg] & Galactic latitude (deg) based on morphology fitting \cr     
 {\asciifamily RA}  &   [deg] & Right ascension (J2000) in degrees transformed from (GLON, GLAT) \cr     
 {\asciifamily DEC}  &   [deg] & Declination (J2000) in degrees transformed from (GLON, GLAT) \cr     
\noalign{\vskip 3pt\hrule\vskip 4pt}
\multicolumn{3}{c}{Morphology} \cr
\noalign{\vskip 4pt}
{\asciifamily GAU\_MAJOR\_AXIS}  &   [arcmin] & FWHM along the major axis of the elliptical Gaussian\cr     
 {\asciifamily GAU\_MAJOR\_AXIS\_SIG}  & [arcmin] &   1$\sigma$ uncertainty on the FWHM along the major axis \cr     
 {\asciifamily GAU\_MINOR\_AXIS} & [arcmin] & FWHM along the minor axis of the elliptical Gaussian\cr     
 {\asciifamily GAU\_MINOR\_AXIS\_SIG} & [arcmin] & 1$\sigma$ uncertainty on the FWHM along the minor axis  \cr     
 {\asciifamily GAU\_POSITION\_ANGLE} &[rd] & Position angle of the elliptical gaussian, defined as the clockwise angle \cr
 \omit \hfil & & between the Galactic plane orientation and the orientation of the major axis  \cr     
 {\asciifamily GAU\_POSITION\_ANGLE\_SIG} &[rd] & 1$\sigma$ uncertainty on the position angle \cr     
\noalign{\vskip 3pt\hrule\vskip 4pt}
\multicolumn{3}{c}{Photometry} \cr
\noalign{\vskip 4pt}
 {\asciifamily FLUX\_3000\_CLUMP}  &   [Jy] & Flux density of the clump at 3\,THz \cr     
 {\asciifamily FLUX\_3000\_CLUMP\_SIG}  &  [Jy] & 1$\sigma$ uncertainty on the flux density of the clump at 3\,THz \cr     
 {\asciifamily FLUX\_857\_CLUMP}  & [Jy] &    Flux density of the clump at 857\,GHz \cr     
 {\asciifamily FLUX\_857\_CLUMP\_SIG}  &  [Jy] &   1$\sigma$ uncertainty on the flux density of the clump at 857\,GHz \cr     
 {\asciifamily FLUX\_545\_CLUMP}  &  [Jy] &   Flux density of the clump at 545\,GHz \cr     
 {\asciifamily FLUX\_545\_CLUMP\_SIG}  &    [Jy] & 1$\sigma$ uncertainty on the flux density of the clump at 545\,GHz \cr     
 {\asciifamily FLUX\_353\_CLUMP}  & [Jy] &    Flux density of the clump at 353\,GHz \cr     
 {\asciifamily FLUX\_353\_CLUMP\_SIG}  & [Jy] &   1$\sigma$ uncertainty on the flux density of the clump at 353\,GHz \cr     
 {\asciifamily FLUX\_3000\_WBKG}  &  [Jy] &   Flux density of the warm background at 3\,THz \cr     
 {\asciifamily FLUX\_3000\_WBKG\_SIG}  & [Jy] &    1$\sigma$ uncertainty on the flux density of the warm background at 3\,THz \cr     
 {\asciifamily FLUX\_857\_WBKG}  & [Jy] &    Flux density of the warm background at 857\,GHz \cr     
 {\asciifamily FLUX\_857\_WBKG\_SIG}  &  [Jy] &    1$\sigma$ uncertainty on the flux density of the warm background at 857\,GHz \cr     
 {\asciifamily FLUX\_545\_WBKG}  &  [Jy] &   Flux density of the warm background at 545\,GHz \cr     
 {\asciifamily FLUX\_545\_WBKG\_SIG}  &  [Jy] &    1$\sigma$ uncertainty on the flux density of the warm background at 545\,GHz \cr     
 {\asciifamily FLUX\_353\_WBKG}  & [Jy] &    Flux density of the warm background at 353\,GHz \cr     
 {\asciifamily FLUX\_353\_WBKG\_SIG}  & [Jy] &     1$\sigma$ uncertainty on the flux density of the warm background at 353\,GHz \cr     
 {\asciifamily FLUX\_QUALITY}  &  [1-3] &  Category of flux density reliability \cr
 {\asciifamily FLUX\_BLENDING}  &  [0,1] &  1 if blending issue with flux density estimate \cr
 {\asciifamily FLUX\_BLENDING\_IDX}  &   &  Catalogue index of the closest source  responsible for blending \cr
 {\asciifamily FLUX\_BLENDING\_ANG\_DIST}  &  [arcmin] &  Angular distance to the closest source responsible for blending \cr
 {\asciifamily FLUX\_BLENDING\_BIAS\_3000}  &  [\%] &  Relative bias  of the flux density at 3\,THz due to blending \cr
 {\asciifamily FLUX\_BLENDING\_BIAS\_857}  &  [\%] &  Relative bias  of the flux density at 857\,GHz due to blending  \cr
 {\asciifamily FLUX\_BLENDING\_BIAS\_545}  &  [\%] &  Relative bias  of the flux density at 545\,GHz due to blending  \cr
 {\asciifamily FLUX\_BLENDING\_BIAS\_353}  &  [\%] &  Relative bias of the flux density at 353\,GHz due to blending \cr
 \noalign{\vskip 3pt\hrule\vskip 4pt}
\multicolumn{3}{c}{Distance} \cr
\noalign{\vskip 4pt}
 {\asciifamily DIST\_KINEMATIC}  &  [kpc] &   Distance estimate [1] using kinematics \cr     
 {\asciifamily DIST\_KINEMATIC\_SIG}  &  [kpc] &  1$\sigma$ distance estimate using kinematics \cr     
 {\asciifamily DIST\_OPT\_EXT\_DR7}  &   [kpc] &  Distance estimate [2] using optical extinction on SDSS DR7\cr     
 {\asciifamily DIST\_OPT\_EXT\_DR7\_SIG}  & [kpc] &   1$\sigma$ distance estimate using optical extinction on SDSS DR7 \cr     
 {\asciifamily DIST\_OPT\_EXT\_DR9}  &   [kpc] &  Distance estimate [3] using optical extinction on SDDS DR9 \cr     
 {\asciifamily DIST\_OPT\_EXT\_DR9\_SIG}  & [kpc] &   1$\sigma$ distance estimate using optical extinction on SDSS DR9 \cr    
 {\asciifamily DIST\_NIR\_EXT\_IRDC}  & [kpc] &   Distance estimate [4] using near-infrared extinction towards IRDCs\cr     
 {\asciifamily DIST\_NIR\_EXT\_IRDC\_SIG}  & [kpc] &     1$\sigma$ distance estimate using near-infrared extinction towards IRDCs\cr      
 {\asciifamily DIST\_NIR\_EXT}  & [kpc] &   Distance estimate [5] using near-infrared extinction \cr     
 {\asciifamily DIST\_NIR\_EXT\_SIG}  & [kpc] &     1$\sigma$ distance estimate using near-infrared extinction \cr      
  {\asciifamily DIST\_MOLECULAR\_COMPLEX}  &  [kpc] &  Distance estimate  [6] using molecular complex association \cr     
 {\asciifamily DIST\_MOLECULAR\_COMPLEX\_SIG}  &  [kpc] &    1$\sigma$ distance estimate using  molecular complex association \cr     
  {\asciifamily DIST\_HKP\_GCC}  &  [kpc] &  Distance estimate [7] from the  {\it Herschel} Key-Programme Galactic Cold Cores \cr     
 {\asciifamily DIST\_HKP\_GCC\_SIG}  &  [kpc] &    1$\sigma$  distance estimate from the  {\it Herschel} HKP-GCC \cr     
 {\asciifamily DIST\_OPTION}  &  [0-7] & Option of the best distance estimate used in other physical properties\cr     
 {\asciifamily DIST\_QUALITY}  &  [0-4] & Quality Flag of the consistency between distance estimates\cr     
 {\asciifamily DIST}  &   [kpc] &  Best distance estimate used  for further physical properties \cr     
 {\asciifamily DIST\_SIG}  & [kpc] &   1$\sigma$ uncertainty on the best distance estimate \cr     
 \noalign{\vskip 3pt\hrule\vskip 4pt}
}}
\endPlancktable
\end{table*}

\begin{table*}
\caption{PGCC catalogue columns (continued).}
\label{tab:pgcc_listing_2}
\nointerlineskip
\setbox\tablebox=\vbox{
\newdimen\digitwidth 
\setbox0=\hbox{\rm 0} 
\digitwidth=\wd0 
\catcode`*=\active 
\def*{\kern\digitwidth} 
\newdimen\signwidth 
\setbox0=\hbox{+} 
\signwidth=\wd0 
\catcode`!=\active 
\def!{\kern\signwidth} 
\newdimen\pointwidth 
\setbox0=\hbox{.} 
\pointwidth=\wd0 
\catcode`?=\active 
\def?{\kern\pointwidth} 
\halign{\hbox to 2.0 in{#\leaderfil}\tabskip=1.0em&
\hfil#\hfil&
#\hfil\tabskip=0pt\cr
\noalign{\doubleline}
\omit \hfil Column Name \hfil & Unit & \hfil Description \cr
\noalign{\vskip 3pt\hrule\vskip 4pt}
\multicolumn{3}{c}{Temperature} \cr
\noalign{\vskip 4pt}
 {\asciifamily TEMP\_CLUMP}  & [K] &   Temperature of the clump with $\beta$ as a free parameter\cr     
 {\asciifamily TEMP\_CLUMP\_SIG}  &  [K] &     1$\sigma$ uncertainty on the clump temperature with $\beta$ free  \cr     
 {\asciifamily TEMP\_CLUMP\_LOW1}  &  [K] &   Lower 68\% confidence limit  of the clump temperature with $\beta$ free \cr     
 {\asciifamily TEMP\_CLUMP\_UP1}  &  [K] &    Upper 68\% confidence limit  of the clump temperature with $\beta$ free  \cr     
 {\asciifamily BETA\_CLUMP}  & &   Spectral index $\beta$ of the clump \cr     
 {\asciifamily BETA\_CLUMP\_SIG}  & &   1$\sigma$ uncertainty (from MCMC) on the emissivity spectral index $\beta$ of the clump  \cr     
 {\asciifamily BETA\_CLUMP\_LOW1}  & &  Lower 68\% confidence limit  of the emissivity spectral index $\beta$ of the clump  \cr     
 {\asciifamily BETA\_CLUMP\_UP1}  & &   Upper 68\% confidence limit  of the emissivity spectral index $\beta$ of the clump  \cr     
 {\asciifamily TEMP\_BETA2\_CLUMP}  & [K] &     Temperature of the clump with $\beta=2$ \cr     
 {\asciifamily TEMP\_BETA2\_CLUMP\_SIG}  & [K] &     1$\sigma$ uncertainty on the temperature of the clump with $\beta=2$ \cr     
 {\asciifamily TEMP\_BETA2\_CLUMP\_LOW1}  &   [K] &   Lower 68\% confidence limit  of the clump temperature with $\beta=2$ \cr     
 {\asciifamily TEMP\_BETA2\_CLUMP\_UP1}  &   [K] &   Upper 68\% confidence limit  of the clump temperature with $\beta=2$ \cr     
 {\asciifamily TEMP\_WBKG}  &   [K] &  Temperature of the warm background with $\beta$ as a free parameter \cr     
 {\asciifamily TEMP\_WBKG\_SIG}  &   [K] &  1$\sigma$ uncertainty on the temperature of the warm background with $\beta$ as a free parameter \cr     
 {\asciifamily TEMP\_WBKG\_LOW1}  &  [K] &  Lower 68\% confidence limit   of the warm background  temperature with $\beta$ free \cr     
 {\asciifamily TEMP\_WBKG\_UP1}  &  [K] &   Upper 68\% confidence limit of the warm background  temperature with $\beta$ free \cr     
 {\asciifamily BETA\_WBKG}  &  &  Spectral index $\beta$ of the warm background \cr     
 {\asciifamily BETA\_WBKG\_SIG}  &  &  1$\sigma$ uncertainty on the spectral index $\beta$ of the warm background \cr     
 {\asciifamily BETA\_WBKG\_LOW1}  & &   Lower 68\% confidence limit of  the emissivity spectral index $\beta$ of the warm background\cr     
 {\asciifamily BETA\_WBKG\_UP1}  & &    Upper 68\% confidence limit of  the emissivity spectral index $\beta$ of the warm background\cr     
 {\asciifamily TEMP\_BETA2\_WBKG}  & [K] &    Temperature of the warm background with $\beta=2$  \cr     
 {\asciifamily TEMP\_BETA2\_WBKG\_SIG}  & [K] &     1$\sigma$ uncertainty on the temperature of the warm background with $\beta=2$  \cr     
 {\asciifamily TEMP\_BETA2\_WBKG\_LOW1}  &  [K] &   Lower 68\% confidence limit of the warm background temperature with $\beta=2$ \cr     
 {\asciifamily TEMP\_BETA2\_WBKG\_UP1}  &  [K] &  Upper 68\% confidence limit of the warm background temperature with $\beta=2$ \cr     
 \noalign{\vskip 3pt\hrule\vskip 4pt}
\multicolumn{3}{c}{Physical properties} \cr
\noalign{\vskip 4pt}
 {\asciifamily NH2}  &  [$\mathrm{cm}^{-2}$]   &  Column density $N_{\rm{H}_2}$ of the clump \cr     
 {\asciifamily NH2\_SIG}  &   [$\mathrm{cm}^{-2}$]  &   1$\sigma$ uncertainty on the  column density of the clump\cr  
 {\asciifamily NH2\_LOW[1,2,3]}  &   [$\mathrm{cm}^{-2}$]  &   Lower 68\%, 95\% and 99\% confidence limit of the  column density \cr  
 {\asciifamily NH2\_UP[1,2,3]}  &   [$\mathrm{cm}^{-2}$]  &   Upper  68\%, 95\% and 99\% confidence limit of the  column density \cr  
 {\asciifamily MASS}  &  [$\mathrm{M}_{\sun}$] &  Mass estimate of the clump \cr     
 {\asciifamily MASS\_SIG}  &   [$\mathrm{M}_{\sun}$] &  1$\sigma$ uncertainty on the mass estimate of the clump \cr     
 {\asciifamily MASS\_LOW[1,2,3]}  &   [$\mathrm{M}_{\sun}$] &  Lower  68\%, 95\% and 99\% confidence limit of the mass estimate \cr     
 {\asciifamily MASS\_UP[1,2,3]}  &   [$\mathrm{M}_{\sun}$] &  Upper 68\%, 95\% and 99\% confidence limit of the mass estimate \cr     
 {\asciifamily DENSITY}  &  [$\mathrm{cm}^{-3}$] &  Mean density of the clump \cr     
 {\asciifamily DENSITY\_SIG}  &   [$\mathrm{cm}^{-3}$] &  1$\sigma$ uncertainty on the mean density estimate of the clump \cr     
 {\asciifamily DENSITY\_LOW[1,2,3]}  &   [$\mathrm{cm}^{-3}$] &  Lower  68\%, 95\% and 99\% confidence limit of the mean density estimate \cr     
 {\asciifamily DENSITY\_UP[1,2,3]}  &   [$\mathrm{cm}^{-3}$] &  Upper 68\%, 95\% and 99\% confidence limit of the mean density estimate \cr     
 {\asciifamily SIZE}  &  [pc] &  Physical size of the clump \cr     
 {\asciifamily SIZE\_SIG}  &   [pc] &  1$\sigma$ uncertainty on the physical size estimate of the clump \cr     
 {\asciifamily SIZE\_LOW[1,2,3]}  &   [pc] &  Lower 68\%, 95\% and 99\% confidence limit of the physical size estimate  \cr     
 {\asciifamily SIZE\_UP[1,2,3]}  &   [pc] &  Upper 68\%, 95\% and 99\% confidence limit of the physical size estimate \cr     
 {\asciifamily LUMINOSITY}  &    [$\mathrm{L}_{\sun}$] & Luminosity of the clump  \cr     
 {\asciifamily LUMINOSITY\_SIG}  & [$\mathrm{L}_{\sun}$] &   1$\sigma$ uncertainty on the luminosity of the clump \cr       
 {\asciifamily LUMINOSITY\_LOW[1,2,3]}  & [$\mathrm{L}_{\sun}$] &   Lower 68\%, 95\% and 99\% confidence limit of the luminosity  \cr       
 {\asciifamily LUMINOSITY\_UP[1,2,3]} & [$\mathrm{L}_{\sun}$] &   Upper 68\%, 95\% and 99\% confidence limit of the luminosity \cr       
 \noalign{\vskip 3pt\hrule\vskip 4pt}
\multicolumn{3}{c}{Flags} \cr
\noalign{\vskip 4pt}
 {\asciifamily NEARBY\_HOT\_SOURCE}  & [arcmin] &    Distance to the closest hot source \cr     
 {\asciifamily XFLAG\_LMC}  & [0,1]  & 1 if part of the LMC  \cr
 {\asciifamily XFLAG\_SMC}  & [0,1]   & 1 if part of the SMC   \cr
 {\asciifamily XFLAG\_ECC}  & [0,1]   & 1 if present in the ECC     \cr
 {\asciifamily XFLAG\_PCCS\_857}  & [0,1]   & 1 if present in the PCCS 857\,GHz band    \cr
 {\asciifamily XFLAG\_PCCS\_545}  & [0,1]   & 1 if present in the  PCCS 545\,GHz band       \cr
 {\asciifamily XFLAG\_PCCS\_353}  & [0,1]   & 1 if present in the  PCCS 353\,GHz band      \cr
 {\asciifamily XFLAG\_PCCS\_217}  & [0,1]   & 1 if present in the  PCCS 217\,GHz band      \cr
 {\asciifamily XFLAG\_PCCS\_143}  & [0,1]   & 1 if present in the  PCCS 143\,GHz band      \cr
 {\asciifamily XFLAG\_PCCS\_100}  & [0,1]   & 1 if present in the  PCCS 100\,GHz band      \cr
 {\asciifamily XFLAG\_PCCS\_70}  & [0,1]   & 1 if present in the  PCCS 70\,GHz band      \cr
 {\asciifamily XFLAG\_PCCS\_44}  & [0,1]   & 1 if present in the  PCCS 44\,GHz band      \cr
 {\asciifamily XFLAG\_PCCS\_30}  & [0,1]   & 1 if present in the  PCCS 30\,GHz band      \cr
 {\asciifamily XFLAG\_PSZ}  & [0,1]   & 1 if present in the  PCCS SZ    \cr
 {\asciifamily XFLAG\_PHZ}  & [0,1]   & 1 if present in the  PCCS HZ    \cr
 {\asciifamily XFLAG\_HKP\_GCC}  & [0,1]   & 1 if present in the  {\it Herschel} HKP-GCC  \cr
 \noalign{\vskip 3pt\hrule\vskip 4pt}
}}
\endPlancktable
\end{table*}

\clearpage

\bibliographystyle{aat}
\bibliography{Planck_2015_XXVIII_astroph.bbl}

\end{document}

%% file: Planck.tex
\def\setsymbol#1#2{\expandafter\def\csname #1\endcsname{#2}}
\def\getsymbol#1{\csname #1\endcsname}

\def\Planck{\textit{Planck}}

\def\alltwentyfifteenresultspapers{\nocite{planck2014-a01, planck2014-a03, planck2014-a04, planck2014-a05, planck2014-a06, planck2014-a07, planck2014-a08, planck2014-a09, planck2014-a11, planck2014-a12, planck2014-a13, planck2014-a14, planck2014-a15, planck2014-a16, planck2014-a17, planck2014-a18, planck2014-a19, planck2014-a20, planck2014-a22, planck2014-a24, planck2014-a26, planck2014-a28, planck2014-a29, planck2014-a30, planck2014-a31, planck2014-a35, planck2014-a36, planck2014-a37, planck2014-ES}}

\newbox\tablebox    \newdimen\tablewidth
\def\leaderfil{\leaders\hbox to 5pt{\hss.\hss}\hfil}
\def\endPlancktable{\tablewidth=\columnwidth 
    $$\hss\copy\tablebox\hss$$
    \vskip-\lastskip\vskip -2pt}

\def\tablenote#1 #2\par{\begingroup \parindent=0.8em
    \abovedisplayshortskip=0pt\belowdisplayshortskip=0pt
    \noindent
    $$\hss\vbox{\hsize\tablewidth \hangindent=\parindent \hangafter=1 \noindent
    \hbox to \parindent{$^#1$\hss}\strut#2\strut\par}\hss$$
    \endgroup}
\def\doubleline{\vskip 3pt\hrule \vskip 1.5pt \hrule \vskip 5pt}

\def\L2{\ifmmode L_2\else $L_2$\fi}

\def\DeltaT{\ifmmode \Delta T\else $\Delta T$\fi}
\def\deltat{\ifmmode \Delta t\else $\Delta t$\fi}
\def\fknee{\ifmmode f_{\rm knee}\else $f_{\rm knee}$\fi}
\def\Fmax{\ifmmode F_{\rm max}\else $F_{\rm max}$\fi}
\def\solar{\ifmmode{\rm M}_{\mathord\odot}\else${\rm M}_{\mathord\odot}$\fi}
\def\Msolar{\ifmmode{\rm M}_{\mathord\odot}\else${\rm M}_{\mathord\odot}$\fi}
\def\Lsolar{\ifmmode{\rm L}_{\mathord\odot}\else${\rm L}_{\mathord\odot}$\fi}
\def\inv{\ifmmode^{-1}\else$^{-1}$\fi}
\def\mo{\ifmmode^{-1}\else$^{-1}$\fi}
\def\sup#1{\ifmmode ^{\rm #1}\else $^{\rm #1}$\fi}
\def\expo#1{\ifmmode \times 10^{#1}\else $\times 10^{#1}$\fi}
\def\,{\thinspace}
\def\lsim{\mathrel{\raise .4ex\hbox{\rlap{$<$}\lower 1.2ex\hbox{$\sim$}}}}
\def\gsim{\mathrel{\raise .4ex\hbox{\rlap{$>$}\lower 1.2ex\hbox{$\sim$}}}}

\def\simprop{\mathrel{\raise .4ex\hbox{\rlap{$\propto$}\lower 1.2ex\hbox{$\sim$}}}}
\def\deg{\ifmmode^\circ\else$^\circ$\fi}
\def\pdeg{\ifmmode $\setbox0=\hbox{$^{\circ}$}\rlap{\hskip.11\wd0 .}$^{\circ}
          \else \setbox0=\hbox{$^{\circ}$}\rlap{\hskip.11\wd0 .}$^{\circ}$\fi}
\def\arcs{\ifmmode {^{\scriptstyle\prime\prime}}
          \else $^{\scriptstyle\prime\prime}$\fi}
\def\arcm{\ifmmode {^{\scriptstyle\prime}}
          \else $^{\scriptstyle\prime}$\fi}
\newdimen\sa  \newdimen\sb
\def\parcs{\sa=.07em \sb=.03em
     \ifmmode \hbox{\rlap{.}}^{\scriptstyle\prime\kern -\sb\prime}\hbox{\kern -\sa}
     \else \rlap{.}$^{\scriptstyle\prime\kern -\sb\prime}$\kern -\sa\fi}
\def\parcm{\sa=.08em \sb=.03em
     \ifmmode \hbox{\rlap{.}\kern\sa}^{\scriptstyle\prime}\hbox{\kern-\sb}
     \else \rlap{.}\kern\sa$^{\scriptstyle\prime}$\kern-\sb\fi}
\def\ra[#1 #2 #3.#4]{#1\sup{h}#2\sup{m}#3\sup{s}\llap.#4}
\def\dec[#1 #2 #3.#4]{#1\deg#2\arcm#3\arcs\llap.#4}
\def\deco[#1 #2 #3]{#1\deg#2\arcm#3\arcs}
\def\rra[#1 #2]{#1\sup{h}#2\sup{m}}

\def\dots{\relax\ifmmode \ldots\else $\ldots$\fi}
\def\WHzsr{\ifmmode $W\,Hz\mo\,sr\mo$\else W\,Hz\mo\,sr\mo\fi}
\def\mHz{\ifmmode $\,mHz$\else \,mHz\fi}
\def\GHz{\ifmmode $\,GHz$\else \,GHz\fi}
\def\mKs{\ifmmode $\,mK\,s$^{1/2}\else \,mK\,s$^{1/2}$\fi}
\def\muKs{\ifmmode \,\mu$K\,s$^{1/2}\else \,$\mu$K\,s$^{1/2}$\fi}
\def\muKRJs{\ifmmode \,\mu$K$_{\rm RJ}$\,s$^{1/2}\else \,$\mu$K$_{\rm RJ}$\,s$^{1/2}$\fi}
\def\muKHz{\ifmmode \,\mu$K\,Hz$^{-1/2}\else \,$\mu$K\,Hz$^{-1/2}$\fi}
\def\MJysr{\ifmmode \,$MJy\,sr\mo$\else \,MJy\,sr\mo\fi}
\def\MJysrmK{\ifmmode \,$MJy\,sr\mo$\,mK$_{\rm CMB}\mo\else \,MJy\,sr\mo\,mK$_{\rm CMB}\mo$\fi}
\def\microns{\ifmmode \,\mu$m$\else \,$\mu$m\fi}

\def\muK{\ifmmode \,\mu$K$\else \,$\mu$\hbox{K}\fi}
\def\microK{\ifmmode \,\mu$K$\else \,$\mu$\hbox{K}\fi}
\def\muW{\ifmmode \,\mu$W$\else \,$\mu$\hbox{W}\fi}
\def\kms{\ifmmode $\,km\,s$^{-1}\else \,km\,s$^{-1}$\fi}
\def\kmsMpc{\ifmmode $\,\kms\,Mpc\mo$\else \,\kms\,Mpc\mo\fi}
\providecommand{\sorthelp}[1]{}

%% file: A37_Catalogue_cold_clumps_authors_and_institutes.tex
\author{\small
Planck Collaboration: P.~A.~R.~Ade\inst{85}
\and
N.~Aghanim\inst{60}
\and
M.~Arnaud\inst{73}
\and
M.~Ashdown\inst{69, 5}
\and
J.~Aumont\inst{60}
\and
C.~Baccigalupi\inst{84}
\and
A.~J.~Banday\inst{94, 8}
\and
R.~B.~Barreiro\inst{65}
\and
N.~Bartolo\inst{29, 66}
\and
E.~Battaner\inst{96, 97}
% ...
}
\institute{\small
APC, AstroParticule et Cosmologie, Universit\'{e} Paris Diderot, CNRS/IN2P3, CEA/lrfu, Observatoire de Paris, Sorbonne Paris Cit\'{e}, 10, rue Alice Domon et L\'{e}onie Duquet, 75205 Paris Cedex 13, France\goodbreak
\and
African Institute for Mathematical Sciences, 6-8 Melrose Road, Muizenberg, Cape Town, South Africa\goodbreak
\and
Agenzia Spaziale Italiana Science Data Center, Via del Politecnico snc, 00133, Roma, Italy\goodbreak
\and
Aix Marseille Universit\'{e}, CNRS, LAM (Laboratoire d'Astrophysique de Marseille) UMR 7326, 13388, Marseille, France\goodbreak
\and
Astrophysics Group, Cavendish Laboratory, University of Cambridge, J J Thomson Avenue, Cambridge CB3 0HE, U.K.\goodbreak
\and
Astrophysics \& Cosmology Research Unit, School of Mathematics, Statistics \& Computer Science, University of KwaZulu-Natal, Westville Campus, Private Bag X54001, Durban 4000, South Africa\goodbreak
\and
CITA, University of Toronto, 60 St. George St., Toronto, ON M5S 3H8, Canada\goodbreak
\and
CNRS, IRAP, 9 Av. colonel Roche, BP 44346, F-31028 Toulouse cedex 4, France\goodbreak
\and
CRANN, Trinity College, Dublin, Ireland\goodbreak
\and
California Institute of Technology, Pasadena, California, U.S.A.\goodbreak
\and
Centre for Theoretical Cosmology, DAMTP, University of Cambridge, Wilberforce Road, Cambridge CB3 0WA, U.K.\goodbreak
\and
Centro de Estudios de F\'{i}sica del Cosmos de Arag\'{o}n (CEFCA), Plaza San Juan, 1, planta 2, E-44001, Teruel, Spain\goodbreak
\and
Computational Cosmology Center, Lawrence Berkeley National Laboratory, Berkeley, California, U.S.A.\goodbreak
\and
Consejo Superior de Investigaciones Cient\'{\i}ficas (CSIC), Madrid, Spain\goodbreak
\and
DSM/Irfu/SPP, CEA-Saclay, F-91191 Gif-sur-Yvette Cedex, France\goodbreak
\and
DTU Space, National Space Institute, Technical University of Denmark, Elektrovej 327, DK-2800 Kgs. Lyngby, Denmark\goodbreak
\and
D\'{e}partement de Physique Th\'{e}orique, Universit\'{e} de Gen\`{e}ve, 24, Quai E. Ansermet,1211 Gen\`{e}ve 4, Switzerland\goodbreak
\and
Departamento de F\'{\i}sica, Universidad de Oviedo, Avda. Calvo Sotelo s/n, Oviedo, Spain\goodbreak
\and
Department of Astronomy and Astrophysics, University of Toronto, 50 Saint George Street, Toronto, Ontario, Canada\goodbreak
\and
Department of Astrophysics/IMAPP, Radboud University Nijmegen, P.O. Box 9010, 6500 GL Nijmegen, The Netherlands\goodbreak
\and
Department of Physics \& Astronomy, University of British Columbia, 6224 Agricultural Road, Vancouver, British Columbia, Canada\goodbreak
\and
Department of Physics and Astronomy, Dana and David Dornsife College of Letter, Arts and Sciences, University of Southern California, Los Angeles, CA 90089, U.S.A.\goodbreak
\and
Department of Physics and Astronomy, University College London, London WC1E 6BT, U.K.\goodbreak
\and
Department of Physics, Florida State University, Keen Physics Building, 77 Chieftan Way, Tallahassee, Florida, U.S.A.\goodbreak
\and
Department of Physics, Gustaf H\"{a}llstr\"{o}min katu 2a, University of Helsinki, Helsinki, Finland\goodbreak
\and
Department of Physics, Princeton University, Princeton, New Jersey, U.S.A.\goodbreak
\and
Department of Physics, University of California, Santa Barbara, California, U.S.A.\goodbreak
\and
Department of Physics, University of Illinois at Urbana-Champaign, 1110 West Green Street, Urbana, Illinois, U.S.A.\goodbreak
\and
Dipartimento di Fisica e Astronomia G. Galilei, Universit\`{a} degli Studi di Padova, via Marzolo 8, 35131 Padova, Italy\goodbreak
\and
Dipartimento di Fisica e Scienze della Terra, Universit\`{a} di Ferrara, Via Saragat 1, 44122 Ferrara, Italy\goodbreak
\and
Dipartimento di Fisica, Universit\`{a} La Sapienza, P. le A. Moro 2, Roma, Italy\goodbreak
\and
Dipartimento di Fisica, Universit\`{a} degli Studi di Milano, Via Celoria, 16, Milano, Italy\goodbreak
\and
Dipartimento di Fisica, Universit\`{a} degli Studi di Trieste, via A. Valerio 2, Trieste, Italy\goodbreak
\and
Dipartimento di Fisica, Universit\`{a} di Roma Tor Vergata, Via della Ricerca Scientifica, 1, Roma, Italy\goodbreak
\and
Dipartimento di Matematica, Universit\`{a} di Roma Tor Vergata, Via della Ricerca Scientifica, 1, Roma, Italy\goodbreak
\and
Discovery Center, Niels Bohr Institute, Blegdamsvej 17, Copenhagen, Denmark\goodbreak
\and
Dpto. Astrof\'{i}sica, Universidad de La Laguna (ULL), E-38206 La Laguna, Tenerife, Spain\goodbreak
\and
European Space Agency, ESAC, Planck Science Office, Camino bajo del Castillo, s/n, Urbanizaci\'{o}n Villafranca del Castillo, Villanueva de la Ca\~{n}ada, Madrid, Spain\goodbreak
\and
European Space Agency, ESTEC, Keplerlaan 1, 2201 AZ Noordwijk, The Netherlands\goodbreak
\and
Facolt\`{a} di Ingegneria, Universit\`{a} degli Studi e-Campus, Via Isimbardi 10, Novedrate (CO), 22060, Italy\goodbreak
\and
Gran Sasso Science Institute, INFN, viale F. Crispi 7, 67100 L'Aquila, Italy\goodbreak
\and
HGSFP and University of Heidelberg, Theoretical Physics Department, Philosophenweg 16, 69120, Heidelberg, Germany\goodbreak
\and
Helsinki Institute of Physics, Gustaf H\"{a}llstr\"{o}min katu 2, University of Helsinki, Helsinki, Finland\goodbreak
\and
INAF - Osservatorio Astrofisico di Catania, Via S. Sofia 78, Catania, Italy\goodbreak
\and
INAF - Osservatorio Astronomico di Padova, Vicolo dell'Osservatorio 5, Padova, Italy\goodbreak
\and
INAF - Osservatorio Astronomico di Roma, via di Frascati 33, Monte Porzio Catone, Italy\goodbreak
\and
INAF - Osservatorio Astronomico di Trieste, Via G.B. Tiepolo 11, Trieste, Italy\goodbreak
\and
INAF/IASF Bologna, Via Gobetti 101, Bologna, Italy\goodbreak
\and
INAF/IASF Milano, Via E. Bassini 15, Milano, Italy\goodbreak
\and
INFN, Sezione di Bologna, Via Irnerio 46, I-40126, Bologna, Italy\goodbreak
\and
INFN, Sezione di Roma 1, Universit\`{a} di Roma Sapienza, Piazzale Aldo Moro 2, 00185, Roma, Italy\goodbreak
\and
INFN, Sezione di Roma 2, Universit\`{a} di Roma Tor Vergata, Via della Ricerca Scientifica, 1, Roma, Italy\goodbreak
\and
INFN/National Institute for Nuclear Physics, Via Valerio 2, I-34127 Trieste, Italy\goodbreak
\and
IPAG: Institut de Plan\'{e}tologie et d'Astrophysique de Grenoble, Universit\'{e} Grenoble Alpes, IPAG, F-38000 Grenoble, France, CNRS, IPAG, F-38000 Grenoble, France\goodbreak
\and
IUCAA, Post Bag 4, Ganeshkhind, Pune University Campus, Pune 411 007, India\goodbreak
\and
Imperial College London, Astrophysics group, Blackett Laboratory, Prince Consort Road, London, SW7 2AZ, U.K.\goodbreak
\and
Infrared Processing and Analysis Center, California Institute of Technology, Pasadena, CA 91125, U.S.A.\goodbreak
\and
Institut N\'{e}el, CNRS, Universit\'{e} Joseph Fourier Grenoble I, 25 rue des Martyrs, Grenoble, France\goodbreak
\and
Institut Universitaire de France, 103, bd Saint-Michel, 75005, Paris, France\goodbreak
\and
Institut d'Astrophysique Spatiale, CNRS (UMR8617) Universit\'{e} Paris-Sud 11, B\^{a}timent 121, Orsay, France\goodbreak
\and
Institut d'Astrophysique de Paris, CNRS (UMR7095), 98 bis Boulevard Arago, F-75014, Paris, France\goodbreak
\and
Institute of Astronomy, University of Cambridge, Madingley Road, Cambridge CB3 0HA, U.K.\goodbreak
\and
Institute of Theoretical Astrophysics, University of Oslo, Blindern, Oslo, Norway\goodbreak
\and
Instituto de Astrof\'{\i}sica de Canarias, C/V\'{\i}a L\'{a}ctea s/n, La Laguna, Tenerife, Spain\goodbreak
\and
Instituto de F\'{\i}sica de Cantabria (CSIC-Universidad de Cantabria), Avda. de los Castros s/n, Santander, Spain\goodbreak
\and
Istituto Nazionale di Fisica Nucleare, Sezione di Padova, via Marzolo 8, I-35131 Padova, Italy\goodbreak
\and
Jet Propulsion Laboratory, California Institute of Technology, 4800 Oak Grove Drive, Pasadena, California, U.S.A.\goodbreak
\and
Jodrell Bank Centre for Astrophysics, Alan Turing Building, School of Physics and Astronomy, The University of Manchester, Oxford Road, Manchester, M13 9PL, U.K.\goodbreak
\and
Kavli Institute for Cosmology Cambridge, Madingley Road, Cambridge, CB3 0HA, U.K.\goodbreak
\and
LAL, Universit\'{e} Paris-Sud, CNRS/IN2P3, Orsay, France\goodbreak
\and
LAPTh, Univ. de Savoie, CNRS, B.P.110, Annecy-le-Vieux F-74941, France\goodbreak
\and
LERMA, CNRS, Observatoire de Paris, 61 Avenue de l'Observatoire, Paris, France\goodbreak
\and
Laboratoire AIM, IRFU/Service d'Astrophysique - CEA/DSM - CNRS - Universit\'{e} Paris Diderot, B\^{a}t. 709, CEA-Saclay, F-91191 Gif-sur-Yvette Cedex, France\goodbreak
\and
Laboratoire de Physique Subatomique et Cosmologie, Universit\'{e} Grenoble-Alpes, CNRS/IN2P3, 53, rue des Martyrs, 38026 Grenoble Cedex, France\goodbreak
\and
Laboratoire de Physique Th\'{e}orique, Universit\'{e} Paris-Sud 11 \& CNRS, B\^{a}timent 210, 91405 Orsay, France\goodbreak
\and
Lawrence Berkeley National Laboratory, Berkeley, California, U.S.A.\goodbreak
\and
Lebedev Physical Institute of the Russian Academy of Sciences, Astro Space Centre, 84/32 Profsoyuznaya st., Moscow, GSP-7, 117997, Russia\goodbreak
\and
Max-Planck-Institut f\"{u}r Astrophysik, Karl-Schwarzschild-Str. 1, 85741 Garching, Germany\goodbreak
\and
McGill Physics, Ernest Rutherford Physics Building, McGill University, 3600 rue University, Montr\'{e}al, QC, H3A 2T8, Canada\goodbreak
\and
National University of Ireland, Department of Experimental Physics, Maynooth, Co. Kildare, Ireland\goodbreak
\and
Niels Bohr Institute, Blegdamsvej 17, Copenhagen, Denmark\goodbreak
\and
Optical Science Laboratory, University College London, Gower Street, London, U.K.\goodbreak
\and
SB-ITP-LPPC, EPFL, CH-1015, Lausanne, Switzerland\goodbreak
\and
SISSA, Astrophysics Sector, via Bonomea 265, 34136, Trieste, Italy\goodbreak
\and
School of Physics and Astronomy, Cardiff University, Queens Buildings, The Parade, Cardiff, CF24 3AA, U.K.\goodbreak
\and
School of Physics and Astronomy, University of Nottingham, Nottingham NG7 2RD, U.K.\goodbreak
\and
Sorbonne Universit\'{e}-UPMC, UMR7095, Institut d'Astrophysique de Paris, 98 bis Boulevard Arago, F-75014, Paris, France\goodbreak
\and
Space Research Institute (IKI), Russian Academy of Sciences, Profsoyuznaya Str, 84/32, Moscow, 117997, Russia\goodbreak
\and
Space Sciences Laboratory, University of California, Berkeley, California, U.S.A.\goodbreak
\and
Special Astrophysical Observatory, Russian Academy of Sciences, Nizhnij Arkhyz, Zelenchukskiy region, Karachai-Cherkessian Republic, 369167, Russia\goodbreak
\and
Sub-Department of Astrophysics, University of Oxford, Keble Road, Oxford OX1 3RH, U.K.\goodbreak
\and
Theory Division, PH-TH, CERN, CH-1211, Geneva 23, Switzerland\goodbreak
\and
UPMC Univ Paris 06, UMR7095, 98 bis Boulevard Arago, F-75014, Paris, France\goodbreak
\and
Universit\'{e} de Toulouse, UPS-OMP, IRAP, F-31028 Toulouse cedex 4, France\goodbreak
\and
Universities Space Research Association, Stratospheric Observatory for Infrared Astronomy, MS 232-11, Moffett Field, CA 94035, U.S.A.\goodbreak
\and
University of Granada, Departamento de F\'{\i}sica Te\'{o}rica y del Cosmos, Facultad de Ciencias, Granada, Spain\goodbreak
\and
University of Granada, Instituto Carlos I de F\'{\i}sica Te\'{o}rica y Computacional, Granada, Spain\goodbreak
% ...
}